\documentclass{tcibook}
\usepackage{fancyhea}
\usepackage{work}
\usepackage{bm}       
\usepackage{graphicx}
\usepackage{hyperref}      

\usepackage{chngcntr}
\counterwithout{figure}{chapter}

\usepackage{subfig}
\usepackage[para]{threeparttable}

\newcommand{\nc}{\newcommand}  

\def\ie{{\it i.e.}}

\def\Acknowledgements{\bigskip  \bigskip \begin{center} \begin{large}
             \bf ACKNOWLEDGEMENTS \end{large}\end{center}}

\def\beq{\begin{equation}}
\def\eeq#1{\label{#1}\end{equation}}
\def\eeqn{\end{equation}}

\newenvironment{Eqnarray}%
   {\arraycolsep 0.14em\begin{eqnarray}}{\end{eqnarray}}
\def\beqa{\begin{Eqnarray}}
\def\eeqa#1{\label{#1}\end{Eqnarray}}
\def\eeqan{\end{Eqnarray}}

\nc{\ra}{\rightarrow}  
\nc{\slsh}{\slash\hspace*{-0.22cm}}
\def\Re{{\cal R \mskip-4mu \lower.1ex \hbox{\it e}\,}}
\def\Im{{\cal I \mskip-5mu \lower.1ex \hbox{\it m}\,}}

\nc{\vev}[1]{ \left\langle {#1} \right\rangle }
\nc{\bra}[1]{ \langle {#1} | }
\nc{\ket}[1]{ | {#1} \rangle }
\nc{\fb}{\,{\rm fb}^{-1}}
\nc{\ev}{{\rm eV}}
\nc{\kev}{{\rm keV}}
\nc{\Mev}{{\rm MeV}}
\nc{\gev}{{\rm GeV}}
\nc{\tev}{{\rm TeV}}
\nc{\mev}{{\rm MeV}}

\def\del{\partial}
\def\Dslash{\not{\hbox{\kern-4pt $D$}}}
\def\dslash{\not{\hbox{\kern-2pt $\del$}}}
\def\pslash{\not{\hbox{\kern-2pt $p$}}}
\def\ETmiss{ \not{\hbox{\kern-4pt $E$}}_T }

\def\msb{{\bar{\ssstyle M \kern -1pt S}}}

\begin{document}

\def\bibname{References}

\bibliographystyle{utphys}  

\raggedbottom

\pagenumbering{roman}

\parindent=0pt
\parskip=8pt
\setlength{\evensidemargin}{0pt}
\setlength{\oddsidemargin}{0pt}
\setlength{\marginparsep}{0.0in}
\setlength{\marginparwidth}{0.0in}
\marginparpush=0pt


\pagenumbering{arabic}

\renewcommand{\chapname}{chap:intro_}
\renewcommand{\chapterdir}{.}
\renewcommand{\arraystretch}{1.25}
\addtolength{\arraycolsep}{-3pt}


\chapter*{Indirect Dark Matter Detection\\ CF2 Working Group Summary}

\renewcommand*\thesection{\arabic{section}}

\begin{center}\begin{boldmath}



\begin{center}

\begin{large} {\bf Conveners: J. Buckley, D.F. Cowen, S. Profumo} \end{large}

A. Archer,
M. Cahill-Rowley,
R. Cotta,
S. Digel,
A. Drlica-Wagner,
F. Ferrer,
S. Funk,
J. Hewett,
J. Holder,
B. Humensky,
A. Ismail,
M. Israel,
T. Jeltema,
A. Olinto,
A. Peter,
J. Pretz,
T. Rizzo,
J. Siegal-Gaskins,
A. Smith,
D. Staszak,
J. Vandenbroucke,
M. Wood

\end{center}



\end{boldmath}\end{center}


\def\lsim{\mathrel{\hbox{\rlap{\hbox{\lower4pt\hbox{$\sim$}}}\hbox{$<$}}}}
\def\gsim{\mathrel{\hbox{\rlap{\hbox{\lower4pt\hbox{$\sim$}}}\hbox{$>$}}}}
\def\sigmavnatural{$\langle\sigma\, v\rangle\sim
10^{−26}{\rm cm}^3 {\rm s}^{-1}$}
\def\apj{Astrophysical Journal}
\newcommand\aj{{Astronomical Journal}}%
\newcommand\araa{{Annual Review of Astron and Astrophys}}%
\newcommand\apjl{{Astrophysical Journal, Letters}}%
\newcommand\apjs{{ApJS}}%
\newcommand\ao{{Appl.~Opt.}}%
\newcommand\apss{{Ap\&SS}}%
\newcommand\aap{{Astronomy and Astrophysics}}%
\newcommand\aapr{{A\&A~Rev.}}%
\newcommand\aaps{{A\&AS}}%
\newcommand\azh{{AZh}}%
\newcommand\baas{{BAAS}}%
\newcommand\jrasc{{JRASC}}%
\newcommand\memras{{MmRAS}}%
\newcommand\mnras{{Monthly Notices of the Royal Astronomical Society}}%
\newcommand\pra{{Phys.~Rev.~A}}%
\newcommand\prb{{Phys.~Rev.~B}}%
\newcommand\prc{{Phys.~Rev.~C}}%
\newcommand\prd{{Physics~Review~D}}%
\newcommand\pre{{Phys.~Rev.~E}}%
\newcommand\prl{{Phys.~Rev.~Lett.}}%
\newcommand\pasp{{Publications of the ASP}}%
\newcommand\pasj{{PASJ}}%
\newcommand\qjras{{QJRAS}}%
\newcommand\jcap{{JCAP}}%
\newcommand\skytel{{S\&T}}%
\newcommand\solphys{{Sol.~Phys.}}%
\newcommand\sovast{{Soviet~Ast.}}%
\newcommand\ssr{{Space~Sci.~Rev.}}%
\newcommand\zap{{ZAp}}%
\newcommand\nat{{Nature}}%
\newcommand\iaucirc{{IAU~Circ.}}%
\newcommand\aplett{{Astrophys.~Lett.}}%
\newcommand\apspr{{Astrophys.~Space~Phys.~Res.}}%
\newcommand\bain{{Bull.~Astron.~Inst.~Netherlands}}%
\newcommand\fcp{{Fund.~Cosmic~Phys.}}%
\newcommand\gca{{Geochim.~Cosmochim.~Acta}}%
\newcommand\grl{{Geophys.~Res.~Lett.}}%
\newcommand\jcp{{J.~Chem.~Phys.}}%
\newcommand\jgr{{J.~Geophys.~Res.}}%
\newcommand\jqsrt{{J.~Quant.~Spec.~Radiat.~Transf.}}%
\newcommand\memsai{{Mem.~Soc.~Astron.~Italiana}}%
\newcommand\nphysa{{Nucl.~Phys.~A}}%
\newcommand\physrep{{Phys.~Rep.}}%
\newcommand\physscr{{Phys.~Scr}}%
\newcommand\planss{{Planet.~Space~Sci.}}%
\newcommand\procspie{{Proc.~SPIE}}%

\newcommand{\sigv}[0]{\ensuremath{\langle\sigma v\rangle}}
\newcommand{\sigvm}[0]{\ensuremath{\langle\sigma v\rangle_{\rm{UL}}}}
\newcommand{\tsigv}[0]{\ensuremath{\langle\sigma v\rangle R^{2}}}

\newcommand{\unit}[1]{\ensuremath{\mathrm{\,#1}}}
\newcommand{\TeV}{\unit{TeV}}
\newcommand{\GeV}{\unit{GeV}}
\newcommand{\MeV}{\unit{MeV}}
\newcommand{\degree}{\unit{^{\circ}}}
\newcommand{\cm}{\unit{cm}}
\newcommand{\kpc}{\unit{kpc}}
\newcommand{\second}{\unit{s}}
\newcommand{\photons}{\unit{photons}}
\newcommand{\photon}{\unit{photons}}
\newcommand{\ph}{\unit{photons}}
\newcommand{\Figref}[1]{Figure \ref{figs:#1}}
\newcommand{\Secref}[1]{Section \ref{sec:#1}}
\newcommand{\Subsecref}[1]{Section \ref{subsec:#1}}
\newcommand{\Eqnref}[1]{Eq.~(\ref{eqn:#1})}

\newcommand{\figref}[1]{figure \ref{figs:#1}}
\newcommand{\secref}[1]{section \ref{sec:#1}}
\newcommand{\subsecref}[1]{section \ref{subsec:#1}}
\newcommand{\eqnref}[1]{eq.~(\ref{eqn:#1})}

\def\to{\rightarrow}
\def\Fermi{\,{\it Fermi}}
\def\gr{$\gamma-$ray}
\def\mug{$\mu\,$G}
\def\es{???}

\newcommand\bea{\begin{eqnarray}}
\newcommand\eea{\end{eqnarray}}

\newcommand\eq[1]{eq.~(\ref{eq:#1})}
\newcommand\fig[1]{fig.~(\ref{fig:#1})}

\newcommand\nn{ \notag \\}

\newcommand{\ov}{\overline}
\renewcommand{\to}{\rightarrow}
\renewcommand{\vec}[1]{\mbox{\bm $#1$}}

\newcommand{\diff}{\mathrm{d}}
\newcommand{\mchi}{m_\chi}
\newcommand{\vrel}{v_\mathrm{rel}}
\newcommand{\vrelt}{\bm{v}_\mathrm{rel}}
\newcommand{\frel}{f_\mathrm{rel}}
\newcommand\pvx[2][{}]{P_{\bm{x}} \left(\bm{#2}_{#1} \right)}

\newcommand{\tanb}{\tan\beta}

\newcommand{\wh}{\widehat}

\newcommand{\br}{\langle}

\newcommand{\kt}{\rangle}

\newcommand{\lcdm}{${\rm \Lambda}$CDM}

\newcommand{\nchi}{n_{\chi}}

\newcommand{\rvir}{r_\mathrm{vir}}
\newcommand{\mvir}{M_\mathrm{vir}}
\newcommand{\rhot}{\tilde{\rho}}
\newcommand{\epst}{\tilde{\epsilon}}
\newcommand{\psit}{\tilde{\psi}}


\vspace*{1cm}

\section*{Executive Summary}

\begin{enumerate}

\item Significant progress in theoretical studies has been made to assess the
detectability of Dark Matter (DM) particle models by various experiments.  These studies have been
conducted both in the framework of UV-complete theories, for example
supersymmetry (SUSY), and in the context of effective theories, such as with
higher-dimensional contact operators. Both methods can be employed to relate
accelerator constraints to the detectability of signatures from weakly
interacting massive particles (WIMPs).  SUSY scans including the latest
constraints from the LHC, and cosmological constraints show that the
\emph{Cherenkov Telescope Array} (CTA) can greatly improve the
current experimental sensitivity over the parameter space of many theoretical DM models.
 
\item CTA, with a critical enhancement provided by the U.S., would provide a powerful new
tool for searching for dark matter, covering parameter space not accessible to
other techniques (direct searches, accelerator).  CTA would provide new
information that will help identify the particle nature of the DM (mass,
spectral signatures of the specific annihilation channel) and could provide a
measurement of the halo profile.  Given the importance of indirect detection
(ID) to Dark matter science, DOE and NSF/Physics contribution to indirect
detection experiment, on the level of its commitment to the G2 Direct Detection
experiments, could result in a U.S. dark matter program with a realistic
prospect for both detecting the dark matter in the lab and identifying it on
the sky.

\item After the LHC has covered the {\sl a priori} natural parameter space for
SUSY, it is a priority in this field to further explore the viable theory
parameter space of both supersymmetric models and other extensions to the
Standard Model.  Indirect detection experiments (gamma-ray and neutrino
measurements in particular) can, in some cases, go beyond the limits of the
{\sl Energy Frontier} of terrestrial accelerators.

\item The field of indirect dark matter detection necessitates a coherent
theoretical and observational effort to pinpoint via astrophysical modeling and
measurements the relevant inputs to detection rates, such as the 
density, velocity distribution and substructure content of dark matter halos.  Multi-wavelength
and cosmic-ray searches are also in need of additional observational and
theoretical efforts, especially in the realm of Galactic propagation and energy
losses for charged species. We recommend, on both fronts, a broad community
based effort to improve measurements (e.g., identifying new Dwarf
galaxies and improving the stellar velocity data on important objects;  
performing key multi-wavelength and cosmic-ray measurements to characterize
the magnetic field structure in the Galaxy and determine secondary-to-primary ratios of light cosmic-ray nuclei)
and conduct more realistic simulation studies (with higher resolution N-body 
simulations, and more detailed modeling of baryonic feedback).  Progress in
this field will allow us to both quantify uncertainties, and to converge
on benchmark models for astrophysical inputs and for cosmic-ray
diffusion models, which will allow for a direct comparison of performance and
expected sensitivity for current and future experiments.

\item The Galactic center is a prime target for DM searches.  Even without
rejection of astrophysical backgrounds, observations of the GC provide strong
constraints. For CTA, i.e. the next generation experiment, angular resolution is key to
improving ability to reject astrophysical backgrounds in the GC.  To achieve
good angular resolution, and good sensitivity to southern sources doubling the
size of the proposed southern CTA telescope is key.  Merging telescopes in a
single site, and improving angular resolution is much more than an incremental
improvement.

\item IceCube with the DeepCore/PINGU/MICA enhancement (in-filling the
  instrument to achieve a significantly lower energy threshold) would
  provide the possibility of detecting a smoking-gun signal of DM (a
  high-energy neutrino signal from the Sun) as well as competitive
  limits on the \emph{spin-dependent} SD nuclear recoil cross section
  compared with planned generation-2 (G2) direct detection
  experiments.

\item  Continued operation of {\em Fermi}, HAWC, and VERITAS over the next $\sim$5
years will result in dramatic improvements in sensitivity to dark matter
covering the energy range from GeV up to the unitarity limit of $\sim$100 TeV.  These
observations are projected to provide between a factor of 2 and
order-of-magnitude improvements compared to existing limits over this range.
Collectively, these experiments would provide the best sensitivity to the
northern-hemisphere (``classical'') dwarf galaxies, even after the first (southern)
CTA array comes on line.

\end{enumerate}

\newpage

\tableofcontents
\hrulefill

\vspace*{2cm}

\section{Overview of the CF2 Report}

As part of the Snowmass process, the {\em Cosmic Frontier Indirect-Detection}
subgroup (CF2) has drawn on input from the Cosmic Frontier and the broader
Particle Physics community to produce this document.  The scope of this report
is to identify opportunities for dark matter science through indirect
detection.  Input was solicited at a number of meetings including the FNAL
community planning meeting, the SLAC CF meeting and finally the Snowmass
CSS2013 meeting.  For this document, we have liberally drawn on work completed
by the Dark Matter Complementarity (CF4) subgroup in the preparation of the
Dark Matter Complementarity document \cite{2013arXiv1305.1605B}, which puts
indirect detection into the larger context of Dark Matter science.

The purpose of this report is to give an overview of the primary scientific
drivers for indirect detection searches for dark matter, and to survey current
and planned experiments that have, as a large part of their scientific program,
the goal of searching for indirect (or astrophysical) signatures of dark
matter.  We primarily address existing experiments with a large U.S. role, or
future experiments where a U.S. contribution is sought (see
Table~\ref{table:experiments} for a summary of indirect detection experiments).
We also address the limitations of this technique, and answer the tough
questions posed by the HEP community through the Snowmass process.

The report is structured as follows: in the following \S\ref{sec:intro} we give an
introduction to dark matter science, and discuss the role of indirect detection
in a comprehensive program for Dark Matter.  \S\ref{sec:input} introduces the theoretical inputs needed for Indirect Dark Matter Detection. In \S\ref{sec:particle} we
describe the dependence on indirect detection signals to the
beyond-standard-model particle physics framework, and in \S\ref{sec:astro} to the halo
model profile.  From updated particle physics models (including the latest LHC
constraints) and halo models (derived from N-body simulations and dynamical
modeling of astrophysical data) we show that, even for conservative
assumptions, the dark matter annihilation signal is within reach for
next-generation experiments.  

We then go through a description of the different
indirect-detection methods, including cosmic-ray
antimatter, gamma-rays and high-energy neutrinos in \S\ref{sec:experiments}.  For each discovery channel, we cover the basic theory for the
indirect signal, and discuss current and future experiments quantifying the
dark matter sensitivity for each case. In
\S\ref{sec:cosmicray} we look at the current state of the art for cosmic-ray
antimatter experiments including positron fraction measurements from PAMELA,
{\em Fermi} and AMS and future positron measurements that might be made with
Atmospheric Cherenkov detectors (\S\ref{sec:positron}).  We discuss
astrophysical backgrounds for these experiments, and describe the potential of
future antideuteron experiments for obtaining less ambiguous results.  We also
describe results of ultra-high-energy (UHE) cosmic-ray measurements that
already provide strong constraints on UHE gamma-rays and non-thermal WIMPs (or
WIMPzillas) (\S\ref{sec:uhecr}).  In \S\ref{sec:gamma} we cover gamma-ray
experiments, demonstrating the potential for existing experiments ({\em Fermi},
VERITAS and H.E.S.S.) in the coming years. \S\ref{sec:futuregamma} we outline the status of future gamma-ray experiments. \S\ref{sec:decay} briefly describes the case of indirect detection of decaying dark matter and the constraints on that scenario from gamma-ray observations.  In \S\ref{sec:neutrino} we then describe
results from current neutrino experiments like Super-K and IceCube, showing
that current limits on high energy neutrinos from dark matter collecting and
annihilating in the sun give the best constraints on the spin-dependent
cross-section (including direct detection experiments).  Predictions for future
enhancements of IceCube (in particular the PINGU upgrade) or Hyper-K point to
the potential of these instruments to make further progress in dark matter
constraints, as well as offering a unique avenue for discovery (\S\ref{sec:futurenu}).    We consider other electromagnetic signatures
of Dark Matter annihilation in galaxies, including radio synchrotron or
inverse-Compton radiation from secondary electrons as well as unique hard X-ray
signatures of non-WIMP dark matter, e.g., sterile neutrinos in
\S\ref{sec:multi}.  Table \ref{table:experiments} schematically summarizes current and planned experiments relevant for indirect detection of dark matter.

A number of the primary experimental techniques for
indirect detection as well as key enabling technologies are described in
\S\ref{sec:technique}. We address the Snowmass ``tough questions'' pertinent to this subgroup in \S\ref{sec:toughquestions}, and briefly comment on the issue of complementarity among dark matter detection in \S\ref{sec:complementarity}. Finally, we conclude in \S\ref{sec:conclusions} by
identifying important opportunities for constraining, detecting or identifying
dark matter that are only possible with indirect detection methods.

\begin{table}[h!]
\caption{Current and planned indirect detection experiments.}
\begin{minipage}{\columnwidth}
\begin{tabular}{|p{0.75 in}|p{0.7 in}|p{0.7 in}|p{2.4 in}|p{1.2 in}|} \hline
{\bf Current:} & & & & \\ \hline
Experiment & Target & Location & Major Support &Comments \\ \hline
AMS & $e^+/e^-$, anti-nuclei & ISS & NASA & Magnet Spectrometer, Running \\ \hline
Fermi & Photons, $e^+/e^-$  & Satellite & NASA, DOE & Pair Telescope and Calorimeter, Running\\ \hline
HESS & Photons, $e^-$ & Namibia & German BMBF, Max Planck Society, French Ministry for Research, CNRS-IN2P3, UK PPARC, South Africa  & Atmospheric Cherenkov Telescope (ACT), Running \\ \hline
IceCube/ DeepCore & Neutrinos & Antarctica & NSF, 
Australia, Belgium, Canada, Germany, Japan, New Zealand, Sweden) & Ice Cherenkov, Running \\ \hline
MAGIC & Photons, $e^+/e^-$ & La Palma & German BMBF and MPG, INFN, WSwiss SNF, Spanish MICINN, CPAN, Bulgarian NSF, Academy of Finland, DFG, Polish MNiSzW  & ACT, Running \\ \hline
PAMELA &  $e^+/e^-$ & Satellite & & \\ \hline
VERITAS & Photons, $e^+/e^-$ & Arizona, USA & DOE, NSF, SAO & ACT, Running \\ \hline\hline
{\bf Planned:} & & & & \\ \hline
CALET & $e^+/e^-$ & ISS & Japan JAXA, Italy ASI, NASA & Calorimeter \\ \hline
CTA & Photons & ground-based (TBD) & {\footnotesize International (MinCyT, CNEA, CONICET, CNRS-INSU, CNRS-IN2P3, Irfu-CEA, ANR, MPI, BMBF, DESY, Helmholtz Association, MIUR, NOVA, NWO, Poland, MICINN, CDTI, CPAN, Swedish Research Council, Royal Swedish Academy of Sciences, SNSF, Durham UK,  NSF, DOE}  & ACT \\ \hline
GAMMA-400 & Photons & Satellite & Russian Space Agency, Russian Academy of Sciences, INFN & Pair Telescope \\ \hline
GAPS & Anti-deuterons & Balloon (LDB) & NASA, JAXA  & TOF, X-ray and Pion detection \\ \hline
HAWC & Photons, $e^+/e^-$ & Sierra Negra & NSF/DOE, CONACyT in Mexico & Water Cherenkov, Air Shower Surface Array \\ \hline
PINGU & Neutrinos & Antarctica & NSF, International Partners & Ice Cherenkov\\ \hline
\end{tabular}
\end{minipage}
\vspace{0.2cm}
\label{table:experiments}
\end{table}

\newpage

\section{Introduction to Dark Matter Indirect Detection}
\label{sec:intro}
\begin{figure}[tbh]
\begin{center}
\includegraphics[width=0.7\hsize]{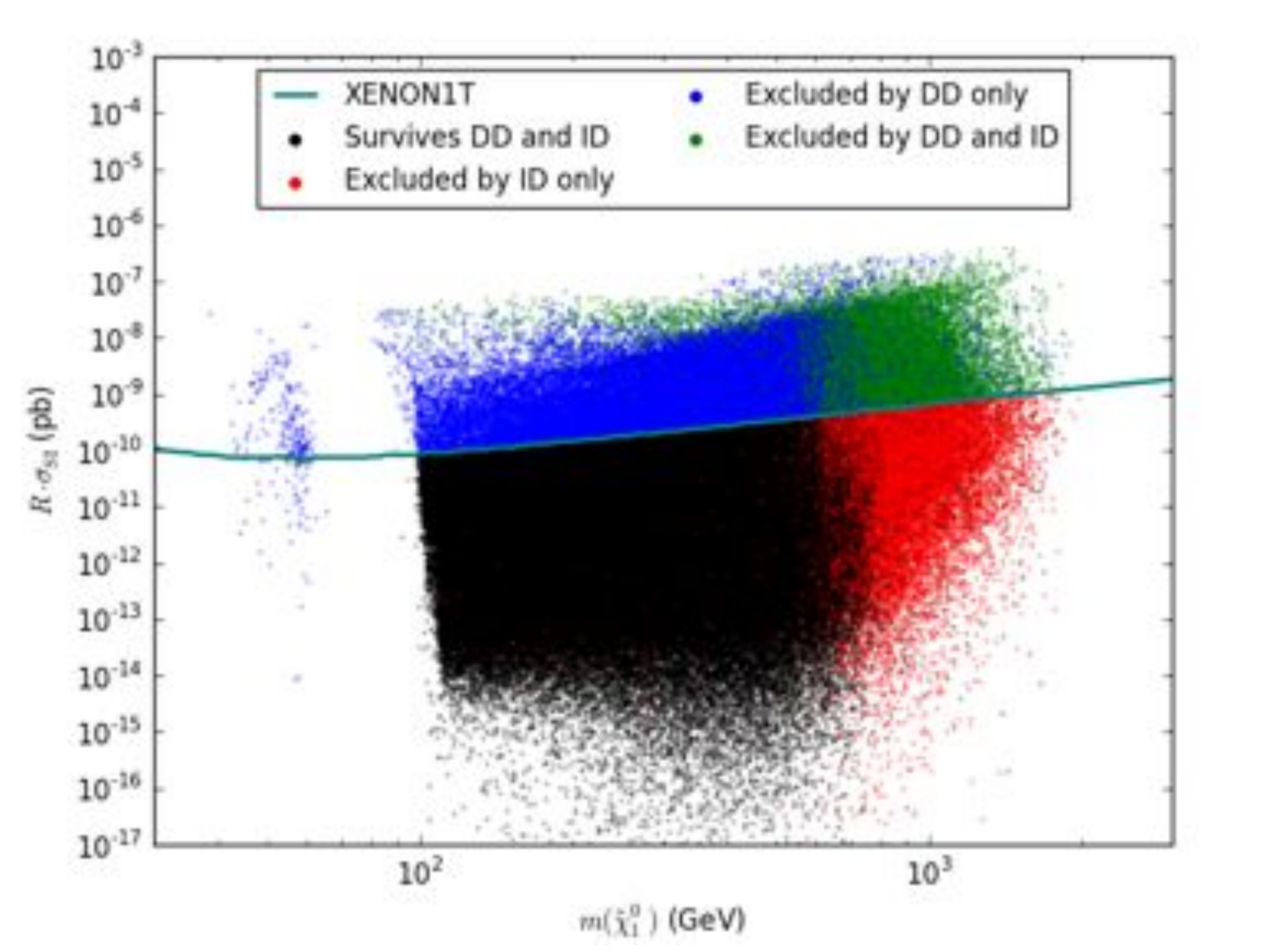}
\caption{An example of the interplay between Direct and Indirect Detection (DD and ID, respectively) for a large collection of supersymmetric dark matter models.}
\end{center}
\end{figure}

A large body of astrophysical evidence points to the fact that the dominant
gravitational mass in the universe is in the form of non-baryonic dark matter.
A 100 GeV thermal relic with weak scale interactions (WIMP) could provide the
cold dark matter needed to explain the observed structure in the universe, and
naturally provide the matter density derived by CMB measurements
\cite{wmap,planck}.
DM has not been directly detected in the laboratory yet, but its gravitational
effects have been observed on spatial scales ranging from the inner kpc
of galaxies out to Mpc in clusters of galaxies, as well as on cosmological
scales. 
Observations of separate distributions of the baryonic and
gravitational mass in galaxy clusters indicate that the DM is likely composed
of particles with a low interaction cross section with ordinary matter.

Numerous well-motivated particle physics models predict new particles in the
100 GeV to TeV scale. Often, these new particles are involved in solving the
so-called hierarchy problem, associated with explaining the large hierarchy
between the Planck and the electroweak scale, and with softening the associated
fine-tuning problem.  
Supersymmetry (SUSY), a theory that provides an elegant solution to the
quadratic divergence of the Higgs mass, also provides a natural candidate for
the DM, the {\emph{neutralino}} or lightest mass eigenstate resulting from a
mixture of the neutral bino, wino and Higgsino \cite{Jungman:1995df}.
Remarkably, a weakly interacting particle with a mass of $m_\chi\sim$100 GeV,
if stable, would have a relic density close to the observed total matter
density inferred from the CMB measurements.  This follows from the fact that
weakly interacting particles fall out of equilibrium earlier in the expansion
history of the universe, avoiding Boltzmann suppression $\sim \exp(-m_\chi
c^2/kT)$, making weakly interacting massive particles (WIMPs) a natural
candidate for the DM.  This feat (sometimes dubbed the ``WIMP miracle'') also
sets the scale for the annihilation rate of dark matter today, under certain
assumptions.  The fact that such a particle interacted with ordinary matter in
the early universe means that such interactions should still take place in the
present epoch. 

In regions of high DM density the annihilation (or decay) of WIMPs into
Standard Model particles including gamma rays, neutrinos, electrons and
positrons, protons and antiprotons, and even antideuterons.  Annihilation
through various channels could produce distinctive signatures such as high
energy neutrinos from the sun, or gamma-rays from the center of galaxies
including our own Milky Way.  Almost any annihilation channel will eventually
produce gamma rays either through pion production (for hadronic channels), or
via final state bremsstrahlung and inverse Compton from leptonic channels.
Moreover, the spectrum from annihilation would be universal (barring the effect
of energy losses and diffusion on the secondary electrons and positrons), with
the same distinctive spectral shape detected in every DM halo. Unlike direct
detection, the detected signal could potentially provide strong constraints on
the particle mass, and help to identify the particle through the different
kinematic signatures predicted for different annihilation channels.

\begin{figure}[tbh]
   \begin{center}
      \includegraphics[width=0.48\hsize]{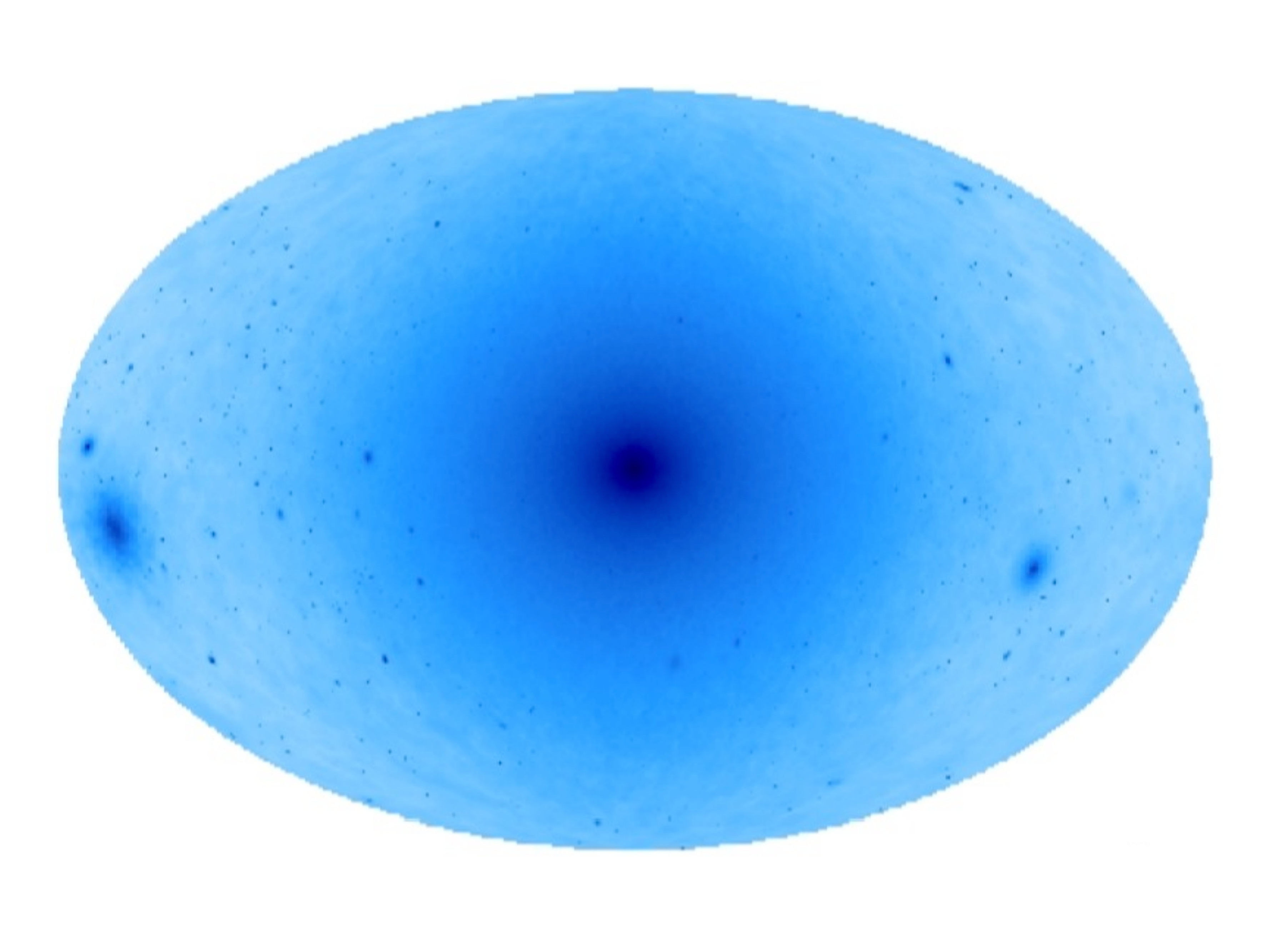}
      \includegraphics[width=0.48\hsize]{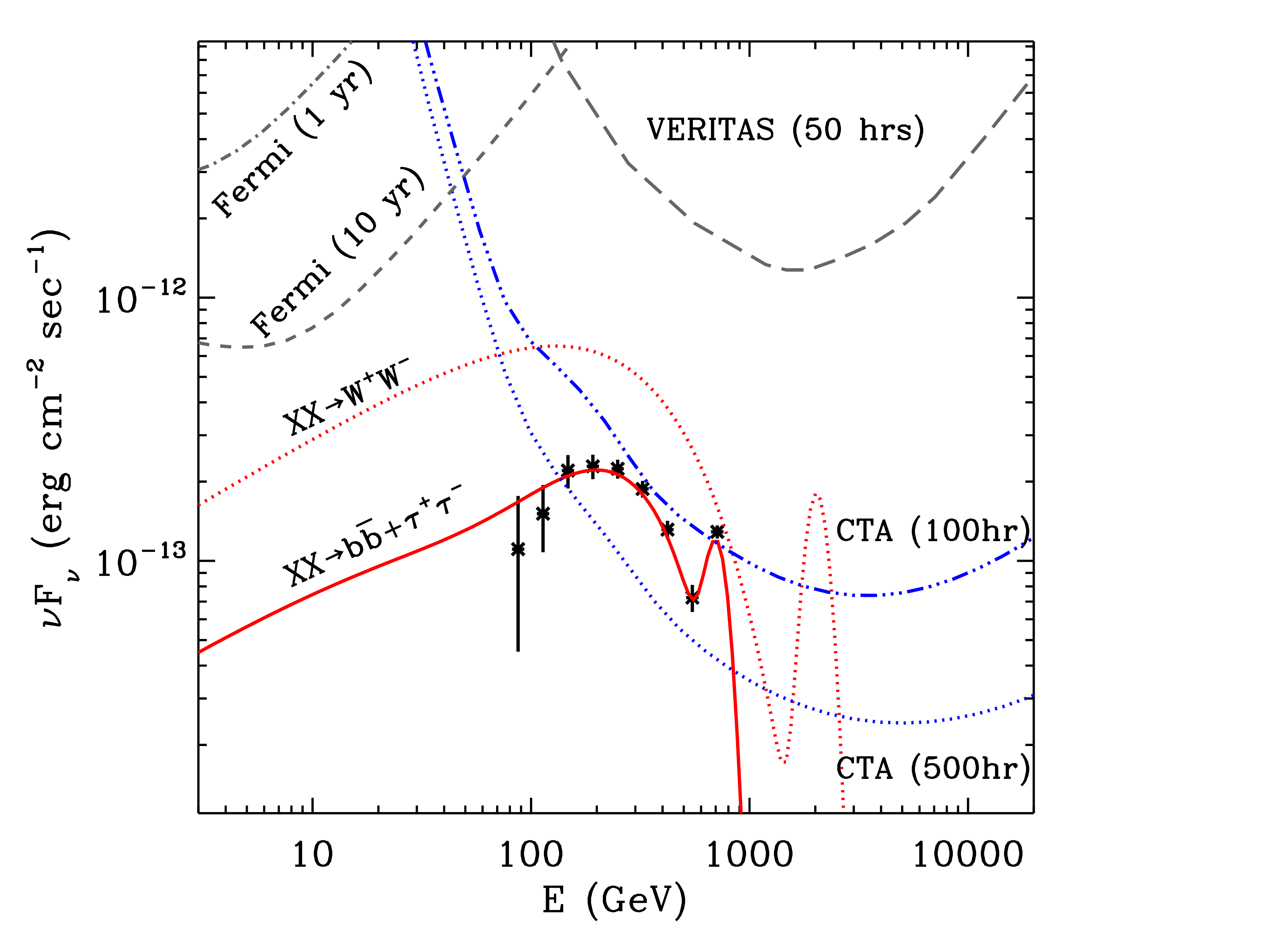}
      \caption{{\it Left:} Angular distribution of the gamma-ray signal from DM annihilation (Galactic coordinates)
	calculated from the line-of-sight integral ($J$-value) of a 
        Milky-Way-sized halo as generated by the VL Lactea
        N-body simulation.  {\it Right:} Simulated gamma-ray spectra as 
        they would be measured using an augmented-CTA instrument.
        The two models correspond to two (arbitrary) annihilation
        channels assuming SUSY model points 
         with an annihilation cross section is close
        to the natural value (with a Sommerfeld boost for the high-mass case)
         and assuming an NFW halo profile for the GC.   Sensitivity curves 
	are calculated for a southern hemisphere instrument with VERITAS
        or H.E.S.S.-like sensitivity, together with the sensitivity curve of the
	augmented CTA instrument.}
\end{center}
\end{figure}

Several tantalizing observations of cosmic and gamma-ray fluxes have been
linked to the possible signals from annihilation or decay of DM particles.  The
bright 511-keV line emission from the bulge of the Galaxy detected by the SPI
spectrometer on the INTEGRAL satellite~\cite{Knodlseder:2003sv}, the excesses
of microwaves and gamma-rays in the inner Galaxy revealed by the WMAP and {\em Fermi}
satellites~\cite{Su:2010qj}, the evidence for a $130$ GeV spectral line in the
{\em Fermi} data~\cite{Bringmann:2012vr,Weniger:2012tx}, or the rise in the positron
fraction above $10$ GeV observed by PAMELA and
AMS-02~\cite{Adriani:2008zr,Aguilar:2013qda}, have been attributed to the
physics associated with
DM~\cite{Boehm:2003bt,Hooper:2003sh,Hooper:2013rwa,Hooper:2011ti,Abazajian:2012pn}.
Currently a dark matter interpretation for these signals is far from clear,
given limited statistics (for the {\em Fermi} line) or large systematics or
astrophysical backgrounds (for the positron, and 511-keV emission).

For almost all of these signals, the flux exceeds the predictions of the most
generic, conservative models, requiring a {\em boost} in either the present-day
cross-section for particle interactions, or in the astrophysical factor, e.g.,
a steepening of the halo profile or a modification of usual assumptions about
dark matter halo substructure.  Some of these scenarios require boost factors
that, while not the most conservative, are also not ruled out by observations.
Indeed, these tentative associations of signals with DM have motivated numerous
important theoretical studies that quantify the possible boosts.  For example,
since the fluxes from DM annihilations depend on the square of the density,
some astrophysical processes (for example the steepening of the halo profile
from the adiabatic growth of the central massive black hole) could change the
inner profile near the Galactic center, so strong boosts near the galactic
center are possible.  Another boost could result if the present day
annihilation cross-section is substantially larger than the cross section at
decoupling, in particular, if there is a strong dependence of cross section on
the relative velocity $v$ between dark matter particles.  Since the DM
particles were relativistic at decoupling, but are moving at non-relativistic
speeds in the halo of the present-day Galaxy, a {\em p-wave} velocity dependent
flux, $\sigma v \sim b v^2$, can result in the right relic density for a light
MeV dark matter particle that annihilates into non-relativistic $e^\pm$ pairs
with $\sigma v \sim 10^{-5}$ pb giving a possible, if not completely natural,
explanation for the observed 511 keV emission from the
bulge~\cite{Boehm:2002yz,Boehm:2003hm}.

Similarly, for the positron excess, one requires significant boosts compared
with the most generic expectations.  In fact, barring the presence of a nearby
DM clump~\cite{Hooper:2008kv,Pieri:2009je}, the thermal cross-section falls
short by about two orders of magnitude to explain the AMS-02 data. Here, the
correct relic abundance and a larger annihilation cross-section can be
reconciled if the annihilation follows $\sigma v \propto 1/v$. This so-called
{\em Sommerfeld enhancement} results from the exchange of new light particles,
and is at the basis for the DM explanation of the rising positron fraction at
GeV
energies~\cite{Hisano:2004ds,arkanihamed2009,Pospelov:2008jd,Cirelli:2007xd}.
These scenarios face stringent constraints from anti-proton, gamma-ray and
synchrotron data, and there is some tension with the higher DM density in the
galactic center predicted by the steep NFW and Einasto profiles favored by
N-body
simulations~\cite{Donato:2008jk,Bertone:2008xr,Bergstrom:2008ag,Cirelli:2009dv,DeSimone:2013fia}.

Most recently, the possible observation of a gamma-ray line in the {\em Fermi}
data from the GC region has sparked considerable interest \cite{wenegerline}.  Since
the excess is offset from the center, there is no prior hypothesis for the
line energy and the line-to-continuum ratio 
exceeds {\it a-priori} expectations, it is difficult to assign an a-posteriori
probability that accurately accounts for trials factors.  Moreover, the {\em Fermi}
team has argued that the line profile is too narrow to match the instrument
energy resolution reducing the inferred significance.  Additional
data is needed to elevate this hint of a signal to a detection;
 a significant increase in the {\em Fermi} GC exposure together with
{\em Fermi} and observations with the new H.E.S.S.-II 
low-threshold ground-based observatory.   Other hints of excesses at $\sim$10 GeV in the Galactic halo \cite{hoopergood11}, or even in the stacked {\em Fermi} dwarf
analysis \cite{fermidwarf} continue to motivate more careful
studies of systematics, and theoretical exceptions.  Many of these results 
are of marginal statistical significance $<$3 sigma, much like hints of
signals seen in accelerator experiments that eventually fade with increased
statistics.  

\section{Theoretical Input to Indirect Detection Signals}\label{sec:input}

Ultimately, the goal of indirect detection experiments is to provide a
measurement that helps to determine both the nature of dark matter and the
distribution of dark matter on the sky.  
But to estimate the signals in different ID detectors, we must make reasonable
assumptions about just such unknown parameters.  For these estimates, we must 
consider specific {\sl benchmark models} for the particle physics,
and must use empirical fits to the best
(gravitational) measurements of the distribution of dark matter in halos.
While it is important to maintain a somewhat agnostic attitude about the exact
nature of beyond-the-standard model physics that might result in a viable
dark-matter candidate, it is important to make generic predictions about the
signals, and to consider some specific well-motivated scenarios to provide
benchmarks by which different ID techniques can be compared, and through which
ID constraints can be compared to accelerator and direct-detection data.  Here
we discuss the particle physics models that might provide a framework for dark
matter, and summarize our understanding about the phase-space distribution of
dark matter particles in halos.

In full generality, the indirect detection signal flux resulting 
from the annihilation of a WIMP with a mass $m_\chi$ to gamma-rays (or
neutrinos)
an be cast
as the product of a particle physics term $\mathcal{P}$ and an astrophysical ter $J$ 
\begin{equation}
\frac{d\phi}{dE\, d\Omega}= 
{\mathcal{P}}{J}.
\end{equation} 
Since the self-annihilation rate is proportional to the square of the
density, the astrophysical term
can be derived by integrating the square of the halo density profile $\rho(r)$ along a line-of-sight
making an angle of $\psi$ with respect to the source direction:
\begin{equation}
 J(\psi)=\int_{\rm l.o.s} \rho^2(r)dl(\psi).
\label{eq:jfactor}
\end{equation}
This line-of-sight integral requires a simple geometrical relationship
between the line-of-sight distance $l$, the radial distance from the center of
the halo $r$ for a particular viewing angle $\psi$.   

The particle physics term can be written 
\begin{equation}
\mathcal{P}=\sum_i {\langle\sigma v\rangle_i\over
 M_\chi^2}\frac{dN_{\gamma,i}}{dE}
\end{equation}
The sum over the index $i$
represents the different possible secondary production mechanisms, and
annihilation channels.
  
To calculate the total signal, it is necessary to convolve the $J$-factor with
the angular distribution with the point-spread-function of the instrument,
convolve the spectrum with the energy resolution function and then to integrate
over solid angle and energy to derive the total flux.   Since the halo may
appear to be quite extended, the cutoff in the solid angle integral is
typically determined so that the ratio of the annihilation signal strength to
the integrated diffuse background is maximized.

While the annihilation rate depends on the density profile, understanding 
the full {\it phase-space density} (including the velocity distribution) is
important for deriving a self-consistent halo model, consistent with dynamical
measurements (e.g., stellar velocities and rotation curves).  Moreover, in some
cases the cross-section may depend on velocity, demanding more care in 
specifying the full distribution function for the dark matter.

We define the phase-space density of dark matter as:
\begin{eqnarray}
  dN = f(\mathbf{x},\mathbf{v},t)d^3\mathbf{x}d^3\mathbf{v},
\end{eqnarray}
such that the number of particles within a phase-space element
$d^3\mathbf{x}d^3\mathbf{v}$ centered on coordinates
$(\mathbf{x},\mathbf{v})$ is $dN$.  The phase-space density of dark
matter affects indirect detection in the following ways:

(1) If the phase-space density is separable,
$f(\mathbf{x},\mathbf{v}) = n(\mathbf{x})f_v(\mathbf{v})$, and dark
matter is self-annihilating, then the annihilation rate of dark matter
in a halo is simply 
\begin{eqnarray} 
\Gamma \propto \langle \sigma_A
  v_{rel} \rangle \int d^3\mathbf{x} n^2(\mathbf{x}), 
\end{eqnarray}
where $\langle \sigma_A v_{rel} \rangle$ is the velocity-averaged
annihilation cross section, and where the average is taken over the
relative velocities of particles with respect to each other.  
In cases
in which the phase-space density is not separable and the cross
section is velocity-dependent, it is not possible to simplify the
expression for the collision term in the 
Boltzmann equation and this must be evaluated by preforming the integral
over the full phase-space distribution function.
However, we
typically assume a velocity-independent (s-wave) annihilation cross
section, so we can estimate the annihilation rate in
the above form.  However, one note of caution is that in simulated
halos (see below), the phase-space density is in general \emph{not}
separable, so to calculate, e.g., p-wave-type cross sections, one
formally must calculate the collision term in the Boltzmann equation
instead of using the above simplified formula.

In contrast to the case for dark matter annihilation, for
 dark matter decay, the signal is proportional
to the line of sight integral of the density (not the density squared), giving
slightly less sensitivity to the detailed halo model.

For annihilation to charged particles, the situation is a bit different since
diffusion and energy loss also affect the observed signal (see \cite{cowsik2013,cowsik2010,ginzburg64}
for details of the calculation).  One convolves the source function
for annihilation with the Green's function for cosmic-ray propagation
including diffusion, and energy loss from inverse-Compton and synchrotron
radiation for electrons (or losses from ionization and nuclear
interactions for protons and deuterons). For electrons, energy losses are
dominated by synchrotron and inverse Compton losses, both of which
have the same dependence dependence on energy (for Compton scattering in
the Thomson regime) and can be combined to give
$dE/dt=-b\,E^2$.  Combining this with diffusion with diffusion constant
$\kappa$, the transport equation can be solved to give the Green's function
which gives the intensity of positrons seen at $\vec{x}$=0, time $t$ and energy
$E$ given an impulse at position $\vec{x}$ and energy $E^\prime$:
\begin{equation}
G(x,t,E,E^\prime)=(4\pi\kappa t)^{-3/2}(1-b\,E\,t)^2 \exp\left(-{x^2\over
4\kappa t}-t/\tau\right)\delta\left(E_0-\frac{D}{-1-b\,E\,t} \right)
\end{equation}
One then folds this with the source function that gives the annihilation rate
\begin{equation}
Q(E,\vec{x})=4\pi\frac{dN_{e^+}}{dE\,d^3x\,dt}=\sum_i {\langle\sigma
v\rangle_i\over M_\chi^2}\frac{dN_{e^+,i}}{dE}\times \rho(r)^2
\end{equation}
and integrates over time (assuming steady state emission) and over the
spatial distribution of dark matter to give the observed, nearly isotropic,
electron spectrum:
\begin{equation}
\frac{dN_{e^+}}{dE_{e^+}}=\int dt \int d^3x \int dE^\prime G(x,t,E,E^\prime) Q(E^\prime,\vec{x})
\end{equation}
This can be accomplished either analytically,
or by using numerical codes like GALPROP that make an effort to include
more detailed information\cite{ms1998}

For cosmic ray electrons, regardless of the source spectrum, there will be a
sharp cutoff in the observed spectrum at energy $E_c\approx 4\kappa\over
b\,d^2$ where $d$ is the distance to the source.  Plugging in typical numbers
for the interstellar magnetic field, the range of 1 TeV electrons is limited to
about $d\sim1~kpc$.   Thus, at high energies, the observed electron or positron
spectrum is dominated by the local dark matter (or cosmic-ray background)
distribution.

\subsection{Particle Physics}

\label{sec:particle}

While a plethora of particle physics models have been envisioned to
provide viable DM candidates, a special category of DM particles
stands out as especially well motivated: that of weakly interacting
massive particles, {\it i.e.} WIMPs. WIMPs have both phenomenological and
theoretical appeal, in that on the one hand a weak-scale mass and
pair-annihilation cross section naturally produces a thermal relic
density in accord with the observed dark matter density, and on the
other hand WIMP candidates are ubiquitous in extensions to the
Standard Model of particle physics where new particles exist at the
electroweak scale.

One such extension that has attracted the interest of
the theory community, for many reasons unrelated to the problem of featuring a
DM candidate, is supersymmetry. 
However, the
recent LHC results have put stringent
limits on some of the particles predicted by weak-scale supersymmetry. For
example, direct searches for strongly interacting squarks and gluinos imply
that the mass of those particles
, if they exist, must
be in the TeV range, with some (although not
dramatic) implications for the phenomenology of supersymmetric DM candidates.
The electroweak sector of the supersymmetric particle content is much less
constrained, at least directly, by LHC results. As part of this sector, the
lightest neutralino -- the supersymmetric WIMP {\em par excellence} -- and its
phenomenology are not severely constrained by LHC results (with the possible
exception, for example, of resonant annihilation channels through the SM Higgs,
or of direct detection cross sections mediated dominantly by SM Higgs
exchange).

Another framework that produces a natural WIMP dark matter candidate
is that of universal extra dimensions, or UED \cite{ued,
uedreview}. In UED, all Standard Model fields propagate in a 5th
compactified extra dimension, giving rise, in the four-dimensional
theory, to a tower of Kaluza-Klein states. Momentum conservation in
the extra dimension leads to a discrete symmetry, in the
four-dimensional picture, known as Kaluza-Klein parity, which
corresponds to the parity assignments $(-1)^N$, with $N$ the
Kaluza-Klein level. As a result, the lightest Kaluza-Klein level-1
particle is stable. For the Higgs mass recently pinpointed at the LHC,
and for minimal boundary conditions, this lightest Kaluza-Klein
particle is the first excitation of the hypercharge gauge boson,
$B^{(1)}$.

In the following sections we limit our detailed estimates to a single benchmark
model corresponding to neutralino dark matter.  However, we expect to obtain
similar results for any thermal WIMP.  In general, any thermal relic must have
interacted with ordinary matter in the universe, and must interact with such
matter in the present epoch, ultimately producing gamma-rays, and in many cases
producing cosmic-ray  electron-positron pairs, protons and antiprotons and
neutrinos.  When one subjects the predictions to cosmological constraints, the
level of the allowed cross section for different models must hover around the
generic total annihilation cross section of \sigmavnatural.  Thus the estimated
sensitivity of experiments to WIMP dark matter is generic.  
Typically, the natural range of dark matter masses is expected to extend from a
few GeV to a few TeV.  However, this argument based on {\it naturalness} in
supersymmetry, the upper limit on generic WIMP masses is significantly higher.
Ultimately, the upper bound on the mass of a thermal relic is set by unitarity
at 120 TeV \cite{griestkamion90} or even higher, if one takes into account
co-annihilations (changing the relic abundance) or a strong velocity dependence
in the cross section \cite{profumo2005}.

\subsubsection{Expected Annihilation Cross-Sections for SUSY WIMPs}

One of the main motivations for R-parity conserving supersymmetry
(SUSY) is the prediction that the lightest SUSY particle (LSP) is
stable and may be identified as a candidate dark matter (DM) particle
if it is both electrically neutral and colorless. Quite commonly the
LSP corresponds to the lightest neutralino, $\chi_1^0$, a linear
combination of the neutral higgsinos, wino and bino, and this be will
assumed in the discussion that follows. While DM searches are
superficially focused on the nature of the LSP, the properties of all
the other super-particles as well as those of the extended SUSY Higgs
sector also come into play. Thus it is impossible to completely
separate searches for DM, \ie, the LSP, from searches for and the
examination of the rest of the SUSY spectrum. However, even in the
simplest SUSY scenario, the MSSM, the number of free parameters
($\sim$ 100) is too large to study in all generality.  The traditional
approach is to assume the existence of some high-scale theory with
only a few parameters (such as mSUGRA\cite {SUSYrefs}) from which all
the properties of the sparticles at the TeV scale can be determined
and studied in detail. While such an approach is quite
valuable~\cite{Cohen:2013kna}, these scenarios are somewhat
phenomenologically limiting and are under increasing tension with a
wide range of experimental data including, in some cases, the $\sim
126$ GeV mass of the recently discovered Higgs boson{\cite
  {ATLASH,CMSH}}.  One way to circumvent such limitations is to
examine instead the far more general 19-parameter
pMSSM{\cite{Djouadi:1998di}}.  We adopt this approach for one of our
benchmarks in the sensitivity estimates presented below.

\subsubsection{Generic WIMPs and contact operator constraints}

A model-independent approach to WIMP phenomenology consists in making
the assumption that the particles mediating dark matter's interactions
with the Standard Model are very heavy, leading to a description in
terms of contact interactions in the context of an effective field
theory (EFT). Depending on the relevant process, and on the spin and
nature (e.g.  Dirac or Majorana) of the DM candidate, it is possible
to write down a complete set of relevant operators of a given
dimension.
Much like Fermi's theory of the weak interaction
the {\emph{strength}} of each operator is expressed in terms of
a quantity $\Lambda$ with the dimensions of mass. Constraints can then
be obtained from a variety of experimental results as a function of
$\Lambda$. A key advantage of this approach is that it captures the
phenomenology of a wide variety of dark matter models (although
naturally not all) and that it is especially simple to compare various
different detection methods.

\subsubsection{Indirect Detection of Non-WIMP DM models}

WIMPs do not nearly exhaust the huge variety of known particle physics
candidates for dark matter. Many {\emph{non-WIMP}}
DM models can be tested by the same experiments relevant for indirect 
WIMP detection (sometimes this is the only practical means of detection).
In other cases, direct detection may provide the only avenue for discovery.
Non-WIMP dark matter is covered at much greater length in the CF3
report.  Here we give a partial list of non-WIMP DM scenarios, and
describe the implications for indirect detection.

\begin{figure}[tbh]
\begin{center}
\includegraphics[width=3.5in]{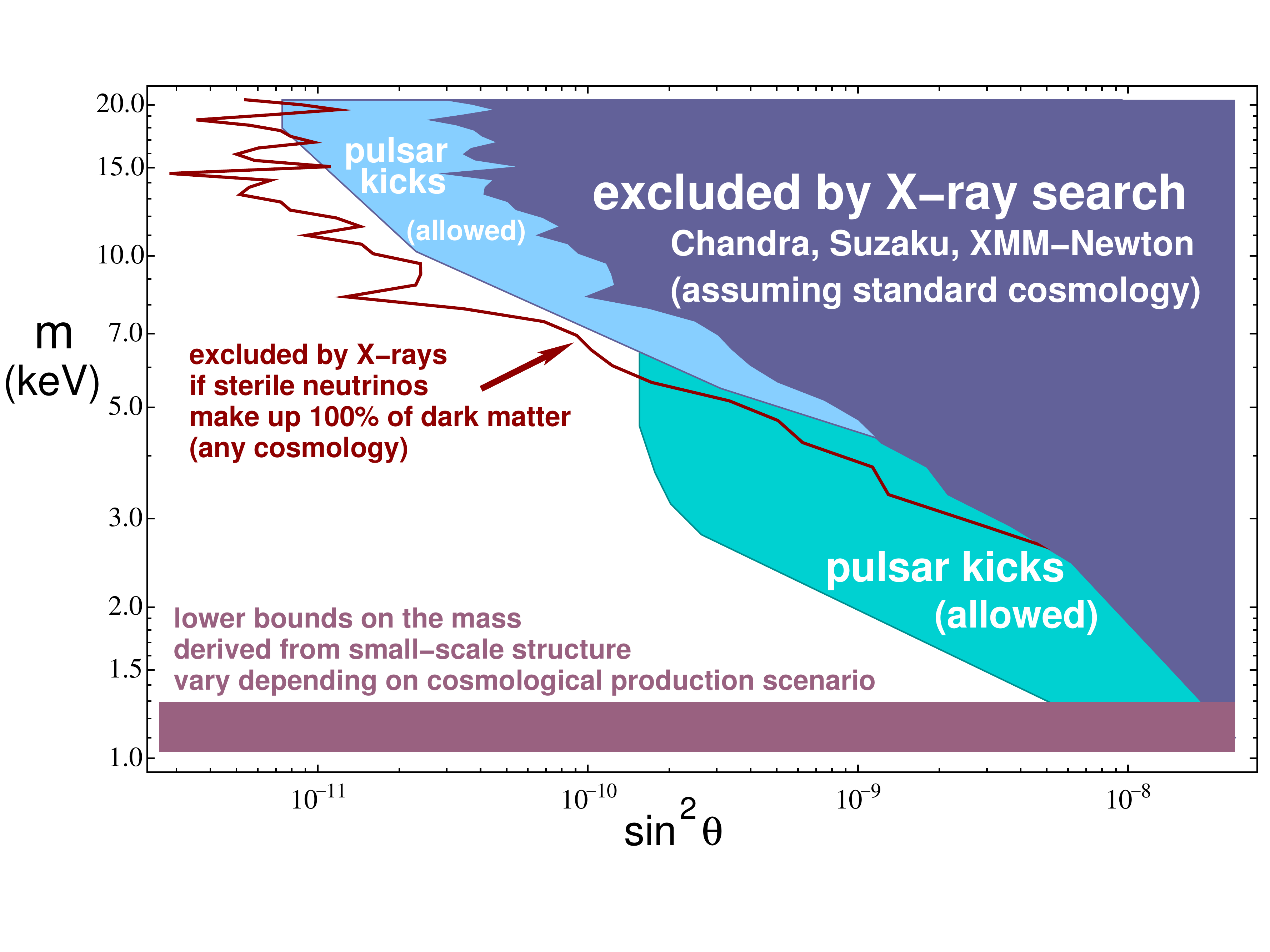}
\end{center}
\caption{Summary of astrophysical constraints on in the $m_{\rm st}-\theta$
plane for sterile neutrino dark matter assuming a standard cosmology below
the temperature when neutrino oscillations occur,
adapted by Kusenko from Ref.\cite{kusenko12}.
}
\label{fig:sterile}
\end{figure}
Sterile neutrinos provide another candidate for non-WIMP dark matter.
\cite{kusenko12}.
Sterile neutrinos are motivated by a variety of arguments besides the
necessity of a DM candidate (for example, the generation of a baryon
asymmetry, and the non-zero value of SM neutrino masses). Sterile
neutrinos generically decay into ordinary SM neutrinos plus a photon,
in a two-body decay that produces nearly monochromatic photons in the
final state, with an energy corresponding to half the sterile neutrino
mass. Preferred sterile neutrino mass ranges makes the relevant
observational window the X-ray to hard X-ray range. Ideal targets are
nearby clusters of galaxies or nearby galaxies such as Andromeda or
local dwarf spheroidal galaxies. Other non-WIMP candidates such as
gravitinos can be unstable, either because of $R$-parity breaking
operators, or because they are not the LSP but they are long-lived
enough to act as dark matter. Gravitino decay can lead, similarly to
the case of sterile neutrinos, to two-body decays featuring a photon
in the final state. The emitted radiation would be monochromatic, producing
a line in the spectrum.
The location of the line in the electromagnetic
spectrum is theoretically rather weakly constrained, and it could
occur anywhere in the X-ray to gamma-ray regime.

Another novel possibility is that some
or all of the cosmological dark matter may be a leftover of a primordial
asymmetry between DM and anti-DM.  The resulting dark matter candidate is
typically referred to as {\emph{asymmetric dark matter}}.
A key benefit of the asymmetric dark
matter scenario is the possibility to relate the DM asymmetry to the
observed baryon asymmetry, via a careful choice of quantum number
attribution to the dark matter candidate.  Although by definition
annihilation does not occur in asymmetric DM models, a residual
population of anti-DM particle could well trigger important indirect
detection signatures. In particular, in certain scenarios asymmetric
dark matter could produce interesting and unique features for example
in antimatter spectra \cite{Chang:2011xn}.

Super-heavy DM candidates could also in principle produce interesting
indirect dark matter signals, especially at ultra-high energy cosmic
ray experiments.
There are a number of proposed scenarios that involve physics beyond
the Standard Model (SM) where the existence of massive particles,
generically called ``$X$'' particles, of mass $M_X \gg 10^{20}$eV,
originate from processes in the early Universe. These scenarios are
called {\em top-down} (TD) scenarios in the context that they may
produce the extreme energy cosmic rays (cosmic rays with energies
around $10^{20}$eV) through the decay of these massive $X$ particles
instead of the conventional {\em bottom-up} scenarios where
acceleration happens in extreme astrophysical environments.

Two general possibilities for the origin of the $X$ particles have
been discussed in the literature: They could be short-lived particles
released in the present Universe from cosmic topological defects such
as cosmic strings, magnetic monopoles, etc., formed in a
symmetry-breaking phase transition in the early
Universe. Alternatively, they could be a metastable (and currently
decaying) particle species with lifetime larger than or of the order
of the age of the Universe. This second scenario arises in models
where the dark matter is a heavy particle such as ``WIMPzillas'' etc. (See
\cite{olinto1} for a list of TD models.)

Since the mass scale $M_X$ of the hypothesized $X$ particle is well
above the energy scale currently available in accelerators, its
primary decay modes are unknown and likely to involve elementary
particles and interactions that belong to physics beyond the
SM. However, irrespective of the primary decay products of the $X$
particle, the observed UHECR particles must eventually result largely
from {\em fragmentation} of the SM quarks and gluons, which are
secondaries to the primary decay products of the $X$ particles, into
hadrons.

The most abundant final observable particle species in the TD scenario
are expected to be photons and neutrinos from the decay of the neutral
and charged pions, respectively, created in the parton fragmentation
process, together with a few percent baryons (nucleons). Therefore,
searches of ultrahigh energy (UHE) photons and neutrinos strongly
constrain TD models.

\subsubsection{Benchmark particle physics models for gamma-rays, neutrinos,
positrons, antiprotons and antideuterons}

The CF2 group recommends that particle physics
benchmarks be adopted to help facilitate the comparison of performance
of different search channels (including but not limited to gamma-rays,
neutrinos, positrons, antiprotons and antideuterons). Such comparison
necessitates not only of particle physics benchmarks, but also of
additional choices for the relevant dark matter density distribution
and for the propagation and energy loss properties of cosmic particles
in the Galaxy, as discussed below.

As a step toward this particle physics benchmarking activity,
some of us have recently begun a detailed study of the indirect detection
signals for neutralino dark matter derived from 
the phenomenological SUSY parameter space (pMSSM).  These pMSSM models are
constrained by observations at the 7 and 8 (and eventually 14) TeV LHC,
supplemented by input from previous DM experiments as well as from precision
electroweak and flavor measurements{\cite {us1,us2a,us2b}}. The pMSSM is the
most general version of the R-parity conserving MSSM when it is
subjected to several experimentally-motivated constraints: ($i$) CP
conservation, ($ii$) Minimal Flavor Violation at the electroweak
scale, ($iii$) degenerate first and second generation sfermion masses,
($iv$) negligible Yukawa couplings and A-terms for the first two
generations. In particular, no assumptions are made about physics at
high scales, e.g., the nature of SUSY breaking, in order to capture
electroweak scale phenomenology for which a UV-complete theory may not
yet exist. Imposing the constraints ($i$)-($iv$) decreases the number
of free parameters in the MSSM at the TeV-scale from 105 to 19 for the
case of a neutralino LSP or 20 when the gravitino mass is included as
an additional parameter when it plays the role of the LSP{\footnote
  {Here we will limit our discussion to the case of neutralino LSPs}}.

Independent of the LSP type, it is not assumed that the thermal
relic density as calculated for the LSP necessarily saturates the
WMAP/Planck value{\cite{Komatsu:2010fb}} to allow for the possibility
of multi-component DM. For example, the axions introduced to solve the
strong CP problem may may make up a substantial amount of DM. The 19
pMSSM parameters and the ranges of values employed in these scans are
listed in Table~\ref{ScanRanges}. To study the pMSSM, many
millions of model points were generated in this space (using
SOFTSUSY{\cite{Allanach:2001kg}} and checking for consistency using
SuSpect{\cite{Djouadi:2002ze}}), each point corresponding to a
specific set of values for these parameters.  These individual models
were then subjected to a large set of collider, flavor, precision
measurement, dark matter and theoretical constraints~\cite{us1}.
Roughly 225k models with neutralino LSPs that survive this initial
selection were used for these studies. 
(A
 model set of similar size has been obtained for the case of gravitino
LSPs.) 
Decay patterns of the SUSY partners and the extended Higgs
sector are calculated using a modified version of
SUSY-HIT{\cite{Djouadi:2006bz}}.

In addition to these two large pMSSM model sets, a
smaller, specialized neutralino LSP set of $\sim$ 10k
{\emph{natural}} models have been generated,
all of which result in the correct Higgs mass $m_h=126\pm 3$ GeV, have an LSP
that {\it does} saturate the WMAP relic density and which produce
values of fine-tuning (FT) better than $1\%$ using the
Barbieri-Giudice measure~\cite{Ellis:1986yg, Barbieri:1987fn}. This
low-FT model set will also be used in the future as part of the
present study.

\begin{table}
\centering
\begin{tabular}{|c|c|} \hline\hline
$m_{\tilde L(e)_{1,2,3}}$ & $100 \gev - 4 \tev$ \\
$m_{\tilde Q(q)_{1,2}}$ & $400 \gev - 4 \tev$ \\
$m_{\tilde Q(q)_{3}}$ &  $200 \gev - 4 \tev$ \\
$|M_1|$ & $50 \gev - 4 \tev$ \\
$|M_2|$ & $100 \gev - 4 \tev$ \\
$|\mu|$ & $100 \gev - 4 \tev$ \\
$M_3$ & $400 \gev - 4 \tev$ \\
$|A_{t,b,\tau}|$ & $0 \gev - 4 \tev$ \\
$M_A$ & $100 \gev - 4 \tev$ \\
$\tan \beta$ & 1 - 60 \\
$m_{3/2}$ & 1 eV$ - 1 \tev$ ($\tilde{G}$ LSP)\\
\hline\hline
\end{tabular}
\caption{Scan ranges for the 19 (20) parameters of the pMSSM with a neutralino (gravitino) LSP. The gravitino mass is scanned with a log prior.
  All other parameters are scanned with flat priors, though we expect this choice to have little qualitative impact on our results~\cite{Cotta:2011ht}.}
\label{ScanRanges}
\end{table}

\begin{figure}[tbh]
\centerline{\includegraphics[width=3.5in]{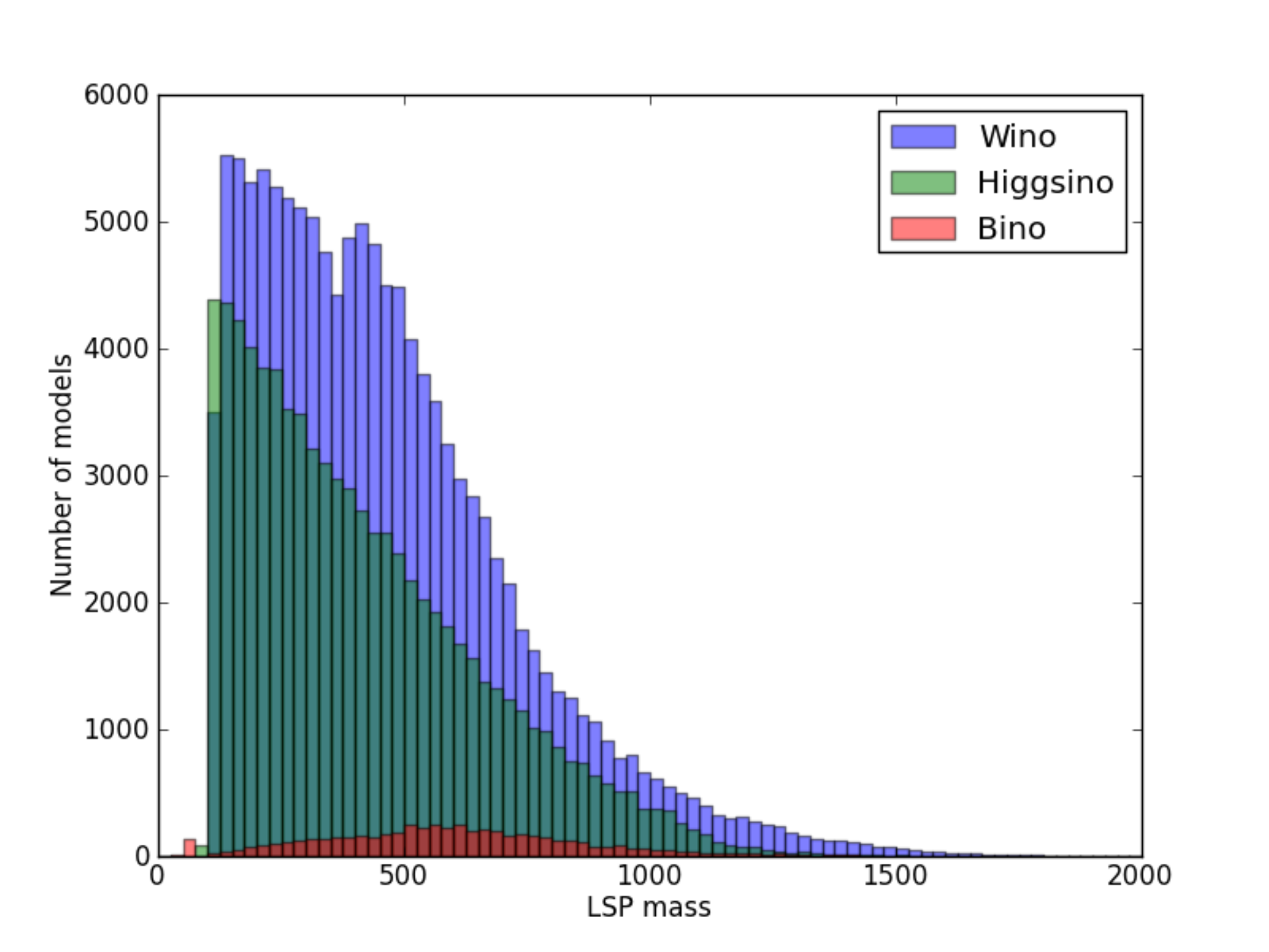}
\hspace{-0.50cm}
\includegraphics[width=3.5in]{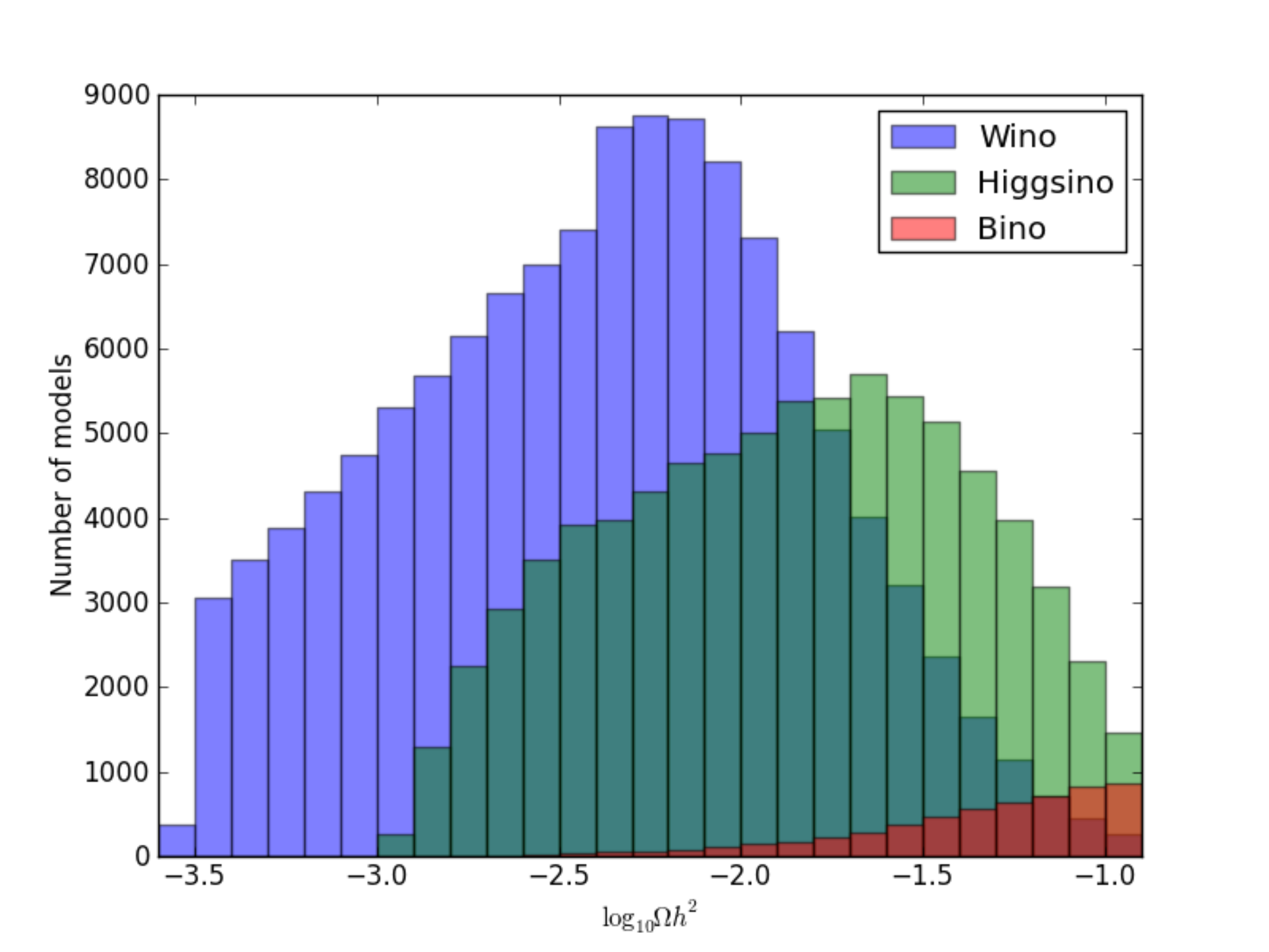}}
\vspace*{-0.10cm}
\caption{Number of models associated with a given LSP mass (left) and predicted relic
  density (right) for neutralino LSPs which are almost pure Wino, Higgsino or Bino-like.}
\label{fig0}
\end{figure}

As a result of the scan ranges going up to masses of 4 TeV, an upper
limit chosen to enable phenomenological studies at the 14 TeV LHC, the
LSPs in these model sets are typically very close to being a pure
electroweak eigenstate since the off-diagonal elements of the chargino
and neutralino mass matrices are at most $\sim M_W$. Figure~\ref{fig0}
shows some of the properties of nearly pure neutralino LSPs (with a
single electroweak eigenstate comprising over 90\% of the mass
eigenstate). In the left panel we see the distribution of the LSP mass
for nearly pure bino, wino, and Higgsino LSPs while in the right-hand
panel we see the corresponding distribution for the predicted LSP
thermal relic density. Note that the masses of all of our neutralino
LSPs lie below $\sim 2$ TeV since the scan ranges only extend to 4 TeV
and the entire SUSY spectrum must be heavier than the LSP and less
than $\sim 4$ TeV (by definition).  The fraction of models that are
nearly pure bino is found to be rather low in this model set since
pure binos generally result in predictions of 
too high of a relic density unless they
co-annihilate with another sparticle, happen to be close to some
($Z,h,A$) funnel region or have a suitable Higgsino admixture.  Note
that only in the rightmost bin of the right panel is the relic density
approximately saturating the WMAP thermal relic value. These LSP
properties will be of particular importance in the discussion that
follows.

Figure~\ref{fig0} shows the histogram of the distribution of  LSP mass (left) and predicted relic
  density (right) for neutralino LSPs which are almost pure Wino, Higgsino or Bino-like.
Essentially every
possible mechanism to obtain (or be below) the WMAP relic density is
seen here: ($i$) The set of models at low masses on the left-hand side
forming `columns' correspond to bino and bino-Higgsino admixtures
surviving due to their proximity to the $Z,h$-funnels. ($ii$) The
bino-Higgsino LSPs saturating the relic density in the upper left are
of the so-called 'well-tempered' variety. ($iii$) the pure{\footnote
  {Here again, `pure' means having an eigenstate fraction $\geq
    90\%$. Points shown as bino-wino, bino-Higgsino, or wino-Higgsino
    mixtures have less than $2\%$ Higgsino, wino, or bino fraction,
    respectively. Mixed points have no more than $90\%$ and no less
    than $2\%$ of each component.}} bino models in the middle top of
the Figure are bino co-annihilators (mostly with sleptons) or are
models near the $A$-funnel region. ($iv$) The green (blue) bands are
pure Higgsino (wino) models that saturate the relic density bound near
$\sim 1(1.7)$ TeV but dominantly appear at far lower relic
densities. Wino-Higgsino hybrids are seen to lie between these two
cases as expected. ($v$) A smattering of other annihilator models are
seen to be loosely distributed in the lower right-hand corner of the
plot.  These pMSSM model sets will be used in subsequent sections, for
calculating anticipated cross sections.

\subsection{Astrophysics}

\label{sec:astro}

To calculate the indirect detection rates one must combine benchmark
models for the particle physics with an assumed form for the halo
profile.  Detailed N-body simulations of structure formation in a
$\Lambda$CDM cosmology provide us with information about the density
profile.  But this work is by no means complete; new measurements of
stellar velocities coupled with more complete simulations of structure
formation including not only CDM but also baryonic physics continue to
change our understanding of halo profiles.  Here we discuss some of
the common parameterizations of the halo profile, that, when
calibrated using astrophysical measurements (e.g., stellar velocities),
provide us with the requisite astrophysical information for the
annihilation rate calculation.

\noindent {\bf Theoretical Input from N-body Simulations:}

\noindent {\it Simulated Halo Density Profiles:}

Most cosmological simulations 
focus on 
dark matter that was non-relativistic at the time of decoupling,
collisionless, and stable (the defining characteristics of WIMPs).
Astronomically, this sort of dark matter is called {\emph{cold dark matter}},
or CDM.  CDM-only simulations now include $>10^{10}$ simulation particles, where
each simulation {\emph{particle}} is actually a tracer for hundreds to thousands
of solar masses and a vast number ($\gsim 10^54$) of dark matter particles
\cite{Diemand:2009bm,Frenk:2012ph,kuhlen2012}.  The two most important results
of these simulations for typical indirect searches are: (1) the gross
properties of the density profile, and (2) the amount and mass function of
substructure.  
As simulations improve, information about the velocity distribution is becoming
available, and may have an impact on the calculation of some indirect signals.

In the absence of baryons, CDM simulations indicate that dark-matter
halos are cuspy ($\rho \sim r^{-1}$ approximately)
\cite{navarro1997,navarro2004,navarro2010}, with their central
concentration dependent on the mass of the halo and the halo formation
time \cite{bullock2001,wechsler2002}.  Small halos tend to form
earlier than big halos, and are much more densely concentrated than
big halos. 
While halos certainly do not show the pronounced disks of baryonic matter
in galaxies and are typically assumed to be spherically symmetric, 
simulations show that
halos are, in fact, somewhat {\emph{triaxial}} or ellipsoidal
with triaxiality increasing in the
center and more pronounced for smaller halos \cite{allgood2006}.
Summarizing, these simulation results imply that annihilation signals should be
centrally concentrated 
and that there may
even be a non-spherically-symmetric shape to the emission, with the emission
shape and normalization 
perhaps somewhat
dependent on the orientation of the halo with
respect to the line of sight.

The major caveat to 
the inferences of these CDM-only simulations is that the centers of halos,
where we expect the highest dark-matter annihilation rates, are also
the places where baryons (here defined as atomic, ionic, and electronic matter)
settle and dominate the gravitational potential well (except in the largest and
smallest dark-matter halos).  There are many arguments as to how baryons should
affect dark-matter halos, but there is as yet no consensus on what actually
happens.  On one hand, when baryons cool, they can drag (by gravity) the dark
matter with them, leading to a steepening of the dark-matter density profile
\cite{blumenthal1984,gnedin2004,tissera2010}.  On the other hand, violent
ejections of gas generated by supernovae or active galactic nuclei reduce the
central potential, and thus the density profile of dark matter
\cite{martizzi2012,governato2012}.  Ref.~\cite{pontzen2012} makes a
heat-engine-type argument that repeated cycles of slow baryon infall and
violent ejection lead to an irreversible trend toward shallower dark-matter
density profiles and central densities.  However, it is not clear if baryon
cooling is slow enough, or gas ejection violent enough, for this argument to
work.  Unfortunately in Galaxy simulations, the effect on the density profile
depends extremely sensitively on the prescription for the sub-grid physics of
star formation/death and the impact of any active galactic nucleus
(AGN) \cite{scannapieco2011}.

\noindent{\it Halo Substructure:}

In addition to informing our knowledge of the gross structure of halos, 
N-body simulations also reveal significant substructure that can have 
a large impact on indirect detection signals.
However,
substructure 
has a different impact on the different types of indirect detection.
For the purpose of the discussion, it is useful to consider the effect
of substructure on (1) photons produced as the primary or secondary products
of dark matter annihilation in the center of our own Galaxy, or more remote
objects (2) cosmic-ray antimatter signals from relatively nearby in our
own Galaxy, and (3) solar neutrino signals produced as the very local halo
is collected in our own Sun.
For gamma-ray observations, the signal is proportional to 
density squared and any clumps along the line of sight can lead to potential
boosts in the signals.  Eventually, for sources with redshifts of Z$\gsim$0.1,
the signals of gamma-rays with energies of $\gsim$ 100 GeV will be affected
by absorption and pair-production as the high energy gamma-rays interact
with diffuse extragalactic background light (EBL).
For cosmic
ray positron experiments, however, only relatively local substructure is important since the range of high energy positrons is limited to several kpc.   

For indirect detection, the amount and impact of substructure is quantified as the {\emph{boost
factor} $B$, which is the ratio of the flux from a halo with all its subhalos
to the flux the halo would have if it only had a smooth, 
virialized
component. 
It is useful to differential substructure into two different categories:
substructure that is self-bound, typically called subhalos; and
substructure that is the un-virialized debris of tidally stripped subhalos.  For
the typical indirect searches, what matters most is the properties of
subhalos---the variations in density induced by tidal debris is not significant
\cite{vogelsberger2009}.  Simulations find that the subhalo mass function is
approximately $dN/dm \sim m^{-1.9}$ down to the resolution limit of the
simulations
 (on the order of 10$^3$ solar masses for present N-body simulations).
It is important to note that the mass function is unconstrained
for masses smaller than this resolution limit, even though substructure
is expected to continue
for up to $\sim 10$ orders of magnitude from this point down to the typical
WIMP free-streaming and kinetic decoupling scale
\cite{springel2008,kuhlen2008}.   
Since the substructure of galaxy clusters includes a number of objects 
more massive than this resolution limit (including Galaxies, Dwarf Galaxies,
and even Galactic substructure), we can, with a high degree of confidence,
attribute a substantial boost factor to these objects.
For smaller subhalos, like those hosting dwarf galaxies, the amount of substructure is probably much lower and there is scant observational data indicating
subhalos.
Current thinking is that
 $B$ is likely of order unity for Milky Way dwarf galaxies, but $B
\sim 100-1000$ for galaxy clusters, depending on the precise form of the
subhalo mass function below simulaion resolution limits
\cite{martinez2009,gao2011,gao2012,pinzke2011}.

Even for CDM-only-models, simulations show that 
substructure can be destroyed by dynamical
friction, especially in the inner parts of galaxies where the
the boost factor is not significant. 
But, once again, the inferences from dark-matter-only simulations may be 
substantially modified by the presence of interacting, baryonic matter.
Specifically, the presence of centrally concentrated baryons greatly
affects tidal stripping of subhalos.  Subhalos passing through disk
galaxies may also 
undergo substantial tidal disruptions.
It is predicted that the amount of
substructure near the centers of halos should be significantly reduced
with respect to simulations that only contain CDM
\cite{gnedin1999,donghia2009b,zolotov2012,brooks2013}.

One significant way that Baryonic matter could change our expectations for
indirect detection signals 
is that dark-matter subhalos could
preferentially be destroyed in the plane of the Milky Way disk
\cite{lake1989,read2008}, boosting the local density as well as
the relative speed between WIMPs and the Sun, and perhaps dramatically
enhancing the solar WIMP signal.  However, more recent, high resolution
simulations do not find evidence of a significant {\emph{dark disk}}.

We note that in the case where dark matter has a strong elastic
self-interaction, the interactions between particles drive the halo toward
isothermality.  This means that there is a core at the center, the velocity
distribution is well modeled by a Maxwell-Boltzmann distribution near the
center, and substructure tends to get erased \cite{rocha2013,vogelsberger2013}.
This implies slightly smaller boost factors due to substructure (although the
subhalo mass function itself is not significantly altered, especially near halo
outskirts), but this effect has not been quantified as yet.  For solar WIMP
searches, this implies that the simplest assumptions made about the local
phase-space density (Maxwellian distribution) are especially reasonable for
strong dark-matter self-interactions.

{\it The Dark Matter Velocity Distribution:}

In the simplest WIMP models the annihilation flux is mostly
{\em s-wave}, and independent of the relative velocity between
annihilating particles, i.e., $\sigma v \sim a$. As mentioned
above, then the flux from DM annihilations varies through the halo tracking the
square of the density, which is expected to be larger at the center of the halo
or in the dwarf Spheroidal satellites of the Milky Way~\cite{Evans:2003sc}.
When the annihilation is velocity dependent, however, the flux is also affected
by the distribution of DM particle velocities, which depends on the location in
the halo. For instance, for the same density profile, a p-wave annihilating DM
gives a shallower distribution of the halo flux~\cite{Ascasibar:2005rw} and
results in a different power spectrum for the diffuse cosmological
signal~\cite{Campbell:2011kf}.  In the case of Sommerfeld enhancement, the flux
is greatly increased towards the center of the halo due to smaller DM
velocities~\cite{Robertson:2009bh}.

The resolution of numerical simulations has increased dramatically in
recent years, but the number of particles is still too small to sample
the possible velocities at each point in the synthetic halo.  However,
there has been some progress in characterizing the velocity
distribution in the solar neighborhood.

To estimate the velocity distribution at the particular location of
the solar system, an average over $100$ randomly distributed sample
spheres centered at $8.5$ kpc was performed in~\cite{Kuhlen:2009vh} to
capture about $10^4$ particles among the billion particles in {\em Via
  Lactea II}, an N-body simulation of a Milky-Way-size galaxy.  The
results provide insight into the non-gaussian shape of the velocity
distribution and on the influence of substructure.

An analysis of a large number of halos from the {\sc Rhapsody} and
{\sc Bolshoi} simulations was used in~\cite{Mao:2012hf} to find an
empirical model to the velocity distribution. Nevertheless, a
subsequent study by the same authors concluded that the results of
different experiments cannot be directly compared, even when
restricting to a single parametrized model of the
VDF~\cite{Mao:2013nda}.

A promising strategy is to use VDF-independent methods, which only compare the
region of $v_\mathrm{min}$ probed by each
experiment~\cite{Frandsen:2011gi,Fox:2010bu}.  We note that for solar WIMP
searches, the expected signal is far less sensitive to sub-parsec variation in
the density and velocity distribution (currently unquantified by simulations)
than direct searches are.
 
\subsubsection{Observational Constraints and Uncertainties in Halo Models}
\label{sec:haloobs}

{\bf Observations of Milky Way dwarf galaxies: } The Milky Way is
known to host at least two dozen satellite galaxies.  About half of
those were discovered in the Sloan Digital Sky Survey using
color-magnitude-diagram filtering techniques
(e.g.,~\cite{willman2005a,belokurov2007,walsh2009}).  The SDSS
galaxies (with the exception of CanVen I \cite{zucker2006}) are
extremely faint, and thus were only detected because they were so
nearby.  This implies that there could be a large number of other
dwarf galaxies in the Milky Way halo that we do not yet have the
sensitivity to find \cite{koposov2008,tollerud2008}. 
Big, deep
wide-field surveys like the Dark Energy Survey and, critically, LSST,
should find more.  In particular,
it is likely that we should find substantially more
galaxies in the southern hemisphere, since this area was not explored
with SDSS.  Even with conservative assumptions about the boost factor,
the resulting stacked Milky Way dwarf analysis (with these additional
objects) will provides annihilation cross
section constraints that are competitive with Galactic Center
constraints, but with no foreground systematics
\cite{GeringerSameth:2011iw,ackermann2011a}.

The predicted gamma-ray signals from Dwarf galaxies are less sensitive
to the detailed distribution of dark matter since, unlike the Galactic Center,
for typical halo profiles most of the emission will be concentrated in an
angular region that is not much larger than the width of the gamma-ray point
spread function.
Still, the dominant uncertainty for the dwarf limits is the dark-matter distribution.
Using 
dynamical modeling, e.g. a Jeans analysis, one may use the line-of-sight velocities of individual
stars in the dwarf galaxies to estimate the mass profile
\cite{strigari2008,martinez2009}.  There are two complicating factors, even if
one assumes spherical symmetry both for the gravitational potential of the
system and the distribution of tracer stars.  First, the enclosed mass of the
system is degenerate with the velocity anisotropy of the stars.  It is
difficult to break that anisotropy without proper motions, and 
those are often
prohibitive to get for individual stars in the dwarfs.  The one place at which
the degeneracy breaks is at the half-light radius.  Thus, we have a good idea
of the average density of dark matter within one (or in the case of multiple
stellar populations, a few \cite{walker2011}) specific fixed radii
 \cite{wolf2010}.
But the detailed density profiles
are not that well constrained
\cite{wolf2012}.  For example,
there is some indication that several of the ``classical''
dwarfs have constant-density cores \cite{walker2011,breddels2013}.  The second complicating
factor is that many of these galaxies simply do not have enough
stars for detailed modeling.   For some of these objects the resulting
density-profile estimates will always be limited by the small number
of measurements, even with the
advent of thirty-meter telescopes and the LSST.  In addition to 
assumptions about isotropy used in modeling, or limitations in the available
data on some objects, other systematics may also enter into the determination
of the details of the halo distribution. 
One such systematic error, in fact a dominant one for many objects, is
the effect of tidal forces on stellar
kinematics, from past interactions of Milky-Way satellites with the 
Milky way disk.   A number of objects now show long tidal tails indicating
close encounters with concentrations of matter in the Milky Way.
The spread in velocities in these tidal tails (sometimes hidden along the 
line of sight direction) 
bias the mass estimates derived using the Jeans equation, which
may only be rigorously applied to equilibrium systems.  We also note that other dynamical
equilibrium methods, distribution function and Schwarzschild modeling, have
also been applied to the Milky Way dwarfs for mass profile inference
\cite{amorisco2011,breddels2013,battaglia2013}.  
While these more detailed analyses of the distribution functions are not
as susceptible to the simplifying assumptions made in a simple Jeans analysis,
it is impossible for any method to fully compensate for the limited number
of velocity measurements or the unknown effects of tidal disruption.

Additional stellar kinematic data will help estimate the magnitude of tidal effects, to determine
whether the galaxies are tidally disturbed to the point that
equilibrium modeling breaks down.  
Another key element of current work has been the development of
equilibrium modeling techniques aimed at properly handling
multiple stellar populations, which has an impact on determining the
appropriate range of models for
both the dark-matter and stellar distributions.
There is an ongoing debate over a single key point, with a significant 
bearing on expected dark matter signals: whether the density profile is
cusped (with an NFW-like profile down to small radii) or cored (with a 
constant density central region 
that may reduce the line of sight integral of $\rho^2(r)$ compared to 
a cusped halo of similar mass.   For most cases, either type of profile
can provide an adequate fit to the data, but for 
several Milky Way dwarf galaxies, there is fairly good evidence
for cored density profiles.
But such cored profiles are not produced by CDM-only N-body simulations,
and the origin of such profiles is still poorly understood.
The two likely candidates
for driving cores are either baryonic feedback or new dark-matter physics.  The
latter is intriguing in the context of this report.

{\bf Observations of the Milky Way:} The brightest DM source for conventional
WIMP dark matter is expected to be the Galactic Center, but 
astrophysical backgrounds and uncertainties in the inner halo profile result in
relatively large uncertainties in the dark matter sensitivity when compared
with other sources, like the dwarf galaxies discussed above.
\cite{kuhlen2008,springel2008a}. 
A large number of astrophysical foregrounds in the direction of the Galactic Center
must either be removed from the region of interest or included (as an
additional contribution to the background) in sensitivity calculations.  Most of these foregrounds
(e.g., emission from supernova remnants, or molecular clouds illuminated by
cosmic-rays) have high energy $E>100 GeV$ spectra steeper than
$E^{-2.2}$, significantly different from the expected spectrum of dark matter
annihilation which, to first order, appears as a hard power-law $\sim E^{-1.5}$
with an exponential cutoff near the neutralino mass.  But there are a few
astrophysical sources with similar spectra, especially at somewhat lower
energies (in the 1-10 GeV regime measured by {\em Fermi}).  At these lower energies,
a particularly troublesome background comes from
pulsars which have energy spectra and Galactic distributions not dissimilar to
dark matter \cite{profumo2008,faucher-giguere2010}. 
Fortunately, all of these foregrounds, whether they are from diffuse cosmic-rays
or point sources, are significantly reduced at the higher energies observed
with ground-based telescopes.  Moreover, ground-based telescopes have very good
angular resolution allowing many of these sources to be identified and removed
from the regions-of-interest for dark matter searches.   We 
discuss in what follows present upper limits on the GC that exclude the astrophysical point
source at the center, and other resolved structures near the center.   Even 
with this relatively minimal subtraction of astrophysical foregrounds, present
dark matter limits fall within an order of magnitude of the natural cross-section.  Thus, while these backgrounds somewhat reduce the sensitivity of GC
observations, the relative strength of the GC signal is high enough to largely
compensate for these backgrounds (at least in determining upper limits).

Implicit in the assumption that the GC is a much stronger source than
other more distant halos is that the 
density profile of dark matter follows a power-law (``cuspy'') profile down to
the inner few kpc.  While this expectation follows from CDM-only N-body
simulations, there is 
little observational evidence 
either directly supporting or contradicting this.  Two factors make it difficult
to obtain good observational constraints on the DM profile in the GC region.
First, since the mass and the
gravitational potential within the solar circle are dominated by baryons 
the impact of the inner DM halo profile on dynamical measurements
(e.g., velocities of molecular clouds or stars) is minimal.  Second, since we are
viewing the GC from within the Milky-Way disk, projection effects and
foregrounds make it particularly difficult to determine the inner rotation curve
for our own Galaxy compared, for example, to other nearby spiral galaxies.
Even local estimates of the dark-matter density at the position of the solar
system have substantial systematic errors.  
Therefore, our understanding of the inner halo-profile of the Milky Way (and
hence our inferences about the brightness of the GC) must be based almost
entirely on N-body simulations.  These simulations show a relatively universal power-law cusp
of dark matter in the centers of dark matter halos on various scales.
While these results are
probably robust down to $\sim$1 kpc scales, to quantify the dark matter
profile in the inner Galaxy one must extrapolate these results
well below the resolution limit of current simulations.  The presence of baryonic matter may lead to large enhancements
or reduction in signals.  Perhaps the only observational reassurance we have in
assuming that the cusp extends down to the inner parsecs of the GC is that
stellar distributions in spiral galaxies also show such power-law cusps, and,
to a crude first-order approximation, these stars are effectively
non-interacting particles that should behave in a similar way to dark matter
particles.  However, other galaxies (including giant elliptical and Dwarf
galaxies) do not show such stellar cusps.
(For a nice review see \cite{strigari2012}.)

{\bf Observations of Galaxy Clusters:} Galaxy clusters are observed to have
density profiles consistent with those predicted in CDM-only simulations,
except within the inner $\sim 50$ kpc.  These inferences are based on X-ray
surface-brightness modeling, stellar kinematics of the brightest cluster
galaxy, and weak and strong lensing
\cite{umetsu2012,coe2012,newman2013a,newman2013b}.  Their concentrations are
approximately what one would predict from CDM simulations.  There are estimates
of the subhalo population (the subhalos hosting observed galaxies)
\cite{natarajan2007,limousin2007}.

The potential multi-wavelength signals from dark matter have been discussed in
the context of galaxy clusters including at gamma-ray
(e.g.,~\cite{2010ApJ...717L..71A, Dugger:2010ys, Huang:2011xr}), radio
\cite{Storm:2012ty, Colafrancesco:2005ji}, and X-ray \cite{Jeltema:2011bd,
Profumo:2008fy} wavelengths.  In particular, clusters are good targets for
searches for secondary inverse-Compton (IC) or synchrotron radiation signals due to the fact that
the high energy electrons and positrons produced in dark matter annihilation or
decay events are typically expected to lose energy through these processes much
faster than the time needed for them to diffuse out of the cluster
\cite{Colafrancesco:2005ji}, making diffusive losses insignificant (compared to
smaller systems like dwarfs).  Gamma-ray observations of cluster also give
comparably strong constraints on the dark matter decay lifetime given the
associated $J$ factors (see e.g. Ref.\cite{Cuesta:2010ex, Dugger:2010ys}).

\subsection{Halo Benchmark Models}

The assumed form for the halo density profiles typically follow from
fits to numerical N-body simulations.  The parameterization of these
fits varies, based, at times, on a parameterization used for stellar
distributions (e.g., the Hernquist models), at other times based on
profiles with desirable analytical features such as a finite total
mass or well-defined phase-space distribution.  The exact
parameterization tends to have an unintended impact on the estimated
flux, since this functional form often is used to extrapolate density
profiles to the inner $\sim$100 parsecs of halos, where stellar
velocity or rotation curve measurements become unreliable, and below
the resolution limit of current N-body simulations.  In order to
address these limitations, we advocate adopting benchmarks that
bracket the range of possibilities, and allow us to marginalize our
estimates over what, for estimates of ID sensitivity, is a nuisance
parameter.  The best choices for such benchmarks are probably a cusped
halo profile (like the NFW profile) and a cored profile.  For
measurements of p-wave, or Sommerfeld-enhanced cross sections, the
velocity distribution should be determined from the density
distribution in a self-consistent manner.  For most of our sensitivity
curves, the NFW profile is the de-facto benchmark distribution.  Here
we describe some other possible choices for the halo profile.

The particular shape of the density profile can be inferred from
observations of stars tracing the gravitational potential of the halo
or from numerical N-body simulations. Although simulations favor a
universal cusped profile, some sub-galactic sized objects are better
described by assuming the presence of a central core. We, hence,
consider both cusped and cored distributions.

\noindent{\it NFW profile}

The Navarro, Frenk, and White (NFW) profile provides a good fit to
DM-only N-body simulations over a wide range of halo
masses~\cite{Navarro:1996gj}.

The NFW profile is given by the equation:
\begin{equation}
\rho_{\rm nfw}(r)=\frac{\rho_0}{(r/r_s)(1+r/r_s)^2}
\end{equation}
where $r_s$ is the scale radius of the halo, $M$ is the total halo mass
and the remaining free parameter, the scale density $\rho_0$ can
also be written in terms of a concentration parameter $c$ defined as the
ratio of the scale virial radius $R_{\rm virial}$ to the scale radius
$r_s$.  The relationship between $\rho_0$ and $c$ can be shown to be:
\begin{equation}
M=4\pi \rho_0 r_s^3\left[\ln(1+c)-{c\over 1+c}\right]
\label{eq:rhonfw}
\end{equation}

To model a dwarf Spheroidal (dSph) satellite of the Milky Way with an NFW
profile, we note that observations are consistent with the known satellites
having a mass of about $10^7 M_\odot$ within their central $300$
pc~\cite{Strigari:2008ib}, while a scale radius $a=0.62$ kpc fits the observed
radial velocity dispersion of the stars~\cite{Evans:2003sc}.  The total mass of
the dSph dark matter halos is difficult to determine, since their extent beyond
the observed stellar distributions is largely unknown~\cite{Walker:2012td}.
Hence, $c=4.8$ in~\eq{rhonfw}, which is substantially lower than the typical
values $c\sim 20$ for halos of this size found in
simulations~\cite{2008MNRAS.391.1940M}.

\subsection{Einasto profile}

In addition to describing the luminosity profiles of early-type
galaxies and bulges and the surface density of hot gas in clusters,
the Einasto profile is as good a fit as the NFW profile, if not
better, to simulated galaxy-sized dark matter halos.

The Einasto density profile is based on the observation that the logarithmic
slope of the density profile, $d\ln\rho/d\ln r$ appears to vary continuously
with radius can be written as 
\begin{equation}
\frac{d\ln\rho}{d\ln r}=-2\left({r\over r_{-2}}\right)^{1/n}
\end{equation}
where $r_{-2}$ is the radius at which the logarithmic slope is $-2$ and $n$ is
a parameter describing the degree of curvature of the distribution.
\begin{equation}
\rho_{\rm einasto}(r)=\rho_s e^{-2n\left[(r/r_s)^{1/n}-1\right]}
\end{equation}
As an example of the parameters for this model, consider the case of the Segue
1 dwarf.  Here the parameters $\rho_s=1.1\times10^8\,M_\odot{\rm kpc}^{-3}$,
$r_s=0.15 kpc$ and $n=3.3$ provide a good fit to the data \cite{versegue}

\subsection{Burkert profile}

The fact that the best NFW fit to the observations of dSphs can be far
less concentrated than expected from simulations suggests that the NFW
profile provides a poor fit to the dynamics of some dSphs. In
particular, detections of distinct stellar sub-populations provide
mass estimates at different radii for Fornax and Sculptor that are
consistent with cored potentials, but largely incompatible with cusped
profiles~\cite{Walker:2011zu,Amorisco:2012rd}.

The Burkert profile is a cored profile that appears to provide a good
fit to the DM distribution in dSph galaxies~\cite{Burkert:1995yz}.
This model was developed by Burkert to capture the observation that
some halo profiles appear to have flat inner density profiles, that role
over into an outer $r^{-3}$.  The Burkert profile is given by the expression
\begin{equation}
\rho_{\rm burkert}(r)=\frac{\rho_0 r_s^3}{(r+r_s)(r_2+r_s^2)}
\label{eq:rhobur}
\end{equation}
where $\rho_0$ is the central density, and $r_s$ is a scale radius.

A scale radius of $r_s=650\mathrm{pc}$ and $\rho_0=1.8\times 10^8
M_\odot$ in~\ref{eq:rhobur}
were found in~\cite{Salucci:2011ee} to fit the kinematics of Draco.

\section{Current and Future Indirect Detection Experiments}\label{sec:experiments}

In this section we overview current experiments with a significant U.S.
involvement, for which indirect detection is a large part of the overall
scientific program.  We do not try to survey all instruments, and omit
a number of important experiments which are primarily led by non-US
international collaborations.   

\subsection{Charged Cosmic-Ray/Antimatter Experiments}

\label{sec:cosmicray}

Dark matter could annihilate through a number of channels (quark-antiquark, W
and Z or heavy leptons) with a similar branching ratio.  These annihilation
channels can ultimately produce cosmic-ray particles including protons and
antiprotons and even deuterons and antideuterons.  Annihilation to lighter
leptons (in particular electrons) is also possible, but will generally be
helicity suppressed unless some other mechanism contributes such as internal
bremsstrahlung, or some new light mediator that could give rise to a Sommerfeld
enhancement.

The transport of cosmic ray particles can be described by a combination of
spatial diffusion, energy gains and losses, losses and gains in the number of a
given particle species due to interactions with the ISM, and escape from the
source region or galaxy.  Since cosmic-rays are charged particles, to a good
approximation, the random magnetic fields in the Galaxy randomize their
directions hiding their directions.  However, a small anisotropy may remain due
to contributions from local sources. 
Unlike
gamma-ray and neutrino observations of dark matter signals that would point
back to the centers of galactic halos or to the sun, the identification of a
dark matter signal from cosmic rays requires the detection of a spectral
feature that stands out against the background.  The best signal to background
ratio is likely to be obtained for cosmic ray antimatter, which is produced in
roughly equal proportions in DM annihilation, but subdominant in the production
at cosmic ray sources such as supernova remnants.

\subsection{Cosmic-Ray Positron Measurements}

\label{sec:positron}

The recent measurement of the cosmic-ray positron fraction
by the AMS-02 magnetic spectrometer
on the International Space Station \cite{Aguilar:2013qda} confirms (with
excellent precision) earlier measurements that showed a positron fraction,
$e^+/(e^+ + e^-)$, increasing with energy from 0.05 at 10 GeV to 0.15 between
200 and 350 GeV.  It is well known that some cosmic-ray positrons must be
created in interstellar space as secondary products of interaction of
cosmic-ray nuclei (mostly protons and helium nuclei) with nuclei of the
interstellar medium (mostly hydrogen and helium).  However, widely used models
of propagation of cosmic rays in the Galaxy\cite{ms1998} predict that such
secondary positrons would give a positron fraction falling in this energy range
from about 0.04 to less than 0.03.  Thus, other explanations for the rising
positron fraction must be considered.  Here we consider several possible
explanations for the positron excess:

\begin{itemize}

\item {\em Dark matter:} 
Perhaps the most interesting explanation for the rising
positron fraction would be that these positrons originate in the decay or
annihilation of dark matter \cite{2009PhRvL.103c1103B,2013arXiv1307.2561H}.
Since such positrons cannot have energy greater than the rest mass of the
dark-matter particles, if the AMS-02 positrons are interpreted as due to dark
matter, those dark-matter particles must have mass greater than 350 GeV.
Furthermore, as pointed out in \S\ref{sec:intro} there must be a large boost
in the dark matter signal caused by an enhancement in the present day
cross-section (e.g., due to a Sommerfeld enhancement), a mechanism to suppress
the large (but unobserved) channel for annihilation to antiprotons
and an enhancement in
the astrophysical factor due, e.g., to a nearby dark matter clump.
Any such boost in the cross-section for annihilation to electrons would produce
an even larger source of electrons and positrons in the galactic center.  These
electrons and positrons would produce a large radio synchrotron and gamma-ray
inverse Compton signal that should be readily detected.  Upper limits on these
signals seem to already strongly contradict a dark matter interpretation for 
the excess.

AMS-02 is capable of identifying cosmic-ray positrons with energy up to $\sim$1
TeV; currently the AMS-02 positron fraction is limited to energies below 350
GeV by the low flux of particles at higher energy.  With several more years of
data, that instrument should be able to extend its determination of the
positron fraction to energies close to 1 TeV.  Despite the difficulties with a
dark matter interpretation discussed above, if the positron fraction should
display an abrupt decrease at some energy, $E$, between 350 GeV and 1 TeV, one
would certainly take a closer look at the possibility of a dark matter source
for the positrons 
since none of the astrophysical explanations for the observed positrons
(summarized below) would produce such an abrupt decrease.   Since such leptonic
annihilation channels must also result in gamma-rays (either as final-state
radiation, internal bremsstrahlung, or inverse-Compton emission) ground-based
gamma-ray observations (with there good sensitivity about a few hundred GeV)
should readily detect the expected gamma-ray signal -- if not with the current
generation experiments, certainly with future gamma-ray instruments.

In any event, lacking an abrupt drop in the positron fraction, it would be very
difficult to attribute the observed positrons to dark matter with any
confidence, since there are several plausible astrophysical mechanisms for
producing them, including primary positrons produced near pulsars, secondary
positrons produced in the normal cosmic-ray sources and accelerated there, and
secondary positrons produced in the interstellar medium with that production
modeled in a manner different from the widely used cosmic-ray propagation
model.

\item {\em Primary positrons from pulsars:} Electrons can be accelerated in a
pulsar magnetosphere and induce an electromagnetic cascade through the
emission, by curvature radiation in the pulsar, strong magnetic field, of
photons above threshold for pair production.  Acceleration of the positrons
thus produced can yield a hard spectrum that could account for the rising
positron fraction.  Many papers have discussed this possible origin of the
positron excess  (e.g.
\cite{2009JCAP...01..025H,2013PhRvD..88b3013C,2013PhRvD..88b3001Y}).

\item {\em Secondary positrons produced in the sources:} In addition to the
secondary positrons produced by cosmic rays interacting in interstellar space,
there are likely to be secondary positrons produced by interactions with
material in the source regions.  These secondary positrons would then
participate in the same acceleration process that produces cosmic-ray protons
and electrons.  The positrons thus accelerated are likely to have a harder
spectrum than the Galactic secondaries calculated by Moskalenko \& Strong and
could thus account for rising positron fraction \cite{blasi09}.  As Blasi stated,
``{\em This effect cannot be avoided though its strength depends on the values of the
environmental parameters during the late stages of evolution of supernova
remnants}.''

\item {\em Modifications of the widely used model of Galactic cosmic-ray
propagation:} 
Four
 approaches, each different from the others, have been
proposed that account for the observed rise in positron fraction above 10 GeV
as being due entirely to secondary positrons produced as the cosmic-ray nuclei
propagate through the interstellar medium.

One approach ignores the details of propagation models and simply assumes
that positrons are produced at the same places as secondary nuclei like boron
(which is the fragmentation product of carbon, oxygen, and other heavier nuclei
interacting with interstellar gas).  From the ratio of production of positrons
to production of boron, one infers an upper limit to the positron fraction.
(It is an upper limit because it ignores energy loss processes suffered by
positrons but not by nuclei.)  The upper limit thus calculated matches the peak
of the positron fraction observed by AMS-02. \cite{katz2010,blum13}

Another approach notes that the Moskalenko \& Strong model of Galactic
cosmic-ray propagation is cylindrically symmetrical, but in fact cosmic-ray
sources are likely to be concentrated in the spiral arms of the Galaxy.  Taking
into account energy loss of primary cosmic-ray electrons from those sources,
while noting that secondary positrons are produced from primary nuclei that are
distributed more uniformly throughout the Galaxy, a model has been developed
that produces a rising positron fraction above 10 GeV \cite{shaviv09}.  (It
must be noted, however, that another calculation of Galactic cosmic-ray
propagation, in which the spiral structure is incorporated, fails to match the
observed positron fraction without invoking some primary positron source.)

A more radical modification of the commonly used Galactic propagation model
invokes the ``Nested Leaky Box'' model \cite{cowsiknested73}.  In this model
boron and other nuclear secondaries are primarily produced in ``cocoons'' around
the sources, from which escape is energy dependent, going as $\sim E^{-0.6}$,
while the positrons, which are produced by interactions of protons and other
nuclei of much higher energy, are mainly produced outside the cocoons, in the
general Galaxy from which escape is energy independent.  In this model, one can
easily reproduce the observed positron fraction over the entire energy range
for which we have observations, from 1 to 350 GeV
\cite{cowsikprd10,cowsikpositrons13}

\item Another possible explanation of the apparent tension between the observed
positron excess and propagation models might lie in the simplifying assumptions
that are currently made about magnetic field structure, diffusion and energy
losses made in cosmic-ray propagation models.  For example, Kistler, Yuksel and
Friedland \cite{kistler12} looked at the implications of the hierarchical structure
of the Milky Way magnetic field on cosmic-ray propagation. They argue that for
positrons, the diffusion approximation to charged-particle propagation is
simply inappropriate.

Although a more detailed discussion lies beyond the scope of this
document, we wish to 
emphasize 
the crucial role played by diffusion
and energy losses uncertainties in cosmic-ray propagation models. We
point the interested reader to, e.g.,
Ref.~\cite{Delahaye:2010ji}. 
The
importance of future cosmic-ray experiments in pinpointing critical
observables such as secondary-to-primary ratios and
unstable-to-stable isotope ratios cannot be overemphasized.

\end{itemize}

\subsubsection{VERITAS Positron and Electron Measurements}

Cosmic ray electrons and positrons provide a unique astrophysical window into
our local Galaxy.  In contrast with hadronic cosmic rays, electrons and
positrons lose energy quickly through IC scattering and synchrotron processes
while propagating in the Galaxy.  These processes effectively place a maximum
propagation distance of $\sim$1 kpc for electrons with $\sim$TeV
energy\cite{kobayashi04}.  Prior to H.E.S.S.\cite{Aharonian:2008aa} all cosmic ray
electron measurements came from balloon-based and satellite-based experiments.
The effective area of IACTs is 5 orders of magnitude larger and thus offers a
method to extend the spectrum to higher energies.  The electron+positron
spectrum is now measured by the {\it Fermi}-LAT\cite{2012PhRvL.108a1103A} from 7
GeV to 1 TeV and by H.E.S.S.\cite{Aharonian:2008aa,hesselectron2} and
MAGIC\cite{magicelectron} from 100 GeV to 5 TeV (a VERITAS result is
forthcoming).  These overlapping results agree within statistical+systematical
uncertainties and give evidence of an additional harder component within the
spectrum that is not accounted for in our standard picture of the local
environment.  This additional piece becomes significant at $\sim$30 GeV and
steepens above $\sim$700 GeV.

IACTs have the ability to extend the positron fraction spectrum to higher
energies.  To measure the positron flux, IACTs can utilize a combination of the
Earth's geomagnetic field with the cosmic ray moon shadow to create a
spectrometer.  This technique was developed by the
Artemis\cite{whipplemoonlight,refArtimis} experiment using data from the
Whipple 10m telescope to search for the anti-proton shadow.  While all IACTs
currently collect data in some moonlight conditions, the positron ratio
measurement is particularly difficult because it requires observations within a
few degrees of bright moon phases (up to $\sim$50$\%$, where the phase
represents the fraction of the moon lit by the sun).  In these conditions the
background rates are very large and the potential to damage the sensitive PMTs
is very real.  To alleviate this, VERITAS has developed filters that cover each
of the four cameras and let in the peak of the Cherenkov spectrum, 250-400~nm,
while rejecting the bulk of the reflected solar spectrum.  MAGIC
\cite{magicpositron}
is also attempting this difficult measurement and estimates that it will
require 50 hours spread over several years to accomplish, assuming that the
missing flux is on the order of a few percent.

\subsubsection{Future Cosmic-Ray Antimatter Measurements}

The cosmic-ray antiproton spectrum has been measured from balloons and
satellites between 60 MeV to 180
GeV~\cite{PhysRevLett.105.121101}. Secondary antiprotons are created
as cosmic rays propagate through the Galaxy, and the flux of these
antiprotons is very sensitive to the propagation processes of cosmic
rays.
Above $\sim 10\ \textrm{GeV}$, the $\bar{p}/p$ ratio from
secondary production is expected to decline. Various new-physics
scenarios can increase the $\bar{p}/p$ ratio above 100 GeV, with
predictions up to $10^{-3}$ for contributions from dark
matter~\cite{2009NuPhB.813....1C} and up to several \% for
contributions from extragalactic sources, such as
antigalaxies~\cite{1984Natur.309...37S}. Antiprotons can also be
produced by primordial black hole
evaporation~\cite{2002A&A...388..676B}. Existing measurements up to
$\sim 180\ \textrm{GeV}$ are consistent with secondary
production. Atmospheric Cherenkov telescopes have the effective area
to make an important contribution in the 200 GeV to few TeV range,
where to date only limits exist.

Electrons and positrons can likewise be produced by secondary
processes during propagation or interaction in SNRs or pulsars, or via
pair production in nearby pulsars. However, they can also be among the
final-state products from annihilation of dark matter particles. The
combined $e^+ + e^-$ spectrum from $7\ \textrm{GeV}$ to $1\
\textrm{TeV}$ has been measured by the {\em Fermi}
satellite~\cite{2009PhRvL.102r1101A}, and up to 6 TeV by
HESS~\cite{Aharonian:2008aa} and MAGIC~\cite{magicpositron}.

Over the next decade, {\em Fermi} and CREST will measure the total $e^+ +
e^-$ spectrum to several TeV~\cite{2008ICRC....2..305S}, and CALET
will measure it to $\sim$20~TeV~\cite{2013AIPC.1516..293R}.  AMS will
extend its positron fraction measurement beyond $350\ \textrm{GeV}$,
measure the total $e^+ + e^-$ spectrum to $1\ \textrm{TeV}$ or beyond,
and provide a precise measurement of antiprotons.  The AMS team estimates that with
an 18-year exposure, the sensitivity to the anti-helium/helium ratio
will reach $10^{-4}$ in the 400 GV to 1 TV range. The sensitivity
improves rapidly with decreasing rigidity, reaching $10^{-9}$ for
energies below 50 GV. AMS will measure the charge-separated $e^+$ and
$e^-$ spectra up to several hundred
GeV.

CTA will complement these satellite- and balloon-based instruments by
providing a high-precision combined $e^+ + e^-$ spectrum from
$\sim$100~GeV to $\sim$100~TeV, well beyond the range feasible for
satellite- or balloon-based experiments due to their limited effective
area. CTA can also provide strong constraints on large-scale
anisotropies (an important signature in distinguishing source models)
in the $e^+ + e^-$ spectrum by comparing flux measurements made in
opposite directions on the sky~\cite{2013ApJ...772...18L}.  CTA will
also measure the charge-separated electron, positron, and antiproton
spectra using the Moon's shadow and the geomagnetic field, a technique
pioneered by ARTEMIS~\cite{2001APh....14..287P} and more recently
pursued by MAGIC~\cite{2009arXiv0907.1026C,magicpositron} and
VERITAS.

The US extension to CTA will is anticipated to be an array of 25-36
mid-size telescopes
(MSTs) with 8$^\circ$~diameter field of view and cameras instrumented
with SiPMs.  The US extension of MSTs provides three important
advantages for these measurements.  First, an increase in the
point-source sensitivity by a factor of 2-3 in the energy range
($\sim$1~TeV) where CTA can contribute best to the positron spectrum
measurement~\cite{2012AIPC.1505..765J}. Second, SiPMs (unlike PMTs)
are not endangered by bright moonlight, so observations can be made
under a wider range of conditions and a greater overall sensitivity
can be achieved.  Third, the improved optical point spread function
and smaller pixel size of the innovative dual-mirror design planned
for the US extension enables improved hadronic/electromagnetic shower
discrimination.  This is important because hadron showers are much
more numerous than electron and positron showers and even a small
fraction of residual background can present a significant challenge to
these measurements.

Instruments on balloons, satellites, and on the ground will provide
important new measurements of charged cosmic particles over the next
decade, extending the surprising discoveries that have generated a
great deal of excitement among particle physicists in the past few
years. These new measurements are essential to distinguishing among
the many competing models. Charged particles are an important
component of indirect dark matter searches, complementing gamma-ray
and neutrino measurements. CTA will contribute unique measurements of
cosmic electrons, positrons, and antiprotons at the highest energies.

\subsubsection{GAPS Anti-Deuteron Search}

\label{sec:antideut}

About a decade ago it was pointed out that antideuterons produced in WIMP-WIMP
annihilations (``primary'' antideuterons) offered a potentially
attractive signature for cold dark matter (CDM) (\cite{donato00} hereafter
DFS). The reason is that the flux of primary antideuterons is fairly flat in
the ~ 0.1 to 1.0 GeV/n energy band, while the ``secondary/tertiary''
antideuterons, those produced in cosmic ray interactions in the interstellar
medium (secondaries) and subsequent reprocessing (tertiaries), have fluxes
that sharply decrease with decreasing energy. The lower antideuteron
background results because of the higher cosmic-ray energy required to create
an antideuteron compared to an antiproton, combined with a cosmic-ray spectrum
steeply falling with energy. In addition, the collision kinematics disfavors
the formation of low-energy antideuterons.

Despite the low astrophysical background and a signature which is rather
generic in many ``beyond-the-Standard-Model'' models, there has been no
dedicated, optimized search for antideuterons. An upper limit on antideuterons
was obtained by the BESS experiment \cite{fuke05}, but it is three orders of
magnitude higher than the interesting range for dark matter searches. The AMS
experiment on the ISS has sensitivity for antideuterons \cite{choutko08},
but other complementary experiments, optimized for making
a clean measurement of antideuterons are being considered.
Just such an optimized search for antideuterons has been
proposed, the General Antiparticle Spectrometer experiment (GAPS)
\cite{hailey09}, and it recently had a successful prototype flight
\cite{mognet13}.

It is important to understand the complementarity of different types of
antideuteron measurements, as well as the complementarity of these measurements
with direct and indirect detection experiments.
There are a number of interesting hints of low-mass neutralinos in a 
number of direct detection experiments (DAMA/LIBRA, CoGent and CDMS).
Since the detected events are very near threshold, at present
it is difficult to make
a convincing case for dark matter, rather than contamination of some
new background.
Because antideuteron searches are very sensitive to SUSY models that provide
low mass neutralino candidates, their overlap with direct detection experiments
is potentially of great utility.
Together with {\em Fermi} observations of dwarf spherical galaxies which
are also 
beginning to place strong limits on low-mass neutralinos (~ few tens of GeV),
antideuteron measurements can provide an important corroborating role in 
the discovery (and identification) of a low-mass dark matter particle.
Antideuteron searches are also
generally sensitive to much higher neutralino masses
than direct detection searches with massive target nuclei, thus also extending 
the reach to higher masses.
AMS and GAPS have mostly complementary kinetic energy ranges, but also some
overlap in the interesting low energy region, which allows the study of both a
large energy range, confirming the potential signals, and the best chance for
controlling the systematic effects. Another very important virtue comes from
the different detection techniques of both experiments. AMS follows the
principle of typical particle physics detectors. Particles are identified by
analyzing the event signatures of different subsequent sub-detectors, also using
a strong magnetic field. The GAPS detector will consist of several planes of
Si(Li) solid state detectors and a surrounding time-of-flight system. The
antideuterons will be slowed down in the Si(Li) material, replace a shell
electron and form an excited exotic atom. The atom will be de-excited by
characteristic X-ray transitions and will end its life by annihilation with the
nucleus producing a characteristic number of protons and pions. 
The approach of using two independent experimental techniques 
is 
important in providing
confidence in any potential primary antideuteron detection.

There are substantial uncertainties in predictions of the
primary antideuteron flux. The dominant uncertainty in the primary flux is due
to propagation uncertainties. During the next years the AMS high precision
observations will tightly constrain the cosmic ray propagation parameters.
However, some degeneracy in the parameters will most likely remain. There is
also an uncertainty due to the halo model employed, but the antideuteron
production is averaged over the halo and over fairly long scale lengths, so
whether or not a cored or un-cored halo profile is used is somewhat reduced in
importance. Recently, increasing attention has been paid to details of the
hadronization process. In the coalescence model, the antineutron and antiproton
combine when their momentum difference is less than a critical value, the
coalescence momentum, $p_0$.  But since $p_0$ is smaller than the QCD phase
transition temperature, there is a strong sensitivity to the hadronization
model.  Recent studies emphasize the need for event-by-event determination of
the production rates and are probing the sensitivities when different
hadronization models are employed \cite{kadastik10,ibarra13,dal12}.
While theoretical progress is made, the best way to pin down
the production uncertainty is unquestionably to make a good accelerator
measurement of $p_0$, given the considerable discrepancy in the various
measurements. One possible source of uncertainty, which only drives up the
expected primary flux, is due to the boost factor. It was previously
fashionable to consider boost factors up to 100 times or more, but recent
simulations seem to have definitively settled this issue, with boost factors in
the range of 1-10 being the largest allowed \cite{lavalle07} . However the
sensitivity curves presented here assume no boosting effect (i.e., boost factor
1), and a factor of 2 or 3 in boost would provide very substantial reach into
discovery space for a number of models presented here.  Similar observations
pertain to the secondary/tertiary background, except that in this case it is
the production uncertainties that dominate \cite{donato08}. Most of the
secondary/tertiary antideuterons are produced in the Galactic disk, so the
propagation is more local and thus less important.  At any rate, for the
antideuteron searches the nominal or optimistic propagation and production
parameters are very promising for dark matter searches. The most pessimistic
numbers would make prospects for detection problematic.

\begin{figure}[tbh]
\begin{center}
\includegraphics[width=0.5\hsize]{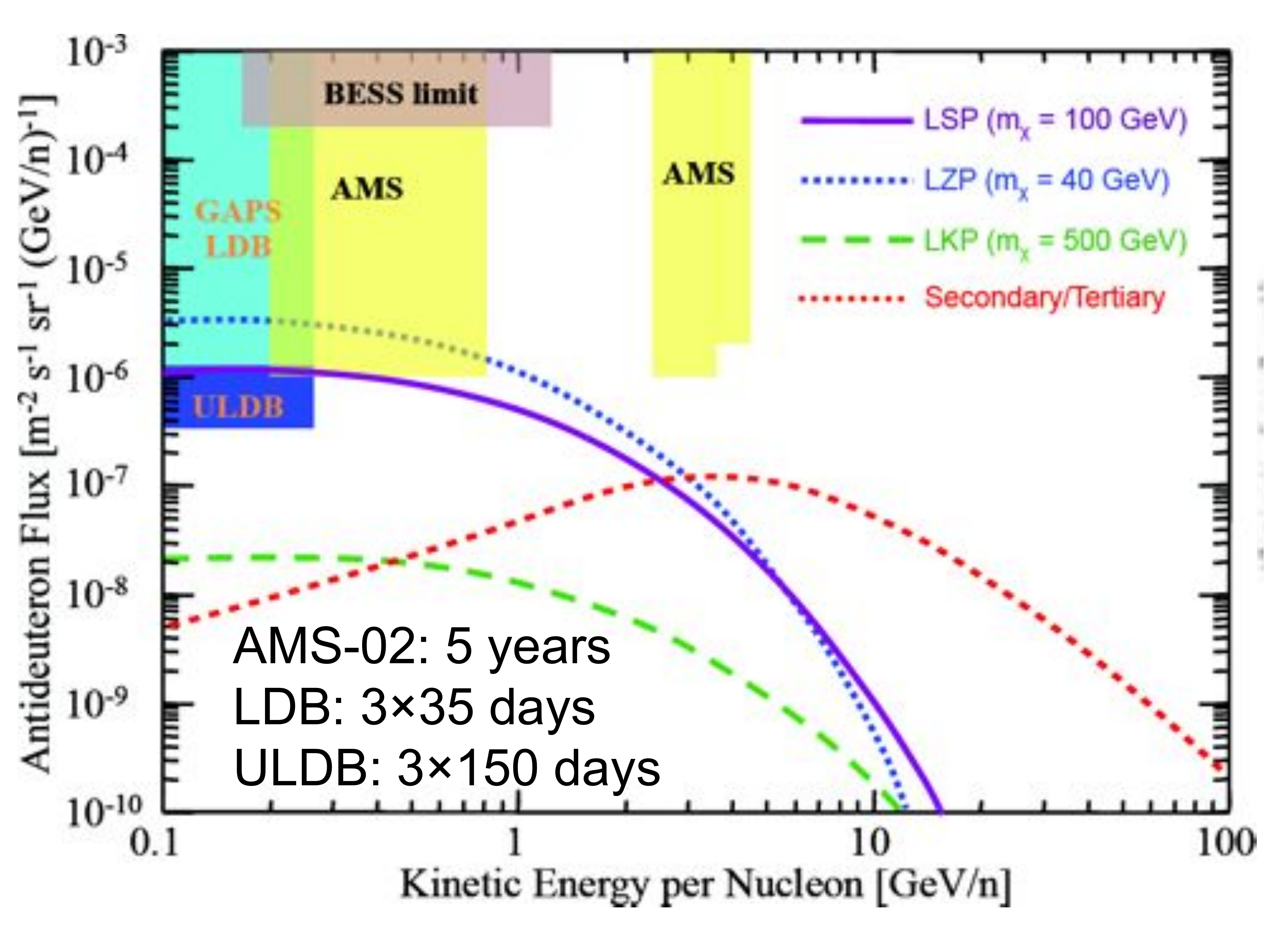}
\caption{LSP, LKP and LZP models along with the antideuteron background
and the sensitivity of GAPS, BESS and AMS (5 years) \cite{gapswhitepaper}.}
\end{center}
\end{figure}

\begin{figure}[tbh]
   \begin{center}
      \includegraphics[width=0.6\hsize]{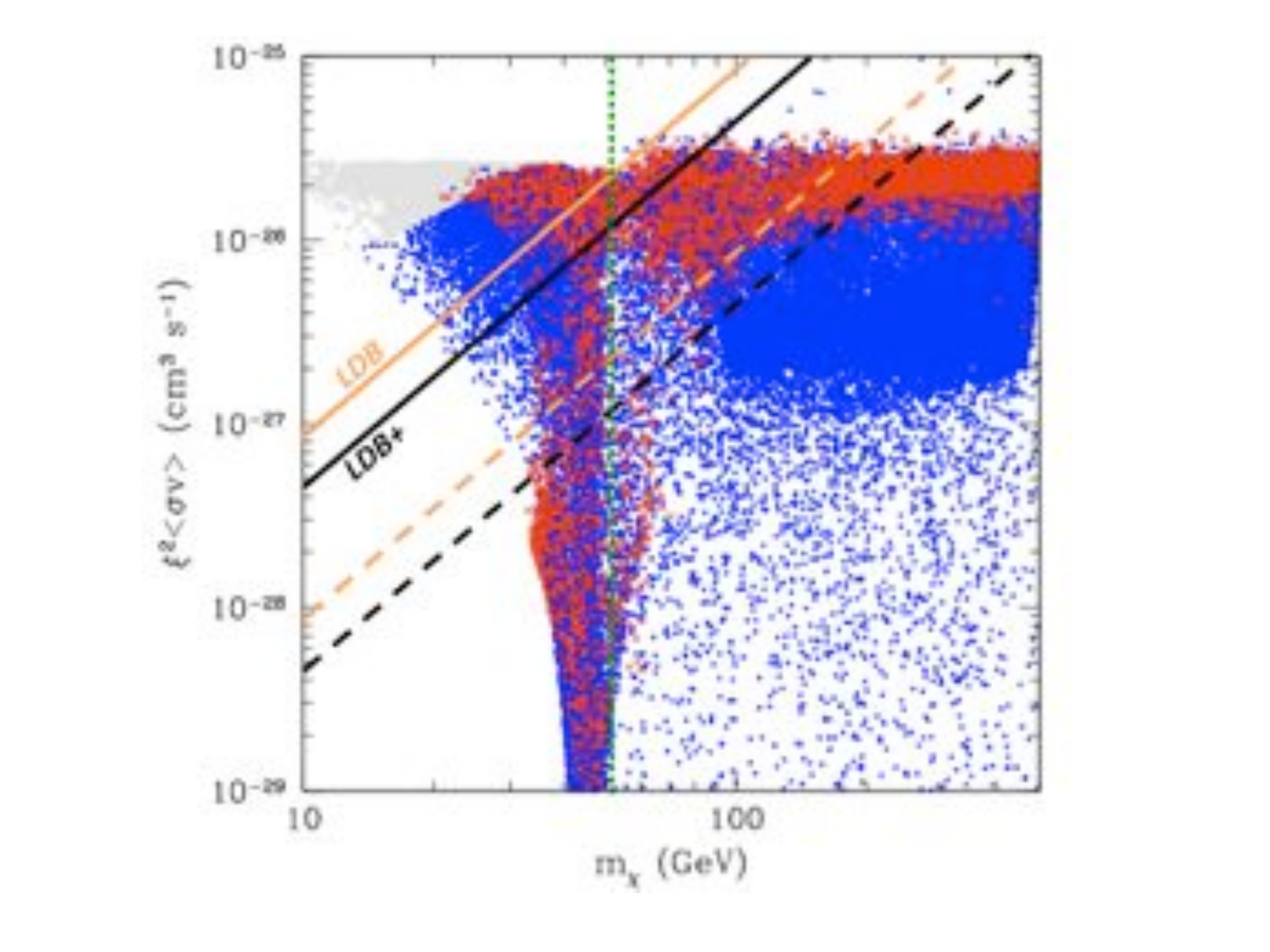}
      \caption{Sensitivity of a dedicated antideuteron experiment (diagonal
lines) for long-duration balloon (LDB) and ultralong duration balloon (ULDB)
flights; the green vertical line separates low mass non-universal gaugino
SUSY models (left) from MSSM models.  Solid (dashed) lines for the sensitivity
correspond to median (high) antideuteron propagation models \cite{gapswhitepaper}.  }
   \label{fig:gapssigmav}
\end{center}
\end{figure}

Figure~\ref{fig:gapssigmav} shows a scan of SUSY models overlaid on the
sensitivity for a dedicated antideuteron search experiment under various
scenarios. 
The experimental curves correspond to a generic GAPS-like experiment for
balloon flights of various duration from a few to 10 months.
This plot illustrates some of the
opportunities as well as challenges of antideuteron searches. An ensemble of
supersymmetric (SUSY) model parameters is shown, all yielding the same
neutralino-annihilation cross-section and mass.
To the right of the vertical
green line are results from a low-energy minimal SUSY model, and to the left
are shown results from a low mass non-universal gaugino model. The latter
admits light neutralinos, which have recently been argued to provide a
candidate for the controversial DAMA/LIBRA, CoGENT and CDMS II signals
\cite{dama10,cdms213,cogent13}. The red dots indicate parameter space in the
WMAP preferred density range, while the blue dots correspond to models in which
the thermally-generated neutralinos are subdominant. The gray models are ruled
out by antiproton searches. It has been argued that it is important to search
the entire parameter space, not just the WMAP preferred range, since there are
many mechanisms to under-produce WIMPS in the early universe and still have
them detectable today \cite{gelmini06}. The solid and dashed lines indicate the
sensitivity of the antideuteron search for the case of nominal (solid) and
maximal (dashed) propagation models. There are several points illustrated by
this figure. Firstly, antideuteron searches are quite sensitive to models with
low energy neutralinos, and they maintain sensitivity up high neutralino
masses.  Secondly, the plot illustrates both the opportunity and the curse of
an antideuteron search, antideuteron propagation and production uncertainties.
This is discussed more below, but we note here that the reach into parameter
space for primary antideuterons is sensitive to details of the propagation (and
less so the production) of the antideuterons in the interstellar medium, which
are still poorly constrained. Thus, for the best case uncertainties,
antideuterons provide a deep reach into parameter space.  Conversely,
compounding the most pessimistic values of the uncertainties can lead to a
reach, in a short balloon observation, that is not much better than has been
obtained for antiproton searches.

\subsection{UHECR Measurements}

\label{sec:uhecr}

Currently the strongest constrains on TD models and therefore heavy
dark matter models with decaying particles come from limits on the
flux of UHE photons from the Pierre Auger Observatory in Malargue,
Argentina \cite{olinto2}

Although the contribution of $X$ particles to the observed UHECR flux
is now constrained to be subdominant, the existence of long-lived $X$
particles continues to be viable below the current observation limits.

\begin{figure}[tbh]
\begin{center}
  \includegraphics[width=3.5in]{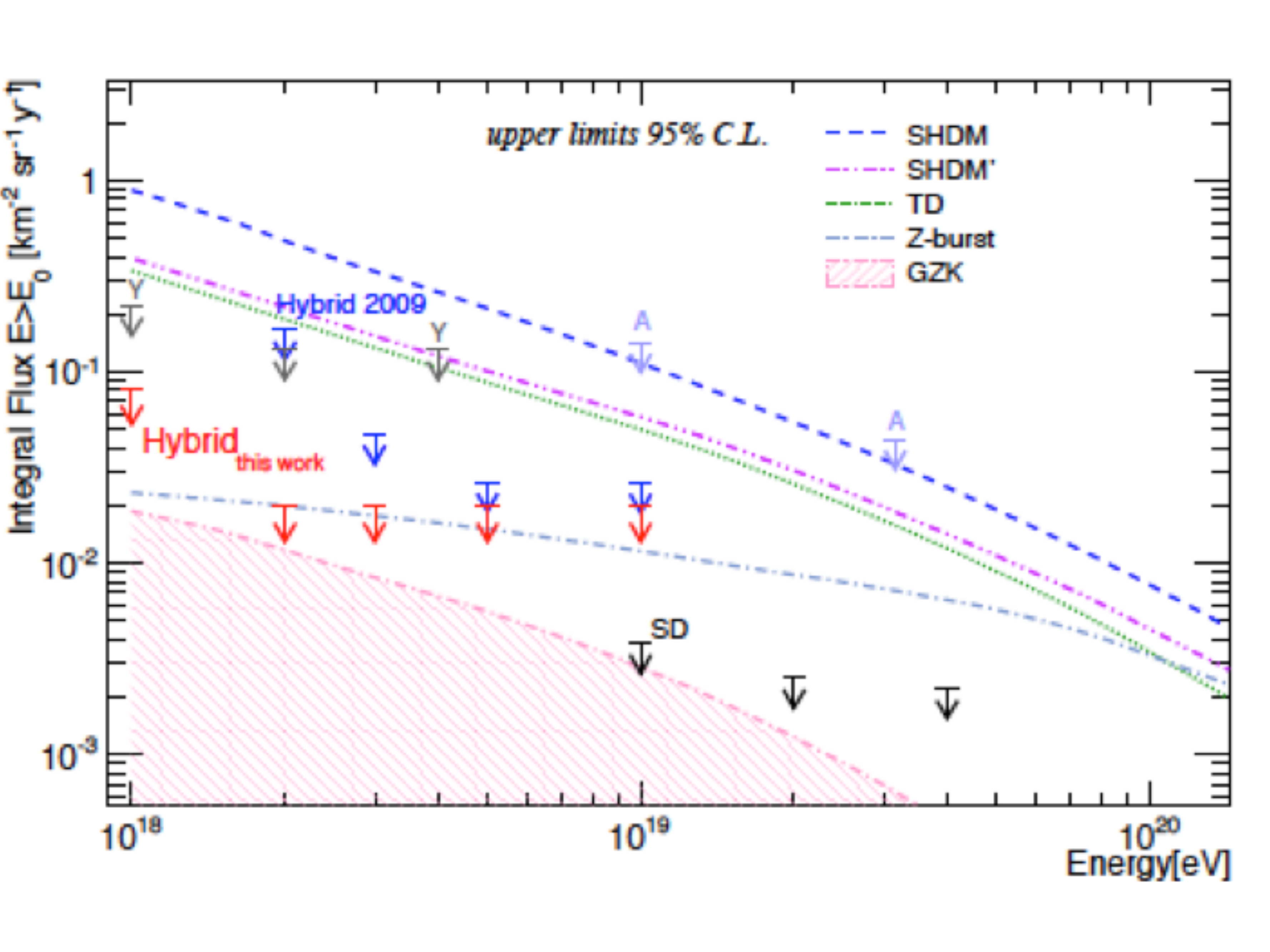}
\end{center}
\caption{ Figure from Ref.~\cite{augerother} showing Auger Surface Detector
  (SD) limits and Hybrid (SD and fluorescence events) limits. Also
  shown are predictions from models of super-heavy dark matter (SHDM),
  TD , Z-bursts, and the expected flux due to the GZK
  (Greisen-Zatsepin-Kuzmin) effect.}
\label{fig:superheavy}
\end{figure}

\subsection{Gamma-ray Experiments}

\label{sec:gamma}

In regions of high DM density the annihilation (or decay) of WIMPs into
Standard Model particles could produce a distinctive signature in gamma rays
potentially detectable with ground- and space-based gamma-ray observatories. In
fact, almost any annihilation channel will eventually produce gamma-rays either
through pion production (for hadronic channels), or final state bremsstrahlung
and inverse Compton from leptonic channels.  Moreover, the spectrum from
annihilation would be essentially universal, with the same distinctive shape
detected in every DM halo. Unlike the signals that can be measured by direct DM
detection experiments, the gamma-ray signature would provide strong constraints
on the particle mass and help to identify the particle through the different
kinematic signatures predicted for different annihilation channels.

The gamma-ray flux from DM annihilations scales with the integral of the square
of the DM density along the line of sight to the source (the $J$-factor
as defined in Eq.~\ref{eq:jfactor}).
Thus, the detectability of the DM signal from a
given target depends critically on its DM density distribution, as well as on
the possible presence of substructure along the line of sight. The ideal
targets for DM annihilation searches are those that have both a large value of
$J$ and relatively low astrophysical gamma-ray foregrounds. These criteria have
motivated a number of Galactic and extragalactic targets including the Galactic
Center (GC), dwarf spheroidal satellite galaxies of the Milky Way (dSphs), and
galaxy clusters. While the sensitivity to the DM halo profile provides the
largest systematic uncertainty, it also provides an avenue for inferring the DM
halo profile from the shape of the gamma-ray emission and connecting the
detected particle to the missing gravitational mass in galaxies.

The GC is expected to be the brightest source of DM annihilations in the
gamma-ray sky by several orders of magnitude. Although the presence of many
astrophysical sources of gamma-ray emission toward the inner Galaxy make
disentangling the DM signal difficult in the crowded GC region (see
\S~\ref{sec:haloobs}), the DM-induced
gamma-ray emission is expected to be so bright there that one can obtain strong
upper limits at the level of the natural cross section \sigmavnatural. In
addition, with the improved angular resolution of CTA, the astrophysical
foregrounds can be more easily identified and separated from the diffuse
annihilation signal. Also, the large concentration of baryons in the innermost
region of the Galaxy might act to further increase the expected DM annihilation
flux by making the inner slope of the DM density profile steeper, a mechanism
known as ``adiabatic contraction'' \cite{gondolosilk99,gondolosilk00,sfw13}.
While the exact role of baryons is not yet well understood, new state-of-the
art numerical studies of structure formation that include baryonic physics
along with the non-interacting DM are beginning to provide valuable insights.

Since the astrophysical foregrounds in the GC make it difficult to unambiguously interpret any detected signal, it is important to identify other sources
that combine the characteristics of high $J$ values, but very low backgrounds.
In the event of a detection from the GC, deeper observations of such sources
could provide an important confirmation of any putative dark-matter signal.
As discussed in \S~\ref{sec:haloobs} the inner halo profile of the GC has
large uncertainties that lead to significant (order-of-magnitude) uncertainties
in the expected signal   A promising class of objects for study are the
dwarf spheroidal satellite galaxies of the Milky Way (dSphs).

dSphs are thus attractive for DM searches in gamma rays for a number of
reasons including
their close proximity, high DM content, and the absence of intrinsic sources of
gamma-ray emission.
These objects are also predominantly
found at high galactic latitudes where the astrophysical foregrounds are much weaker.
 Because they are highly DM-dominated, the DM mass on small
spatial scales ($\sim$100 pc) can be directly inferred from measurements of
their stellar velocity dispersions. The uncertainty of the line of sight
distribution of DM for these systems is therefore much less than for other
candidates. Additionally, smaller DM subhalos may not have attracted enough
baryonic matter to ignite star-formation and would therefore be invisible to
most astronomical observations from radio to X-rays. All-sky monitoring
instruments sensitive at gamma-ray energies, like {\em Fermi}-LAT, may detect the DM
annihilation flux from such subhalos \cite{belikov12}, while follow-up
observations with CTA would characterize the distinctive spectral cut-off that
would eventually determine the DM particle mass.

\subsubsection{{\em Fermi}}

The Large Area Telescope (LAT), the primary instrument on the {\it Fermi
Gamma-ray Space Telescope}, launched in 2008, possesses
unprecedented sensitivity
in the GeV energy range.  In the first weeks of operation, the LAT collected
more gamma rays with energy $>100$ MeV than all previous missions combined.
The LAT is a pair-conversion gamma-ray detector.  It has a modular arrangement
of silicon strip trackers interleaved with tungsten converter foils (1.5
radiation lengths on-axis) above calorimeters with a hodoscopic arrangement of
CsI crystals (8.5 radiation lengths on-axis).  A 4$\times$4 array of these
tracker/calorimeter modules is surrounded by tiled plastic scintillators for
charged-particle rejection.  The LAT is approximately 1.8 m square, with a
squat aspect ratio.  Positron-electron pairs from gamma rays that convert in
the tungsten are followed through the silicon strip tracker and into the
calorimeter, where the light yield in the CsI crystals is recorded.  This
information is used in ground processing to precisely reconstruct the direction
and energy of the incident gamma ray.  The flux of celestial gamma rays is
several orders of magnitude smaller than that of cosmic rays at the orbit of
{\it Fermi} and the design of the LAT and the ground processing enables very
efficient rejection of this background. The LAT is sensitive to gamma rays from
20 MeV to greater than 300 GeV with $\sim$10\% resolution over much of that
range.  The effective collecting area peaks at about 8000 cm$^2$ at 10 GeV and
the field of view is 2.4 sr.  The per-photon angular resolution ranges from
several degrees at 100 MeV to $\sim 0.1$ deg at the highest energies (Atwood et
al. 2009).  The scanning and rocking pattern of the attitude control system of
{\it Fermi} allows the LAT to observe the entire sky every 3 hours.  The
characteristics of the LAT make it an excellent instrument for the discovery of
faint new gamma-ray sources.

Searches for gamma-ray signals from annihilation or decay of massive
dark matter particles were a principal scientific driver for the {\it
  Fermi} mission.  Searches in the LAT data have been undertaken in a
variety of ways, to try to maximize sensitivity and to overcome
limitations from astrophysical foregrounds.  The cross section limits
obtained for annihilation necessarily depend on the assumed decay
channels, e.g., through $b\bar{b}$, $\tau^+\tau^-$, $W^+W^-$, with
gamma rays also produced via interactions of secondaries.  The limits
also depend on the assumed spatial distribution of particle dark
matter.  Theoretically well-motivated predictions for the
distributions have a range of ``cuspiness'' for the cores; in addition
substructure on the smallest scales can be very important to the
overall annihilation rate because the rate depends on the square of
the density.  Such `boost factors' are typically quite uncertain and
limits for annihilation cross sections are generally conservatively
quoted for unit substructure boosts.  For the searches, templates of
the expected gamma-ray emission for the assumed mass and annihilation
channel and density distribution are compared to the LAT data.  As
discussed further below, annihilation can also be directly into two
gamma rays.  This would produce a spectrally-distinct signal (a narrow
line) but the branching ratios are expected to be very small.

\begin{figure}[tbh]
   \begin{center}
      \includegraphics[width=0.48\hsize]{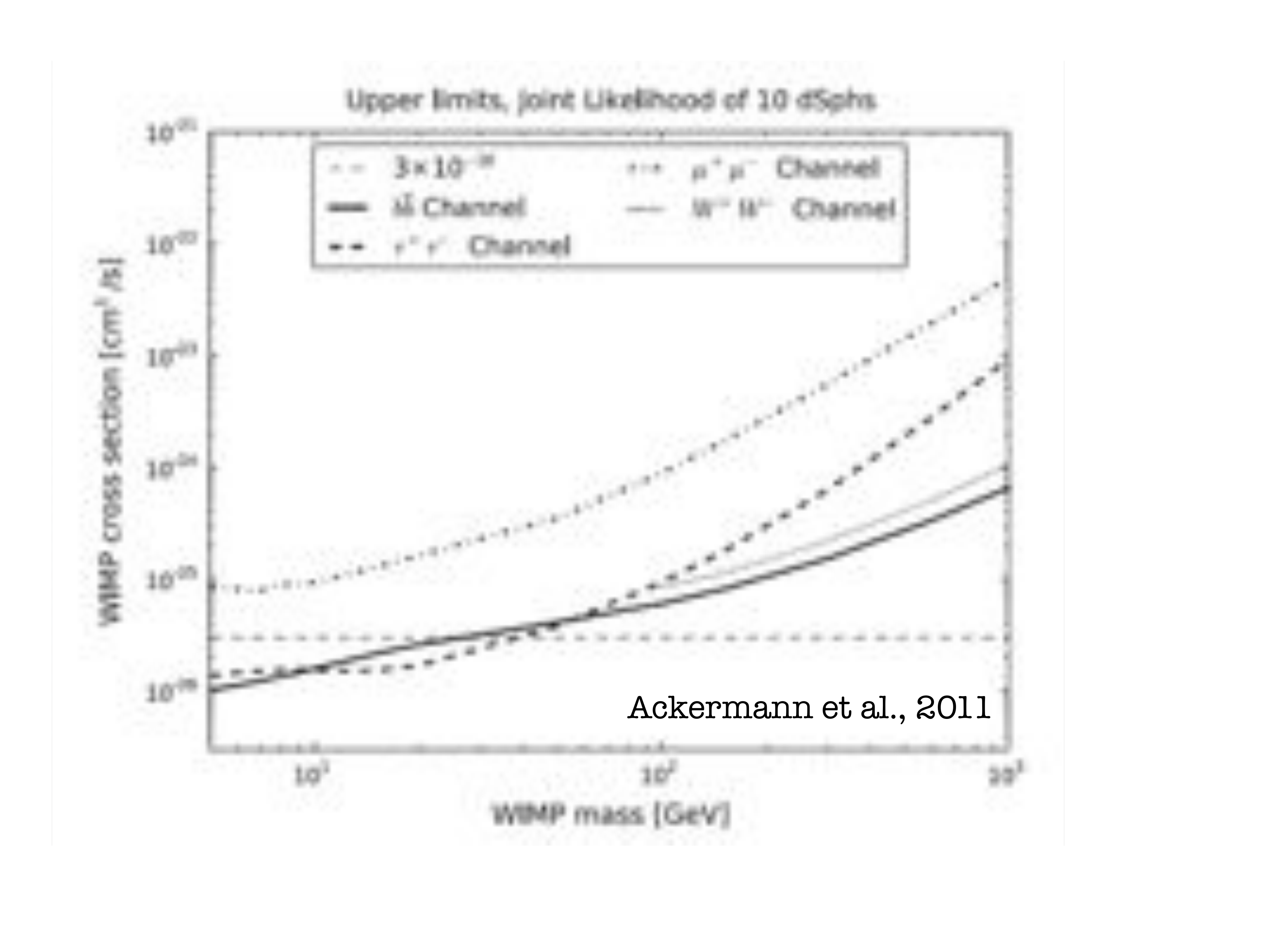}
      \includegraphics[width=0.48\hsize]{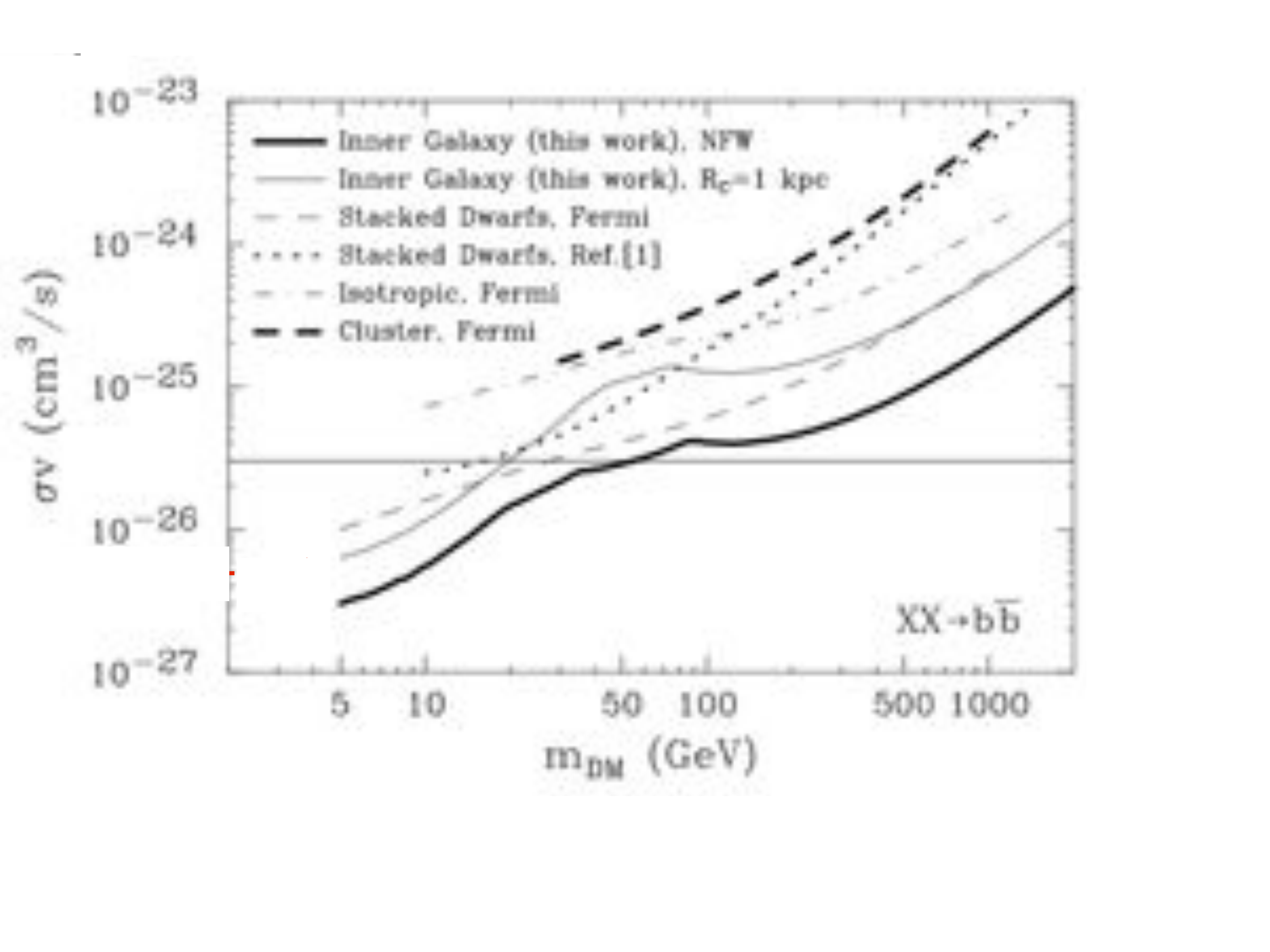}
      \caption{{\it Left:} Constraints on the total annihilation cross section
derived by combining {\em Fermi} data on a number of dSphs \cite{fermidwarf} Dwarf stacking.  {\it Right:} Compilation of {\em Fermi} DM constraints (courtesy D. Hooper).
}
\end{center}
\end{figure}

The density of dark matter is greatest in the Galactic center (GC)
region and the annihilation fluxes are expected to be greatest from
the GC.  However, the astrophysical foregrounds are extremely bright
and challenging to model toward the GC.  Conservative limits on the
annihilation cross section have been obtained using the entire
observed gamma-ray intensity as an upper limit for the dark matter
signal.  Less conservative limits have relied on (uncertain) modeling
of the foreground astrophysical signals and/or masked the
lowest-latitude regions where the foreground is brightest.

Searches for gamma-ray lines in the GC region suffer much less from
foreground confusion.  An analysis of publicly-available LAT data has
reported evidence for a line-like feature near 130 GeV from the GC
region.  (The actual region analyzed was optimized for sensitivity to
the assumed distribution of dark matter.)  The implied cross section
was much greater than expected theoretically.  This finding has
stimulated a number of follow-up studies, characterizing the potential
systematics and looking for similar line-like signals from other
directions, including galaxy clusters and the Sun.

The distribution of dark matter is expected to be clumpy on a wide
range of scales and in particular the halo of the Milky Way should
contain a number of massive sub-halos, which would themselves be
sources of annihilation gamma rays.  Some searches have looked for
unidentified LAT sources that have the right spectral and spatial
distributions.  The most constraining limits have come from stacking
analyses of observations of dwarf spheroidal (dSph) galaxies in the
halo of the Milky Way.  These are the largest sub-halos and are
expected to be dominated by dark matter.  Additionally, they host
essentially no on-going star formation and so have little non-thermal
emission from astrophysical processes.  The limits for annihilation
cross sections from a joint analysis of the best-characterized dSphs
are now below the thermal cross section limit for some energy ranges
and annihilation channels. 

A novel search for a dark matter signal in the isotropic gamma-ray
background has been developed using the angular power spectrum to
distinguish various extragalactic and Galactic foregrounds.  The
ultimate sensitivity of the method should be sufficient to detect a
dark matter contribution to the isotropic diffuse spectrum at the
several percent level.

\subsubsection{VERITAS}

\begin{figure}[tbh]
   \begin{center}
      \includegraphics[width=0.9\hsize]{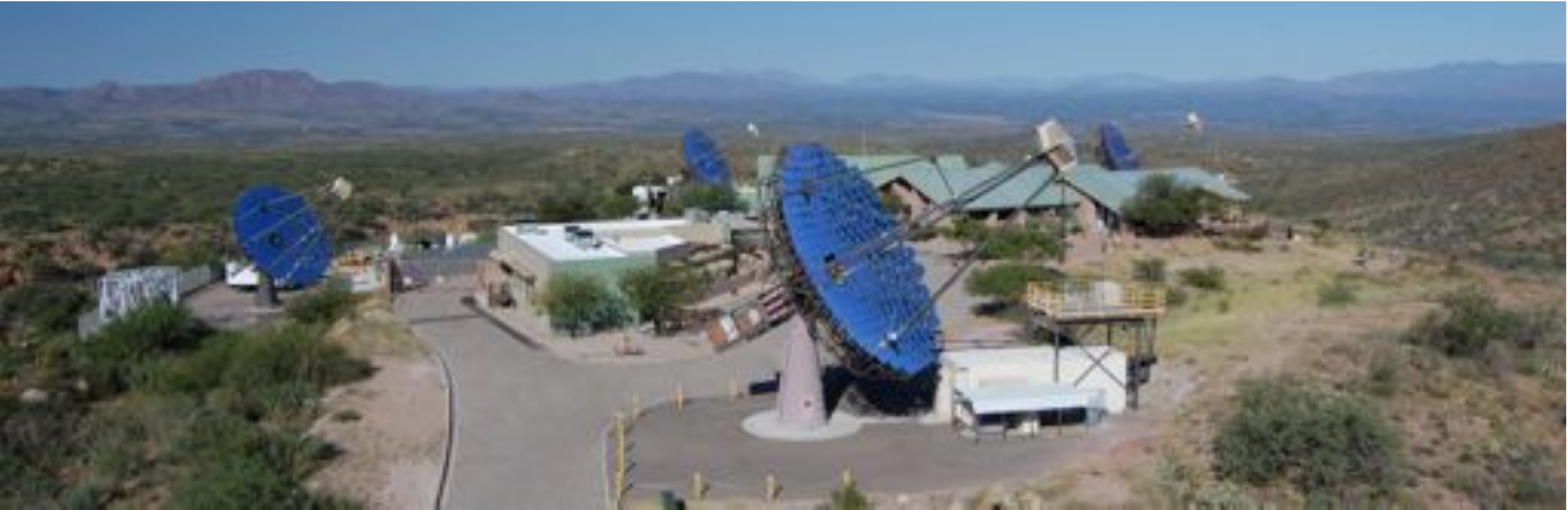}
      \caption{The VERITAS IACT array.
        }
   \label{fig:veritasphoto}
\end{center}
\end{figure}

The VERITAS IACT array\cite{veritas09}, located in Southern Arizona, is an
additional key contributor to the indirect search for particle dark
matter.  VERITAS is comprised of four, 12 meter diameter Davies-Cotton
optical reflectors which focus light from gamma-ray air showers onto
four 499 pixel photomultiplier tube( PMT) cameras. The array, with a
total field of view of 3.5$^{\circ}$, is sensitive in the range of 100
GeV to 50 TeV, easily covering a large range of parameter space
favored by the SUSY models discussed previously. Since the
commissioning of the array in 2007, VERITAS has accrued approximately
350 hours of observations on a range of putative dark matter targets
including dSph galaxies, galaxy clusters and the GC. The analysis of these
observations is ongoing, however, the results already published from
subsets of the observations have resulted in some of the strongest DM
constraints available from indirect searches. While VERITAS focuses on
a range of potential dark matter targets such as the Galactic Center,
and galaxy clusters; one of VERITAS' key contributions to indirect
matter searches thus far has been in the study of dwarf spheroidal
galaxies (dSphs).

\begin{figure}
\begin{center}
  \includegraphics[width=5.5in]{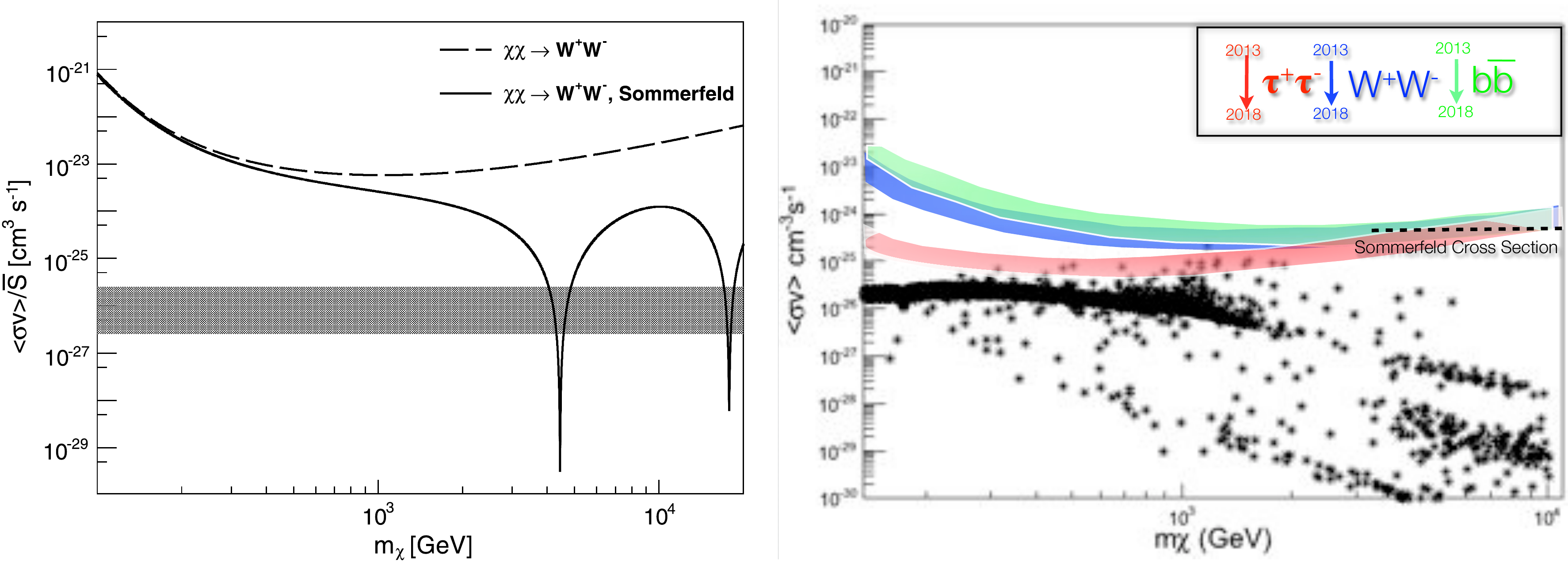}
\end{center}
\vspace*{-0.35in}
\caption{VERITAS Segue 1 dwarf limits \cite{versegue} and projections for
proposed dwarf program}
\label{fig:versegue}
\vspace*{-0.15in}
\end{figure}

The first VERITAS dSph results \cite{verdwarfs10} were composed of observations
of the dSph galaxies Draco, Ursa Minor, Bootes I, and Willman I. This result
was based on a relatively small observational exposure (10-15 hours per source)
and was significantly improved by the result of \cite{versegue} in which over
50 hours of observation on the dSph Segue I was presented (Figure 2). While
these limits do not constrain the most conservative realizations of minimal
SUSY, the VERITAS dSphs limits (in particular the Segue I limits) provide much
stronger constraints on alternative models of SUSY (for example, invoking a
Sommerfeld enhancement at very high mass from W and Z exchange, or at a lower
mass from a new scalar mediator \cite{arkanihamed2009}).  Models with a
kinetically enhanced cross section are already constrained by current VERITAS
observations, and may be all but excluded over the next few years of VERITAS
observations.

Of particular importance, VERITAS dSph observations have now provided strong
constraints on models of dark matter annihilation invoked to reproduce the
PAMELA positron excess\cite{Adriani:2008zr}. In these models, DM annihilates
exclusively into $\mu^{+}\mu^{-}$ (leptophillic models). In order for these
models to explain the PAMELA excess, a boost factor $B$ is required (both
astrophysical and particle physics boosts are convolved into $B$. The VERITAS
Segue I observations places strong constraints on $B$, limiting the allowed
parameter space that can be utilized to explain the PAMELA excess within a DM
framework. These results are especially relevant in light of the recent results
from the AMS experiment \cite{Aguilar:2013qda}. As with the constraints on
$\langle\sigma v\rangle$, these limits will improve with additional
observations.

\begin{figure}[tbh]
\begin{center}
  \includegraphics[width=5.5in]{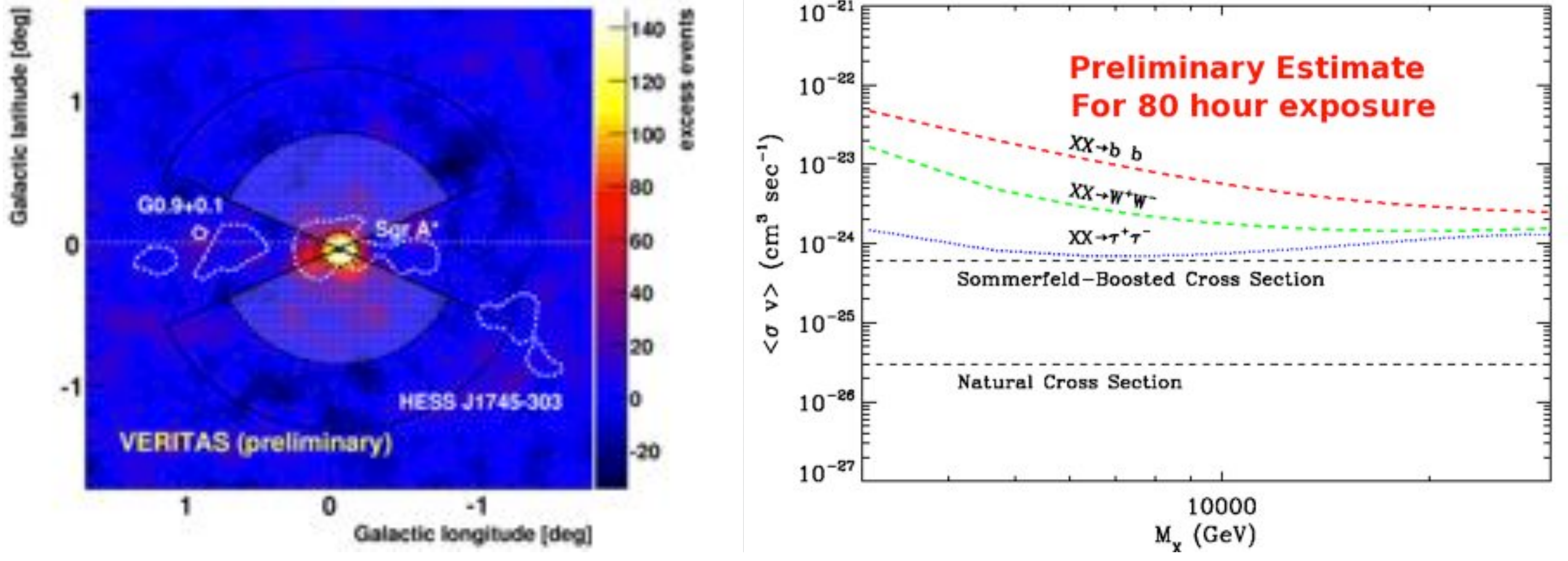}
\end{center}
\caption{\emph{Left:}VERITAS GC exposure showing
the bright source at Sgr A$^*$, and DM signal and background regions \cite{beilickeconf}.  \emph{Right:}
VERITAS projected sensitivity for an 80 hr exposure (JB)}
\vspace*{-0.15in}
\label{fig:vergc}
\end{figure}

As observations for the VERITAS DM program are ongoing and will continue
through the operational lifetime of the array, these constraints will only be
improved with time.  Recently, VERITAS has dramatically increased the time
allocation for Dark Matter observations with a goal of acquiring $\sim$250
hours of data on Dwarf galaxies every year.  This change represents an
order-of-magnitude increase in observing time compared to that of previous
published DM results. Together with the recent upgrade of VERITAS
\cite{veritasupgrade} (replacing the PMTs to realize a 50\% increase in light
collection efficiency), this will substantially improve DM sensitivity.   

As shown in Figure~\ref{fig:versegue}, the anticipated dwarf galaxy limits with
5 years with VERITAS are surprisingly good, pulling within an order of
magnitude of the natural cross-section, a result not significantly different
from some predictions for CTA (e.g., the analysis of Ref. ~\cite{Doro:2012xx}
for the baseline CTA instrument without the U.S. extension).  There are a
number of reasons why VERITAS (a current generation experiment) might be
competitive with the baseline CTA instrument.   First, VERITAS has made
observations of dSph's one of the highest priority science programs and should
accumulate very deep dwarf exposures long before CTA is fully on-line and when
other higher-priority scientific programs could dominate the CTA schedule (U.S.
participation in CTA might change these priorities).  Thus the large exposure
time assumed in these limits reflects the higher priority of dark-matter
science in the U.S. community compared with the European-led CTA consortium.
Another significant factor comes from the fact that, of the currently known
dSphs, the best candidates are in the northern hemisphere.  In particular, the
Segue 1 Dwarf that appears to have a higher $J$-factor than any of the
presently known dwarfs (a factor of 2 larger than any of the new Sloan sources,
and a factor of $\gsim$10 larger than the best ``classical'' dwarfs (see
Table.~\ref{tab:dwarfs}).  These limits also include a new advanced analysis
technique which makes use of an optimal weighting of each event based on the
reconstructed energy and angular position \cite{GeringerSameth:2011iw}.  Thus,
with 5 years of continued (post-upgrade) observations, the extensive dark
matter search program with VERITAS offers one of the strongest avenues
available for the possible detection of WIMP dark matter and, in the absence of
detection, can severely constrain many conservative SUSY models (see
Figures~\ref{fig:versegue},\ref{fig:vergc}).

Since the GC transits at large zenith angles at the VERITAS site, the energy
threshold is raised compared with southern hemisphere experiments like H.E.S.S., or
the future Southern-CTA array.  However, the effective area increases at large
zenith angle (LZA) and the VERITAS team has developed a new LZ analysis
technique that exploits the large effective area, providing VERITAS
observations of the GC region a similar sensitivity to H.E.S.S. but in $\sim$1/6th
of the observing time for energies above a few TeV \cite{beilickeconf}.   At
lower energies (below a TeV) the H.E.S.S. limits on the GC now come within an order
of magnitude of the natural cross section, better than any of the current dSph
limits \cite{hessgclimit}.   As shown in Fig.~\ref{fig:vergc} with 5 years of observation, VERITAS
will extend the cross-section energies to higher WIMP masses, in some cases
reaching the natural cross-section when non-perturbative boosts in the cross
section are taken into account.

In addition to VERITAS, the other major ground-based instruments MAGIC and H.E.S.S have performed observations of dSphs to obtain DM upper limits.  While we do not discuss all of these results in detail, these measurements are
summarized in Table~\ref{tab:dwarfs} below.

\begin{table}[ht]
\caption{J-factors for Dwarf Galaxies} 
\centering 
\begin{tabular}{c c c c} 
\hline\hline 
Source & J-factor & Profile & Reference  \\
& [GeV$^2$ cm$^{-5}$] & \\ [0.5ex] 
\hline 
Draco & $7.1 \times 10^{17}$ &  NFW & \cite{2011MNRAS.418.1526C} \\
Willman I & $8.4 \times 10^{18}$ & NFW & \cite{2010ApJ...720.1174A}\\
Segue I & $1.7 \times 10^{19}$ &  Einsasto & \cite{2011JCAP...06..035A}\\
Ursa Minor & $2.2 \times 10^{18}$ &  NFW & \cite{2011MNRAS.418.1526C}\\
Bo\"otes I & - & - &-\\
Sagittarius & - & -&-\\
Sculptor & $8.9 \times 10^{17}$ & NFW& \cite{2011MNRAS.418.1526C} \\
Carina & $2.8 \times 10^{17}$ &  NFW & \cite{2011MNRAS.418.1526C}\\
\hline 
\end{tabular}
\label{tab:dwarfs} 
\end{table}

 \begin{threeparttable}[ht]
  \caption{Flux Upper Limits from IACTs} 
  \centering
  \begin{tabular}{c c c c} 
  \hline\hline 
  Source & VERITAS  $\Phi^{u.l.}$ & MAGIC $\Phi^{u.l.}$ & H.E.S.S. $\Phi^{u.l.}$  \\
 &cm$^{-2}$s$^{-1}$ & cm$^{-2}$s$^{-1}$&cm$^{-2}$s$^{-1}$ \\ [0.5ex] 
  \hline 

   Draco            &$0.49\times 10^{-12} $  \tnote{a} \cite{2010ApJ...720.1174A}  &  $1.1 \times 10^{-11}$ \tnote{b} \cite{2008ApJ...679..428A}   & -  \\
   Willman I       & $1.17 \times 10^{-12}$ \tnote{c} \cite{2010ApJ...720.1174A}           & $9.87 \times 10^{-12}$ \tnote{d} \cite{2009ApJ...697.1299A}   & - \\
   Segue I         & $7.6 \times 10^{-13} $ \tnote{e} \cite{2012PhRvD..85f2001A}                   & $2 \times 10^{-12} $ \tnote{f} \cite{2011ICRC....5..149A}     &-\\
   Ursa Minor      & $ 0.40 \times 10^{-12} $ \tnote{g} \cite{2010ApJ...720.1174A}         & -                                                                     &  -\\
   Bo\"otes I      & $2.19 \times 10^{-12} $ \tnote{h} \cite{2010ApJ...720.1174A}          & -                                                                     & - \\
   Sagittarius     & -                                                                             & -                                                                     & $3.6 \times 10^{-12} $ \tnote{i} \cite{2008APh....29...55A}\\
   Sculptor        & -                                                                             &-                                                                      & $5.1 \times 10^{-13}$ \tnote{j} \cite{2011APh....34..608H} \\
   Carina  & -                                                                             & -                                                                     & $1.6 \times 10^{-13} $  \tnote{k} \cite{2011APh....34..608H}  \\[1ex] 
   \hline 
   \end{tabular}
   \begin{tablenotes}
      \item[a] $E_{min}= 340$ GeV
      \item[b] $E_{min}= 140 $ GeV
      \item[c] $E_{min}= 320$ GeV
      \item[d] $E_{min}= 100$ GeV
      \item[e] $E_{min}= 300$ GeV
      \item[f] $E_{min}= 200$ GeV
      \item[g] $E_{min}= 380$ GeV
      \item[h] $E_{min}= 300$ GeV
      \item[i] $E_{min}= 250$ GeV
      \item[j] $E_{min}= 220$ GeV
      \item[k] $E_{min}= 320$ GeV
   \end{tablenotes}
   \label{table:nonlin4} 
 \end{threeparttable}

  \begin{threeparttable}[ht]
    \caption{dSph upper limits on $\langle \sigma v \rangle$ for VERITAS, MAGIC and H.E.S.S. collaborations.} 
     \centering 
     \begin{tabular}{c c c c} 
     \hline\hline 

      Source & VERITAS $\langle \sigma v \rangle^{u.l.}$ & 
      MAGIC $\langle \sigma v \rangle ^{u.l.}$ &
      H.E.S.S. $\langle \sigma v \rangle^{u.l.}$\\
       & cm$^3$ s$^{-1}$ & cm$^3$ s$^{-1}$  & cm$^3$ s$^{-1}$ \\ [0.5ex] 
       \hline 
      Draco  & $2 \times 10^{-23}$ \tnote{a} \cite{2010ApJ...720.1174A} &
        $10^{-23}$ \tnote{b} \cite{2008ApJ...679..428A}&-  \\
      Willman I & $10^{-23}$ \tnote{c} \cite{2010ApJ...720.1174A} &- &-\\
      Segue I & $ 2 \times 10^{-24}$ \tnote{d} , $10^{-23}$ \tnote{e} \cite{2012PhRvD..85f2001A}      & $2 \times 10^{-24}$ \tnote{f} \cite{2011JCAP...06..035A}      &-  \\
        Ursa Minor & $10^{-23}$ \tnote{g} \cite{2010ApJ...720.1174A}
      &-      &-      \\
      Bo\"otes I      & $10^{-22}$ \tnote{h} \cite{2010ApJ...720.1174A}       &-      &-      \\
      Sagittarius     &-&-& $10^{-25}$ \tnote{i} , $3 \times 10^{-26}$ \tnote{j} \cite{2008APh....29...55A}\\
      Sculptor        &-&-& $4 \times 10^{-23}$ \tnote{k} \cite{2011APh....34..608H}\\
      Carina  &-&-& $2 \times 10^{-22}$ \tnote{l} \cite{2011APh....34..608H}\\
        \hline 
    \end{tabular}
    \begin{tablenotes}
    \item [a] For a composite neutralino spectrum and $m_{\chi} \sim 1$ TeV \\
    \item [b] For Model $D^{\prime}$ given by \cite{2004EPJC...33..273B} and with $m_{\chi} \sim 200$ GeV \\
    \item [c] For a composite neutralino spectrum and $m_{\chi} \sim 900$ GeV \\
    \item [d] For $\tau^+ \tau^-$ and $m_{\chi} \sim 300$ GeV \\
    \item [e] For $b \bar b$ and $m_{\chi} \sim 1$ TeV \\
    \item [f] For $\tau^+\tau^-$ and $m_{\chi} \sim 500$ GeV\\
    \item [g] For a composite neutralino spectrum and $m_{\chi} \sim 1$ TeV\\
    \item [h] For a composite neutralino spectrum and $m_{\chi} \sim 800$ GeV\\
    \item [i] For pMSSM + Cored profile and with $m_{\chi} \sim 200$ GeV\\
    \item [j] For Kaluza-Klein + Cored profile with $m_{\chi} \sim 300$ GeV\\
    \item [k] For self annihilation into $WW$ and $ZZ$ pairs and with $m_{\chi} \sim 2$ TeV \\
      \item [l] For self annihilation into $WW$ and $ZZ$ pairs and with $m_{\chi} \sim 3$ TeV\\
     \item [m]
  Note: All values for $\langle \sigma v \rangle^{u.l.}$ are approximate minimum values taken from plots of $\langle \sigma v \rangle^{u.l.}$ for a range of values of $m_{\chi}$.
  \end{tablenotes}
  \label{table:nonlin2} 
 \end{threeparttable}

\subsection{Future Gamma-Ray Experiments}\label{sec:futuregamma}

\subsubsection{Anticipated {\em Fermi} 10-Year Data}

While the {\em Fermi} LAT data reach the natural cross section below 30 GeV, they
do not presently constrain any of the pMSSM benchmark models.  Here we
estimate the improvements expected over a 10 year
mission lifetime.  In the low-energy, background dominated regime, the
LAT point source sensitivity increases as roughly the square-root of
the integration time.  However, in the high-energy, limited background
regime (where many pMSSM models contribute), the LAT sensitivity
increases more linearly with integration time.  Thus, 10 years of data
could provide a factor of $\sqrt{5}$ to 5 increase in
sensitivity. Additionally, optical surveys such as
Pan-STARRS~\cite{Kaiser:2002zz}, the Dark Energy
Survey~\cite{Abbott:2005bi}, and LSST~\cite{Ivezic:2008fe} could
provide a factor of 3 increase in the number of Milky Way dSphs
corresponding to an increased constraining power of $\sqrt{3}$ to
3~\cite{tollerud2008}.  Ongoing improvements in LAT event
reconstruction, a better understanding of background contamination,
and an increased energy range are all expected to provide additional
increases in the LAT sensitivity. Thus, we find it plausible that the
LAT constraints could improve by a factor of 10 compared to current
constraints.

In Figure~\ref{fig4} we display the boost required to constrain the
various pMSSM models at $95\%$ CL based on the \Fermi-LAT dwarf
analysis employing only the first 2 years of data colored-coded by
either the annihilation cross section or the LSP thermal relic
density. Here we see that the LAT analysis does not currently
constrain any of our pMSSM models. However, as discussed above with
more dwarfs and longer integration times we would expect an $\sim
10$-fold improvement in the sensitivity and thus all models with boost
factors less than 10 would become accessible. We will assume this
$\sim 10$-fold improvement in sensitivity assumption for the analysis
that follows.

\begin{figure}[tbh]
\centerline{\includegraphics[width=3.5in]{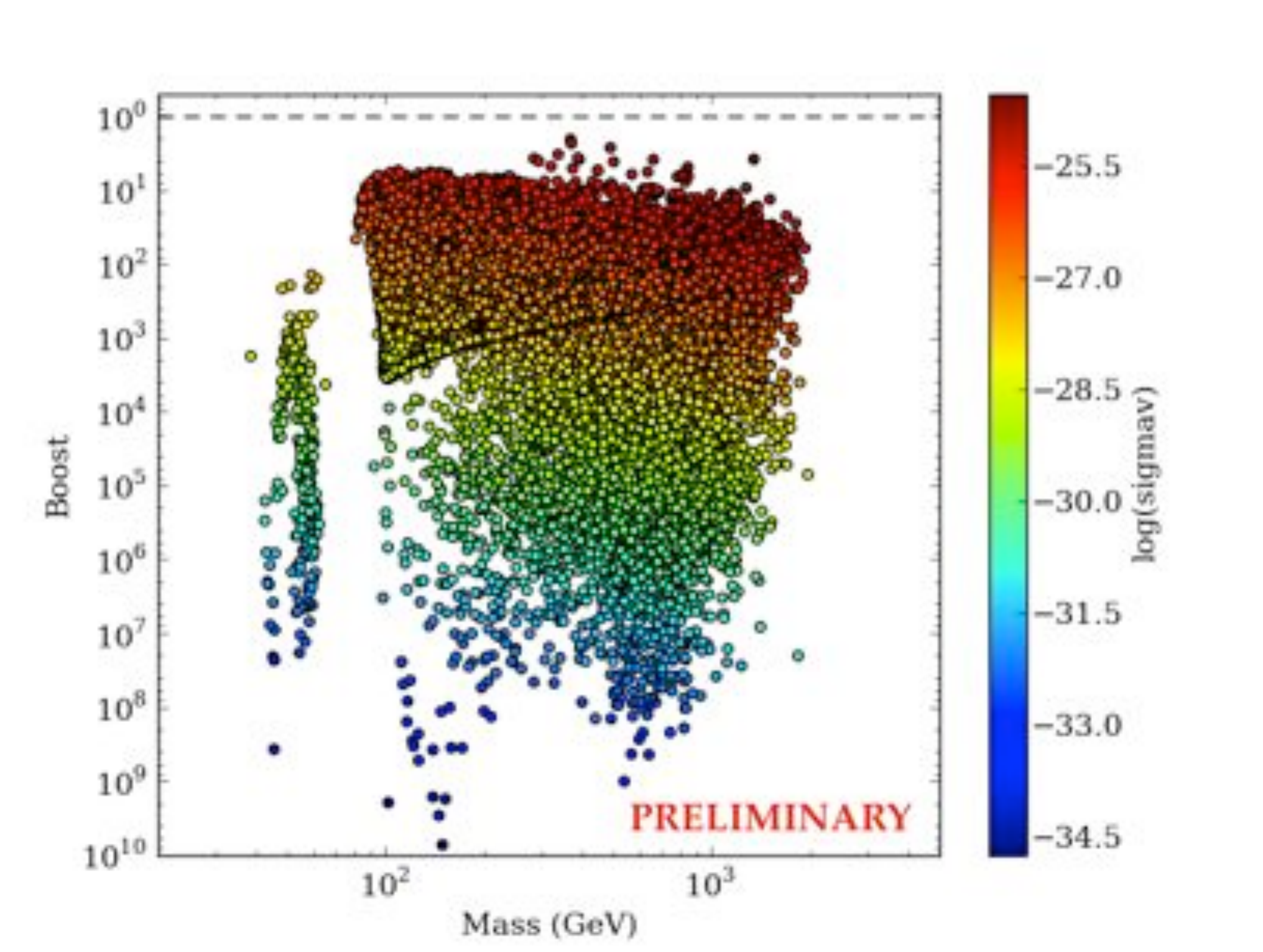}
\hspace{-0.50cm}
\includegraphics[width=3.5in]{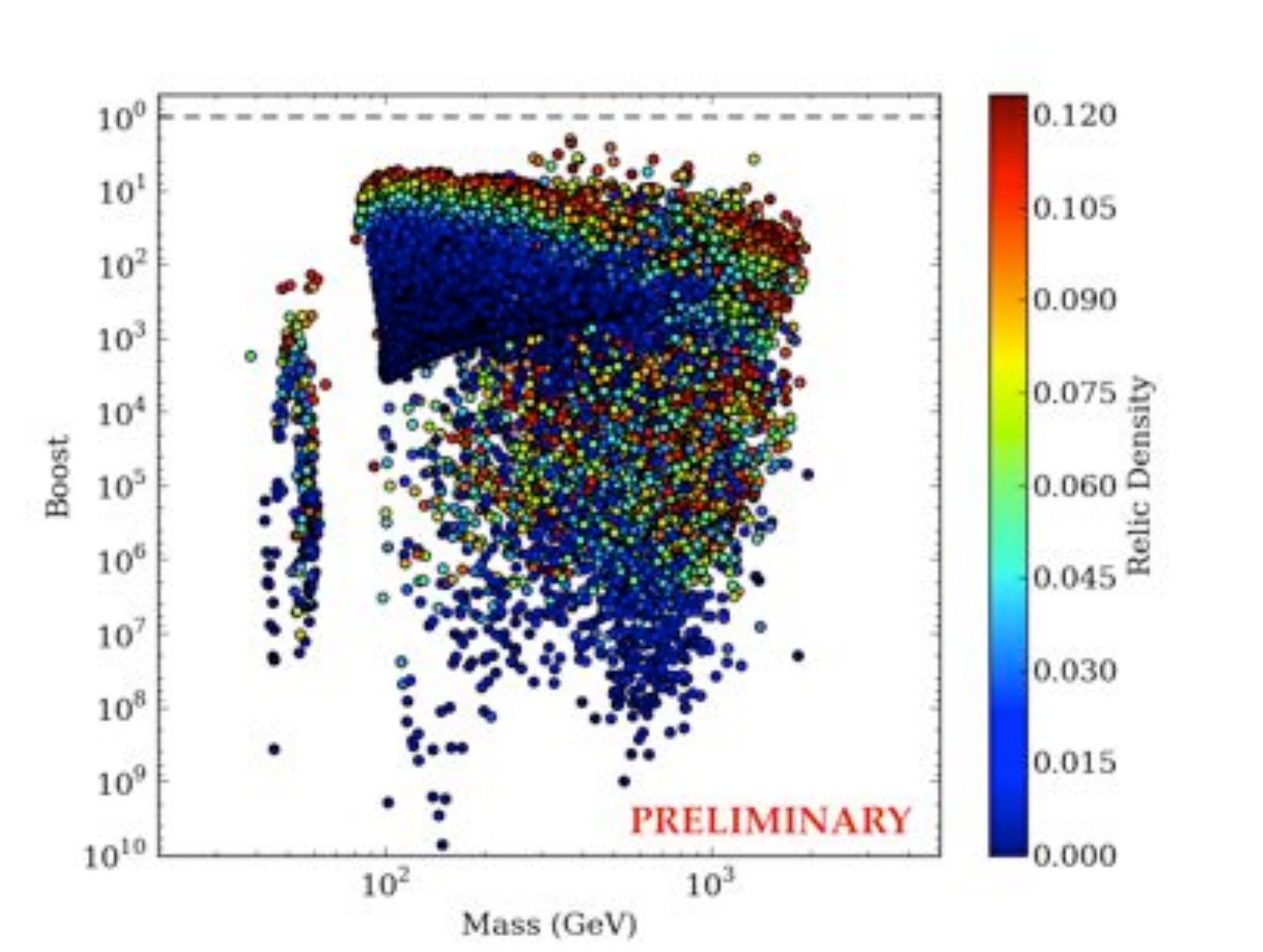}}
\vspace*{-0.10cm}
\caption{(Left) \Fermi-LAT boost factor vs. LSP mass for the pMSSM
  model set. The true cross section, \sigv, for each model is plotted
  on the color scale. (Right) Here the corresponding relic density for
  each model is plotted on the color scale.}
\label{fig4}
\end{figure}
\subsubsection{The Augmented Cherenkov Telescope Array (CTA)}

\begin{figure}[tbh]
\begin{center}
\includegraphics[width=3.5in]{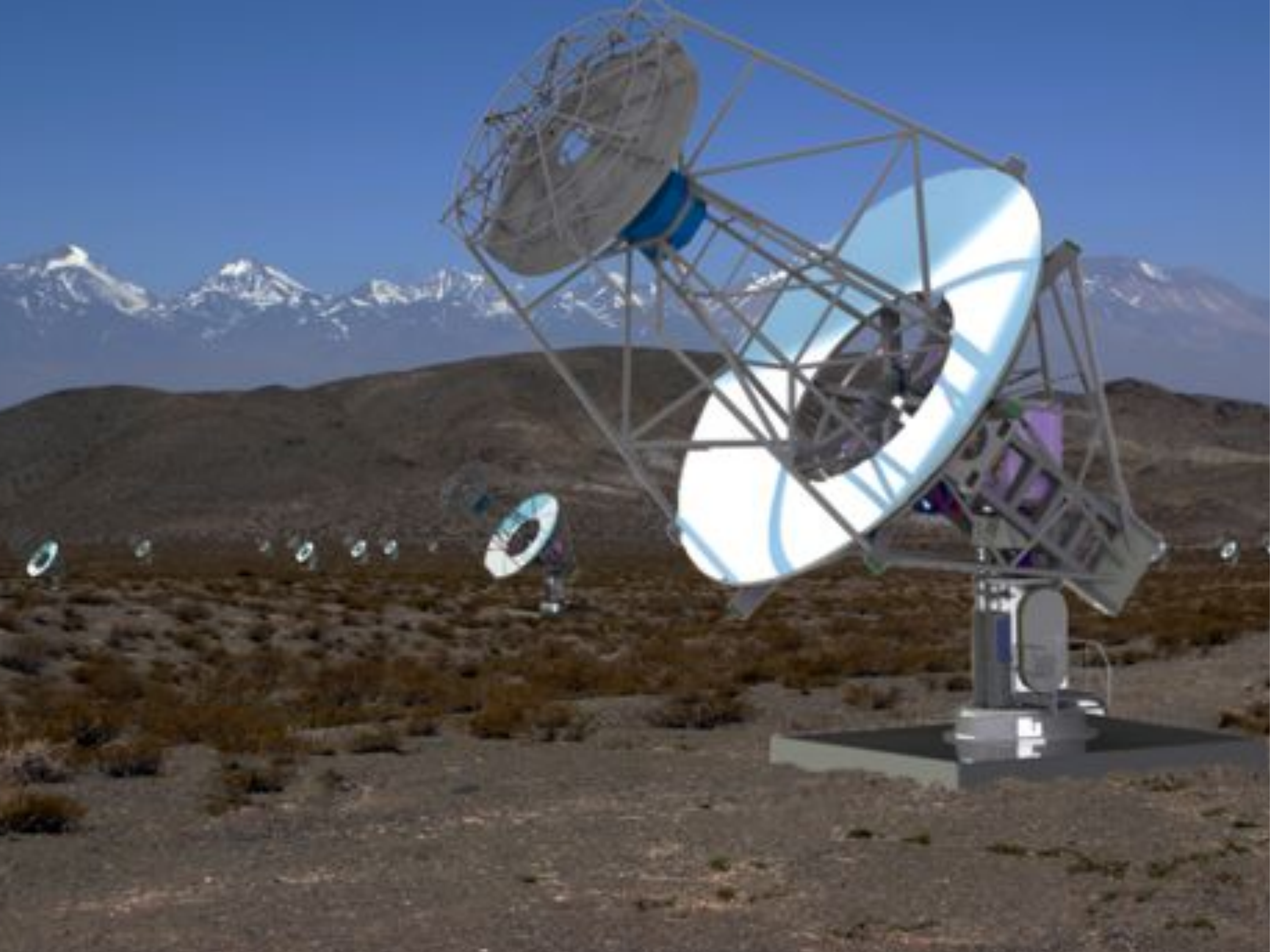}
\end{center}
\caption{ Artist's concept of an array of SCT telescopes as being
  developed by the U.S. CTA group to extend the baseline CTA array
  providing a factor of 2 to 3 in sensitivity, improved angular
  resolution and larger field of view compared with the baseline CTA
  concept.  }
\label{fig:cta_concept}
\end{figure}

\begin{figure}[tbh]
   \begin{center}
      \includegraphics[width=0.45\hsize]{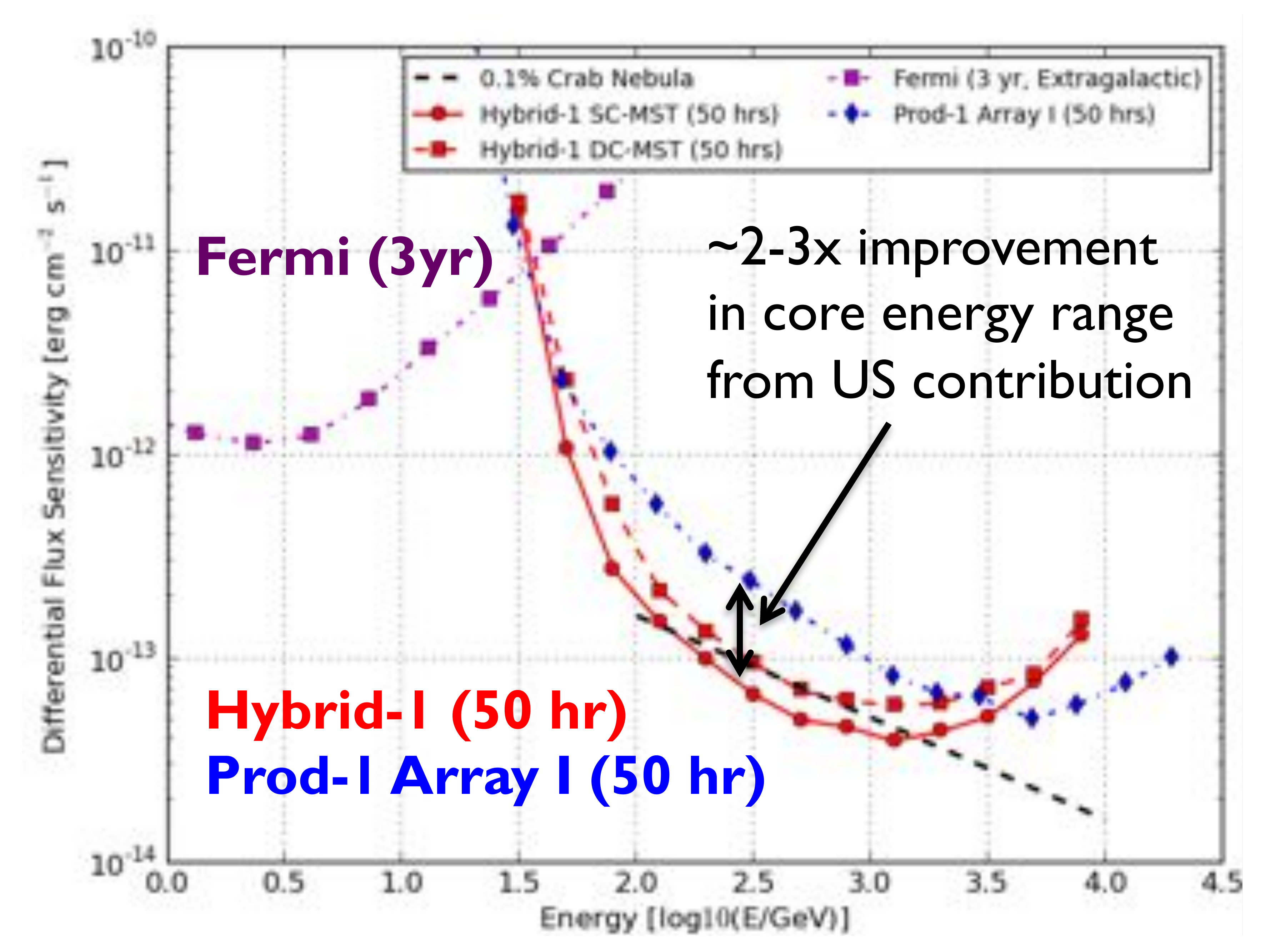}
      \includegraphics[width=0.45\hsize]{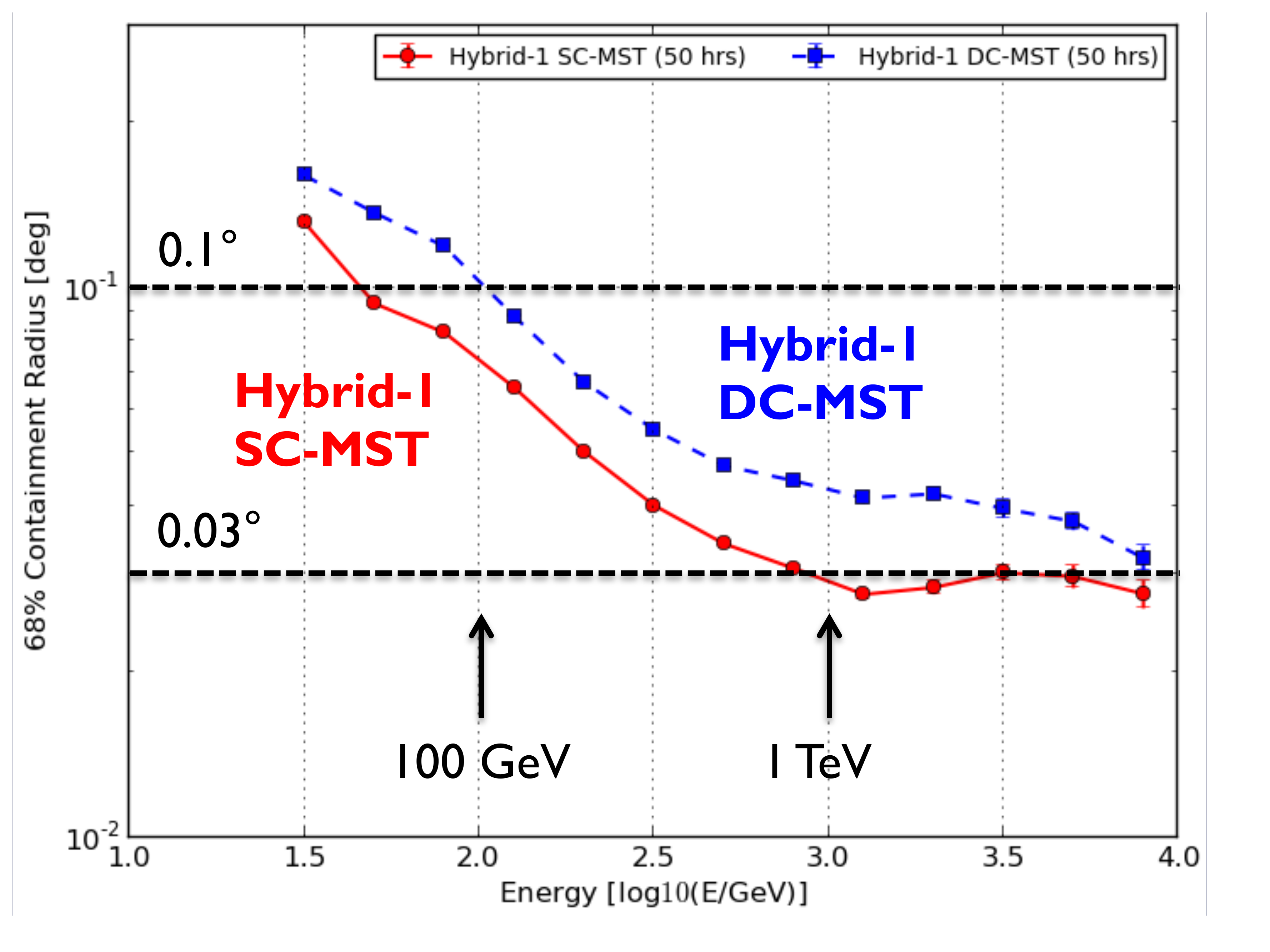}
      \caption{Projected sensitivity of a candidate CTA configuration, see the text for details.
        }
   \label{fig:ctasens}
\end{center}
\end{figure}

\begin{figure}[tbh]
   \begin{center}
      \includegraphics[width=0.7\hsize]{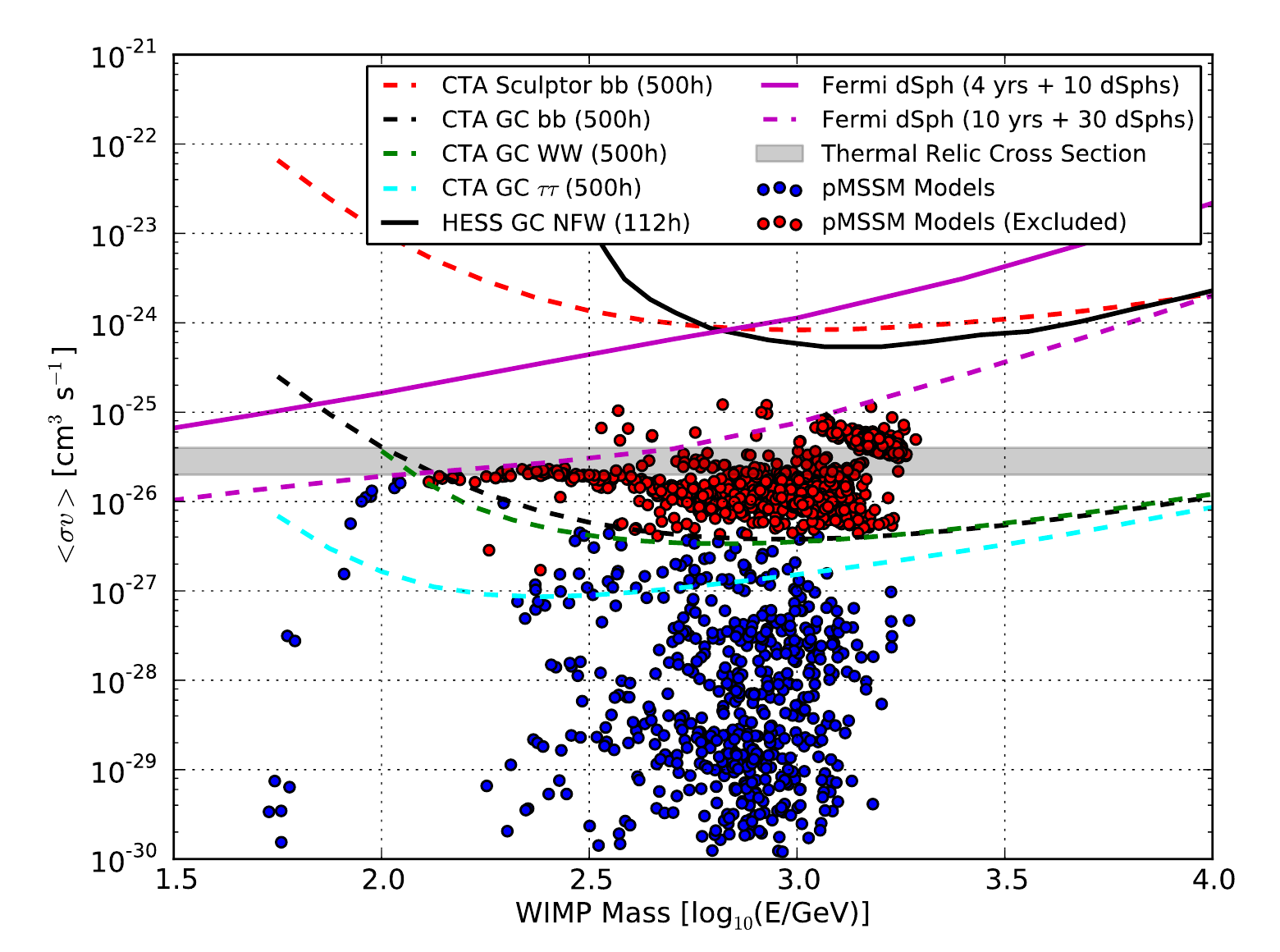}
      \caption{A comparison of pMSSM predictions and projected CTA, H.E.S.S. and {\em Fermi} sensitivities.
        }
\label{fig:cta_pmssm_sigmav}
\end{center}
\end{figure}

The planned (baseline)
Cherenkov Telescope Array (CTA) \cite{cta13} will have
great sensitivity over the energy range from a few tens of GeV to 100 TeV. To
achieve the best sensitivity over this wide energy range CTA will include three
telescope types: Large Size Telescope (LST, 23 m diameter), Medium Size
Telescope (MST, 10-12 m) and Small Size Telescope (SST, 4-6 m). Over this
energy range the point-source sensitivity of CTA will be at least one order of
magnitude better than current generation imaging atmospheric Cherenkov
telescopes such as H.E.S.S., MAGIC, and VERITAS. CTA will also have an angular
resolution at least 2 to 3 times better than current ground-based instruments,
improving with energy from 0.1$^\circ$ at 100 GeV to better than 0.03$^\circ$
at energies above 1 TeV.

The potential of CTA for DM searches and testing other exotic physics
has been studied in detail by \cite{Doro:2012xx} using the projected
performance of several alternative array configurations
\cite{Bernlohr:2012we} with 18--37 MSTs and different combinations of
SSTs and LSTs.  For this study we use the projected sensitivity of a
candidate CTA configuration with 61 MSTs on a regular grid with 120~m
spacing \cite{Jogler:2012}.  This telescope layout is similar to the
anticipated layout of the baseline MST array with 18--25 telescopes
with an additional US contribution of 36 MSTs.  This configuration has
a gamma-ray angular resolution that can be parameterized as a function
of energy as $\theta \simeq 0.07^\circ(E/100$~GeV)$^{-0.5}$ and a
total effective area above 100 GeV of $\sim$10$^{6}$ m$^{2}$.  We
define the GC signal region as an annulus centered on the GC that
extends from 0.3$^\circ$--1.0$^\circ$ and calculate the sensitivity of
CTA for an integrated exposure of 500 hours that is uniform over the
whole region.  An energy-dependent model for the background in the
signal region is taken from a simulation of residual hadronic
contamination.  The uncertainty in the background model is calculated
for a control region with no signal contamination and a solid angle
equal to five times the signal region (14.3~deg$^{2}$).

Figure~\ref{fig:ctasens} shows the projected sensitivity of our
candidate CTA configuration.
We evaluate
the sensitivity to a WIMP particle annihilating through
three possible final states: $b\bar{b}$, $W^+W^−$, and
$\tau^+\tau^−$. For the Sculptor dSph, one of the best dSph candidates
in the south, CTA could reach $\sim 10^{-24} {\rm cm}^3 {\rm }s^{-1}$
at 1 TeV which is comparable to current limits from H.E.S.S.
observations of the GC halo.   For an observation of the GC utilizing
the same 0.3$^\circ$ to 1.0$^\circ$ annular search region as the
H.E.S.S. analysis CTA could rule out models with cross sections
significantly below the thermal relic cross section down to $\sim
3\times 10^{-27} {\rm cm}^3 {\rm s}^{-1}$. Overlaid in the figure are
WIMP models generated in the pMSSM framework that satisfy all current
experimental constraints from collider and direct detection searches.
Approximately half of the models in this set could be
excluded at the 95\% C.L.  in a 500 hour observation of the Galactic
center.

We compute the sensitivity of CTA using a two-dimensional likelihood
analysis that models the distribution of detected gamma-rays in energy
and angular offset from the GC.  Following the same procedure as the
\Fermi-LAT analysis, we compute the model boost factor as the ratio of
the model cross section with the 95\% C.L. upper limit on the
annihilation cross section of a DM model with the same spectral shape.
For models with LSP masses below 1~TeV, the sensitivity of CTA is
dominated by events at low-energy ($E\lsim 300$~GeV).

The optimal DM search region for CTA will be limited by the CTA FoV of
8$^\circ$ to the area within $2^\circ$ to $3^\circ$ of the GC.  On these
angular scales the DM signal is predominantly determined by the DM distribution
in the inner Galaxy ($R_{GC} <1$~kpc).  We model the Galactic DM distribution
with our benchmark NFW profile with a scale radius of 20 kpc normalized to
0.4~GeV~cm$^{-3}$ at the solar radius.  This model is consistent with all
current observational constraints on the Galactic DM halo in the absence of
baryons.  Because the annihilation signal is proportional to the square of the
DM density, the projected limits for CTA depend strongly on the assumptions
that are made on the shape and normalization of the Galactic DM halo profile,
and the role of baryons in either washing out or enhancing the signal.  The
projected limits presented here could be reduced by as much as a factor of 10,
or improved by an even larger factor given these uncertainties.

Figure \ref{fig:cta_pmssm_sigmav} shows the distribution of the CTA
boost factor versus LSP mass for all pMSSM models and the subset of
models that have an LSP relic density consistent with 100\% of the DM
relic density.  CTA can exclude $\sim$20\% of the total model set and
$>$50\% of the models in the subset with the DM relic density.  Here
we see that $\sim 19$\% of the models would be excluded by CTA if no
signal were to be observed.

\subsubsection{HAWC}

The High Altitude Water Cherenkov (HAWC) observatory, nearing
completion at Sierra Negra in the state of Puebla, Mexico, consists of
a 20,000 square meter area of water Cherenkov detectors: water tanks
instrumented with light-sensitive photomultiplier tubes. The
experiment is used to detect energetic secondary particles reaching
the ground when a 100 GeV to 100 TeV cosmic ray or gamma ray interacts
in the atmosphere above the experiment.  Gamma-ray primaries may be
distinguished from cosmic-ray background by identifying the
penetrating particles characteristic of a hadronic particle shower. As
of summer 2013, the instrument is over 30\% complete and is performing
as designed. The full instrument will be completed in summer 2014.

\begin{figure}
  \begin{center}
    \includegraphics[width=0.7\hsize]{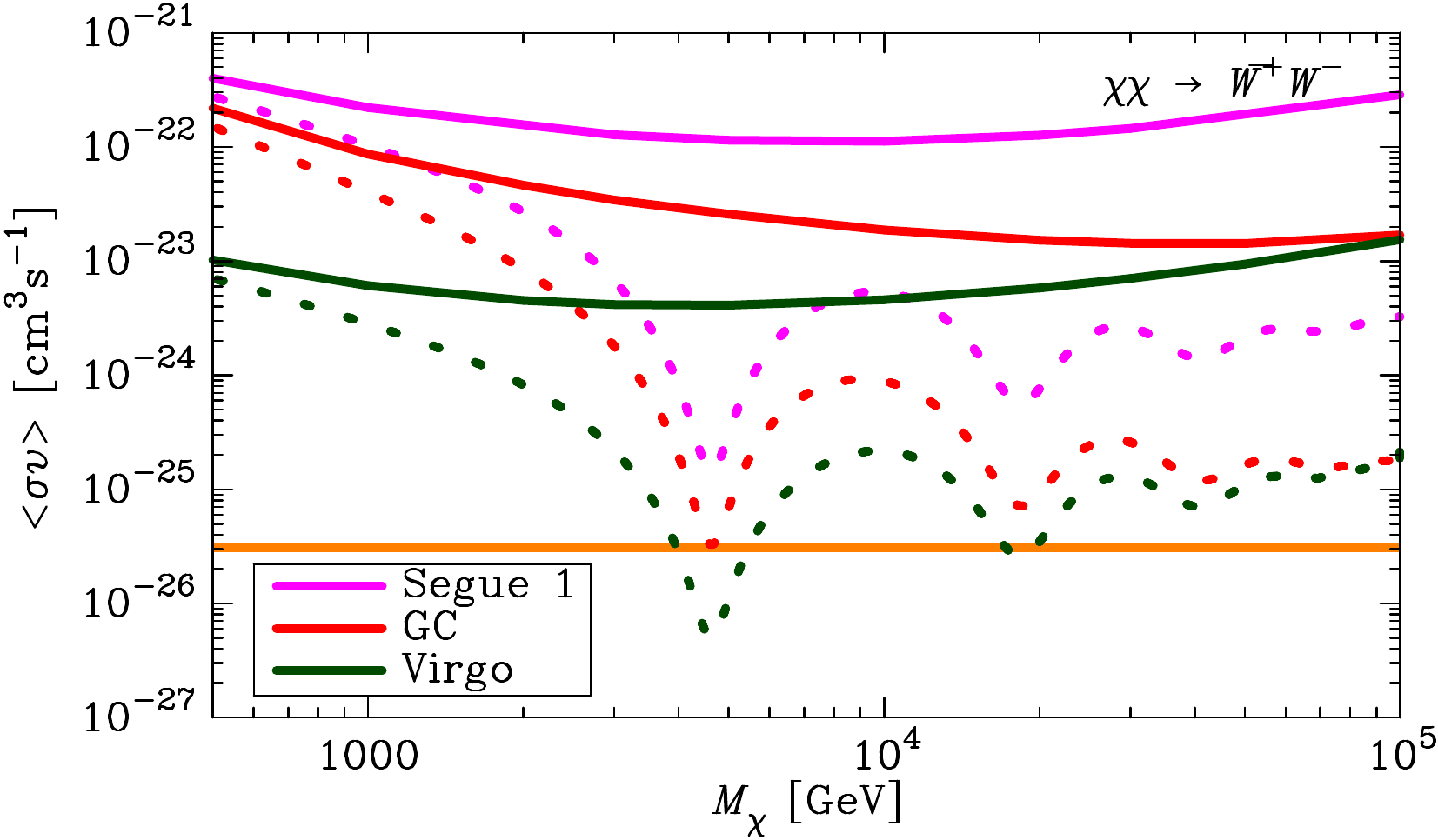}
  \end{center}
  \caption{HAWC sensitivity to the W$^+$W$-$ annihilation channel
    (solid lines) and the same including nonperturbative effects (a
    Sommerfeld enhancement from W exchange)}
\label{fig:hawcww}
\end{figure}

HAWC will complement existing Imaging Atmospheric Cherenkov Telescopes
and the space-based gamma-ray telescopes with its extreme high-energy
sensitivity and its large field-of-view. The instrument has peak
sensitivity to annihilation photons from dark matter with masses
between about 10 and 100 TeV. Much like {\em Fermi}, HAWC will survey the
entire Northern sky with sensitivity roughly comparable to existing
IACTs and will search for annihilation from candidates that are not
known a priori, such as baryon-poor dwarf galaxies. Furthermore, HAWC
can search for sources of gamma rays that are extended by 10 degrees
or more and can constrain even nearby sub halos of dark matter.

\subsection{Constraints on Dark Matter Decay from Gamma-Ray observations}\label{sec:decay}

While this discussion is somewhat beyond the scope of the present work, in some
dark matter particles it is possible that the dark matter is in the form of
an unstable, but very long-lived particle.  Example of decaying dark 
matter include neutralinos with slight R-parity violation as predicted by models when the gaugino masses 
are dominated by anomaly mediated supersymmetry breaking.
\cite{1998JHEP...12..027G,1999NuPhB.557...79R}.  One such candidate of
some recent interest is the TeV-scale decaying Wino that could explain
the PAMELA/AMS excess \cite{2013arXiv1305.0084I}.  Figure~\ref{fig:figdecay}
gives some examples of constraints on decaying dark matter from indirect
detection experiments.

\begin{figure}[tbh]
\begin{center}
\includegraphics[width=0.45\hsize]{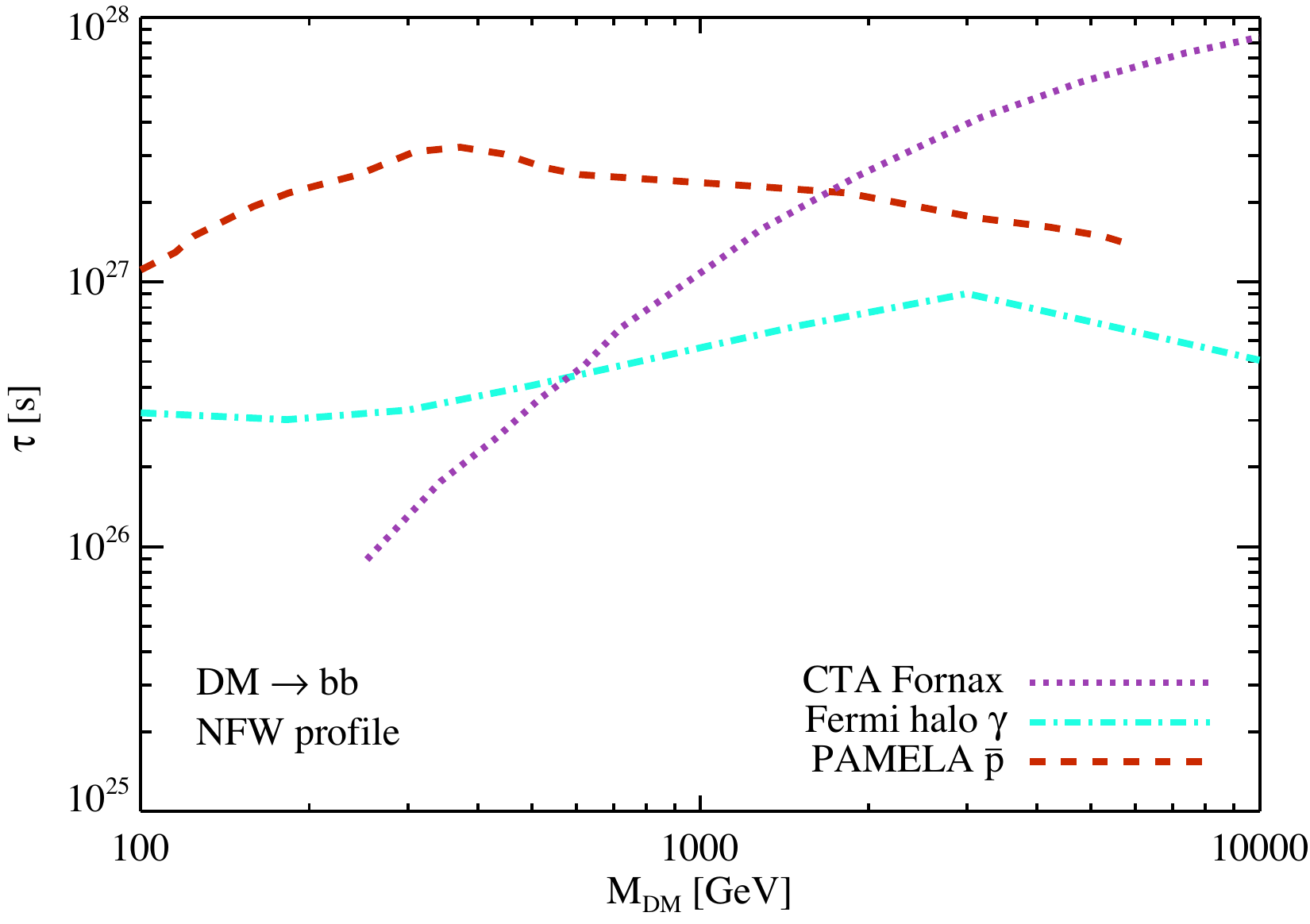}
\includegraphics[width=0.45\hsize]{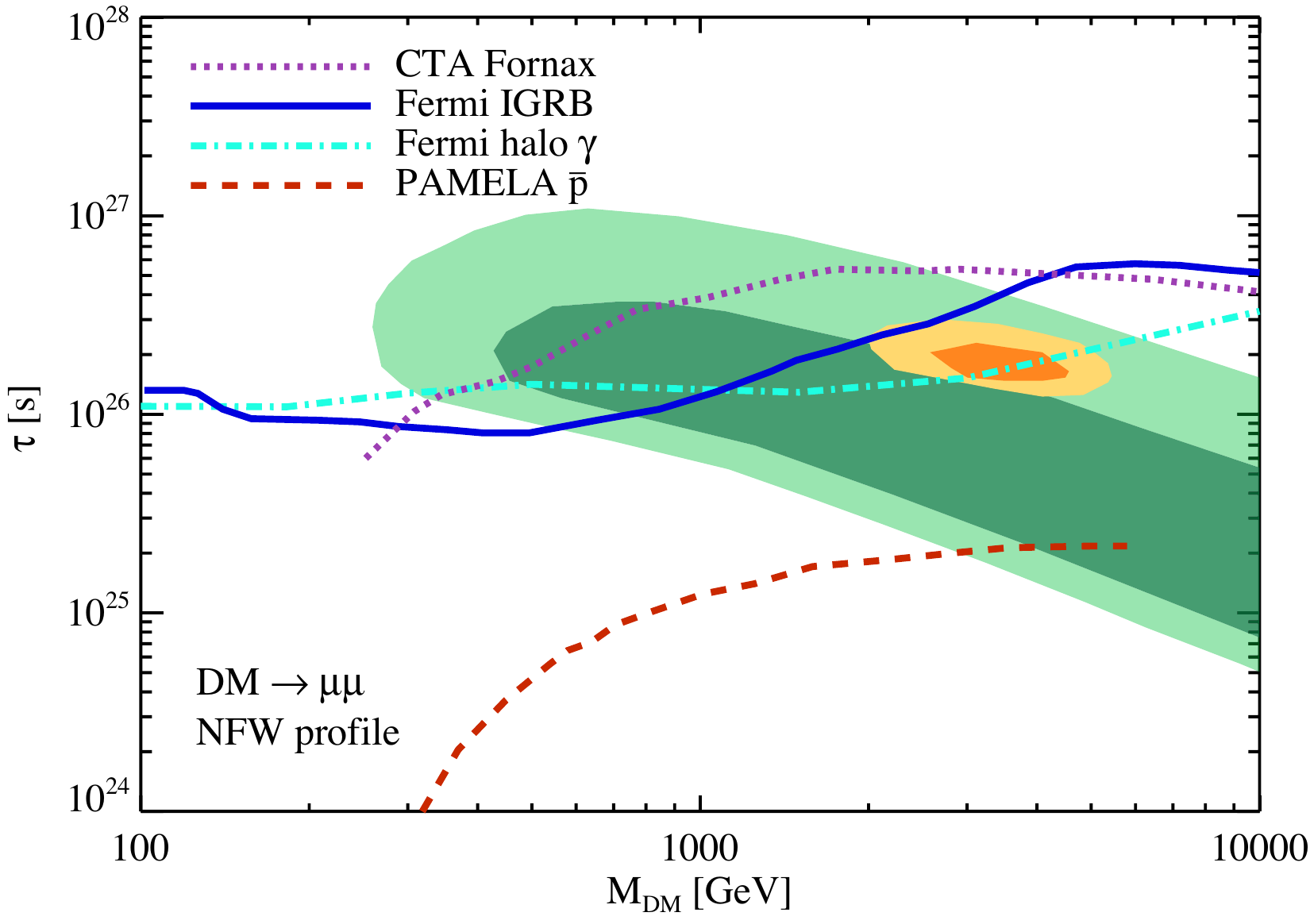}
\caption{Top: Current and projected constraints on the $M_{\rm
    DM}$--$\tau$ parameter space for decay to $b\bar{b}$ from the
  \emph{Fermi} LAT analysis of gamma-ray emission from the Milky Way
  halo~\cite{Ackermann:2012rg}, from PAMELA observations of the
  anti-proton flux~\cite{Cirelli:2013hv}, and projected sensitivity
  for CTA observations of the Fornax cluster in gamma
  rays~\cite{Cirelli:2012ut}.  The curves indicate lower limits on or
  projected sensitivity for CTA to the DM particle lifetime at 95\%
  C.L. (99\% C.L. for the \emph{Fermi} Milky Way halo analysis),
  assuming a NFW density profile.  Bottom: As in top panel, for decay
  to $\mu^{+}\mu^{-}$.  Also shown are constraints from the
  \emph{Fermi} LAT measurement of the Isotropic Gamma-ray Background
  (IGRB) published in~\cite{Abdo:2010nz} as derived
  in~\cite{Cirelli:2012ut}, and regions consistent with the positron
  fraction measurements by PAMELA and \emph{Fermi}, and the PAMELA
  measurement of the antiproton flux, at 95.45\% C.L. and 99.999\%
  C.L. (dark and light green regions, respectively).  Regions
  remaining consistent when the $e^{+}+e^{-}$ fluxes measured by
  \emph{Fermi}, H.E.S.S., and MAGIC are included in the fit are also
  shown at 95.45\% C.L. and 99.999\% C.L. (orange and yellow regions,
  respectively)~\cite{Cirelli:2012ut}. }
\label{fig:figdecay}
\end{center}
\end{figure}

We note that {\em Fermi} observations of galaxy clusters give comparable to or better limits on the dark matter particle lifetime to those shown for the {\em Fermi} Milky Way halo analysis \cite{Dugger:2010ys, Huang:2011xr}.

\subsection{Neutrino Experiments}

\label{sec:neutrino}

Like their gamma-ray counterparts, neutrino telescopes have sensitivity to WIMP
annihilations in the Galactic Center, Halo, from galaxy clusters and from
dSph galaxies.  In addition, neutrino telescopes are sensitive to WIMP annihilations in
the core of the earth or sun, regions that are inaccessible to gamma-ray
telescopes.  In fact, searches for WIMP annihilations in the solar core avoid
many of the model dependencies suffered by searches from other candidate
locations using neutrinos or photons.  The WIMP source in the sun has built up
over solar time, averaging over the galactic dark matter distribution as those
WIMPs that scattered elastically with solar nuclei and lost enough momentum
became gravitationally trapped.  The main assumption one needs to make is that
a sufficient density of WIMPs has accumulated in the solar core that
equilibrium now exists between WIMP capture and annihilation.  Then, given a
WIMP mass and decay branching ratios, one can predict the signal in a neutrino
telescope unambiguously.  As seen in Fig.~\ref{fig:SIConstraints}, the results
from neutrino experiments are and will continue to be very competitive with
direct-detection results, especially for spin-independent scattering.  At WIMP
masses above about 1~TeV, large neutrino telescopes such as IceCube probe a
region of parameter space that is inaccessible to the
LHC~\cite{Silverwood:2012tp}.

\subsubsection{Solar WIMP Model Uncertainties}

In the scenarios for production of secondaries by annihilation of halo
WIMPs, we assumed that the dark matter's distribution was determined
by something that did not depend on any of its particle properties, so
that the only particle properties that matter for the annihilation
rate are the annihilation cross section, and indirectly, the mass
(because the mass density, which we can measure or estimate with
astronomical observations, has to be divided through by the particle
mass to get a number density).  For WIMP capture and annihilation in
the Sun, the other particle properties of WIMPs matter in estimating
the annihilation rate, and hence, the flux of neutrinos emerging from
the annihilations.

Fundamentally, the annihilation rate depends on the capture rate of
WIMPs by the Sun.  WIMPs that scatter off nuclei in the Sun can lose
energy to solar nuclei.  Depending on the reduced mass of the
WIMP-nuclear system and the typical initial speed of WIMPs, WIMPs are
scattered to speeds below the escape velocity of the Sun.  Captured
WIMPs keep re-scattering off solar WIMPs in their ever-tightening
orbit, until they come to equilibrium with solar nuclei at the center
of the Sun.  This implies a dense core of WIMPs at the center of the
Sun, where WIMPs can thus efficiently annihilate.  The equation that
governs the relationship between WIMP creation and destruction
processes is:
\begin{eqnarray} 
  \dot{N} = C - C_AN^2 - C_E N,
\end{eqnarray}
where $N$ is the number of WIMPs in the Sun, and
\begin{eqnarray}
  \label{eq:capture} C = \sum_i \int d^3\mathbf{r}_\odot
  d^3\mathbf{v}d^3\mathbf{v}_i d\Omega f_i(\mathbf{r}_\odot,
  \mathbf{v}_i)f(\mathbf{r}_\odot,\mathbf{v}|\mathbf{v} - \mathbf{v_i}|
  \frac{d\sigma_i}{d\Omega} \|_{v_f < v_{esc}} 
\end{eqnarray} is the capture rate
of WIMPs in the Sun from elastic WIMP-nuclear scattering.  Here $i$ denotes a
nuclear species, $d\sigma_i/d\Omega$ is the elastic scattering cross section,
and the $f$'s are the phase-space densities of nuclei ($i$) and the local WIMP
population.  The final WIMP speed $v_f$ must be less than the escape velocity
from the Sun $v_{esc}$ for the WIMP to be trapped.
The capture rate is quite sensitive to the WIMP mass,
because for WIMPs with $m_\chi \gg m_i$, WIMPs only lose a small bit of kinetic
energy to nuclei.  Thus, for heavy WIMPs, only particles initially moving
slowly with respect to the Sun may be captured.  Note that this
quantity is independent of $N$, the number of captured WIMPs in the Sun, since
we are considering the process by which WIMPs are gathered within the Sun from
the local Galactic population.  In addition, $C_A$ is the annihilation term,
defined such that the annihilation rate is: 
\begin{eqnarray} 
  \Gamma =
  \frac{1}{2}C_A N^2, \end{eqnarray} and \begin{eqnarray} C_A = \langle \sigma
  v_{rel} \rangle \int d^3\mathrm{r}_\odot n^2(r)/N^2. 
\end{eqnarray}
Here, $\mathbf{r}_\odot$ denotes the position of the WIMPs or nuclei
with respect to the center of the Sun.  Note that we have implicitly
assumed self-annihilation in this equation.  If the WIMP mass $m_\chi
< m_i$, then WIMPs can {\it gain} energy by elastic scattering off
nuclei in the Sun.  In fact, they can scatter above $v_{esc}$ and
hence ``evaporate'' from the Sun.  See \cite{gould1990} for a detailed
calculation of $C_E$.  In brief, we see that $\dot{N} \propto N$ for
evaporation because the evaporation depends on the WIMP-nuclear
scattering rate, and hence on one power of the number of WIMPs in the
Sun.  For spin-independent scattering, the evaporation term is only
relevant for WIMPs below $m_\chi \lsim 3.3$ GeV (slightly less than
the mass of helium); solar WIMP searches have no sensitivity below
that mass.  (This sensitivity is bracketed from above at $m_\chi \sim
1$~TeV because the daughter neutrinos of such massive WIMPs experience
absorption as they pass through the solar interior.)

For typical values of the WIMP annihilation cross section ($\langle
\sigma_A v_{rel} \rangle \approx 3\times 10^{-26} \hbox{
  cm}^3{s}^{-1}$) and WIMP-nuclear elastic cross sections 
(that depend on the local Galactic WIMP velocity distribution) capture
and annihilation are in equilibrium, and
\begin{eqnarray}
  \Gamma = \frac{1}{2}C_A.
\end{eqnarray}
Thus, in this special
case, the annihilation rate grows only linearly with the local Galactic WIMP
number density.
(More detailed discussions of this process can be found in Refs.~
\cite{press1985,gaisser1986,griest1987,gould1987,gould1987b,gould1990,edsjo1995,edsjo1997,peter2009b,sivertsson2012}).

The velocity distributions also affect WIMP capture in the Sun, because WIMPs
that are slow relative to the Sun are much more likely to scatter below the
Sun's escape speed than will fast particles.  In particular, if the Milky Way
has a ``dark disk'' or any un-virialized dark-matter structures that
approximately co-rotate with the Sun, the capture rate in the Sun will be
dramatically enhanced \cite{bruch2009}.

For solar WIMP annihilation
we need to know the 
local density structure  (on scales of
Astronomical Units, not the kpc scales that N-body simulations resolve).
It is highly unlikely that we are sitting in a subhalo---the
local volume filling factor of subhalos is at most $10^{-4}$
\cite{diemand2005,vogelsberger2009,kamionkowski2010}.  However,
because of the long equilibrium time in the Sun, the annihilation rate
depends on the time integral of the capture rate.  In this case, small
variations in the density structure of our local patch of the Milky
Way could affect the WIMP annihilation rate in the Sun
\cite{koushiappas2009}.

Of course, for dark-matter particles that have velocity-dependent annihilation
cross sections, the velocity distribution matters as well.  Velocity
distributions tend to show large departures from the traditional
Maxwell-Boltzmann distribution, and imprints of the accretion history of the
host halo \cite{vogelsberger2009,kuhlen2010,lisanti2011,mao2013}.  It is
anticipated that the hyper-local velocity distribution is highly streamy, but
it is unclear how much that streaminess will affect solar WIMP searches---it is
probably not enormously significant \cite{schneider2010}. 
While these local
anisotropies (in velocity and position space) are unlikely to affect this or
other indirect searches, they may well have a significant effect on 
direct detection experiments. 
So while some of these uncertainties may have an impact on the reach of
neutrino searches, they do not significantly change the great
potential for discovery for neutrino experiments, in the nearly
unambiguous detection of a high-energy neutrino signal from the sun.

\subsubsection{IceCube/DeepCore}
\label{sec:IceCubeDeepCoreStatus}

IceCube has taken physics-quality data since 2006 and was completed in
early 2010.  Optimized for detecting neutrinos at TeV to PeV energies,
IceCube has unparalleled sensitivity to neutrinos from WIMPs at masses
of 1~TeV and above.  Since 2009 IceCube has included DeepCore, an
in-fill sub-array with increased module density providing improved
sensitivity to low energy neutrinos and extending IceCube's reach to
WIMP masses as low as roughly 20~GeV. The current configuration of the
IceCube detector is illustrated in Fig.~\ref{fig:IceCubeDetector}.
\begin{figure}[tbh]
  \begin{center}
      \includegraphics[width=0.9\hsize]{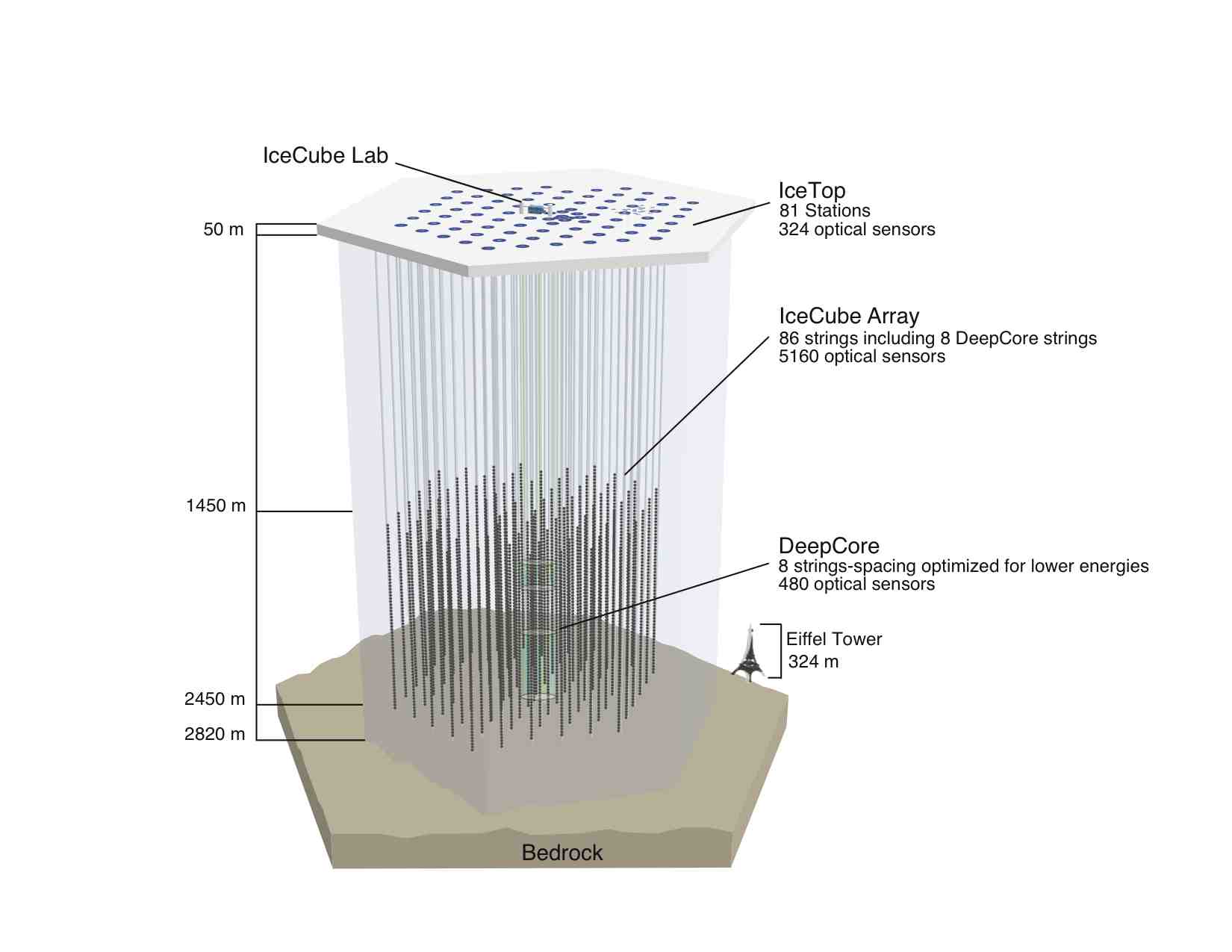}
      \caption{The IceCube neutrino telescope comprises 86 strings
        and 5,160 modules buried in the icecap at the
        South Pole, Antarctica.  The modules have been deployed at
        depths ranging from 1450~m to 2450~m below the surface.
        DeepCore, at the bottom center of IceCube, has modules at
        higher density than the surrounding array for improved
        sensitivity to lower energy neutrinos and hence lower mass
        WIMPs.}
   \label{fig:IceCubeDetector}
\end{center}
\end{figure}
The IceCube Collaboration continues to take data and has performed or
is performing searches for neutrino signals from WIMPs in the center
of the earth~\cite{Achterberg:2006jf}, the solar
core~\cite{Aartsen:2012kia}, the galactic halo~\cite{Abbasi:2011eq}
and center~\cite{Abbasi:2012ws}, galaxy clusters, and dwarf spheroidal
galaxies~\cite{IceCube:2011ae} (see Fig.~\ref{fig:SIConstraints}).
\begin{figure}[tbh]
   \begin{center}
      \includegraphics[width=0.7\hsize]{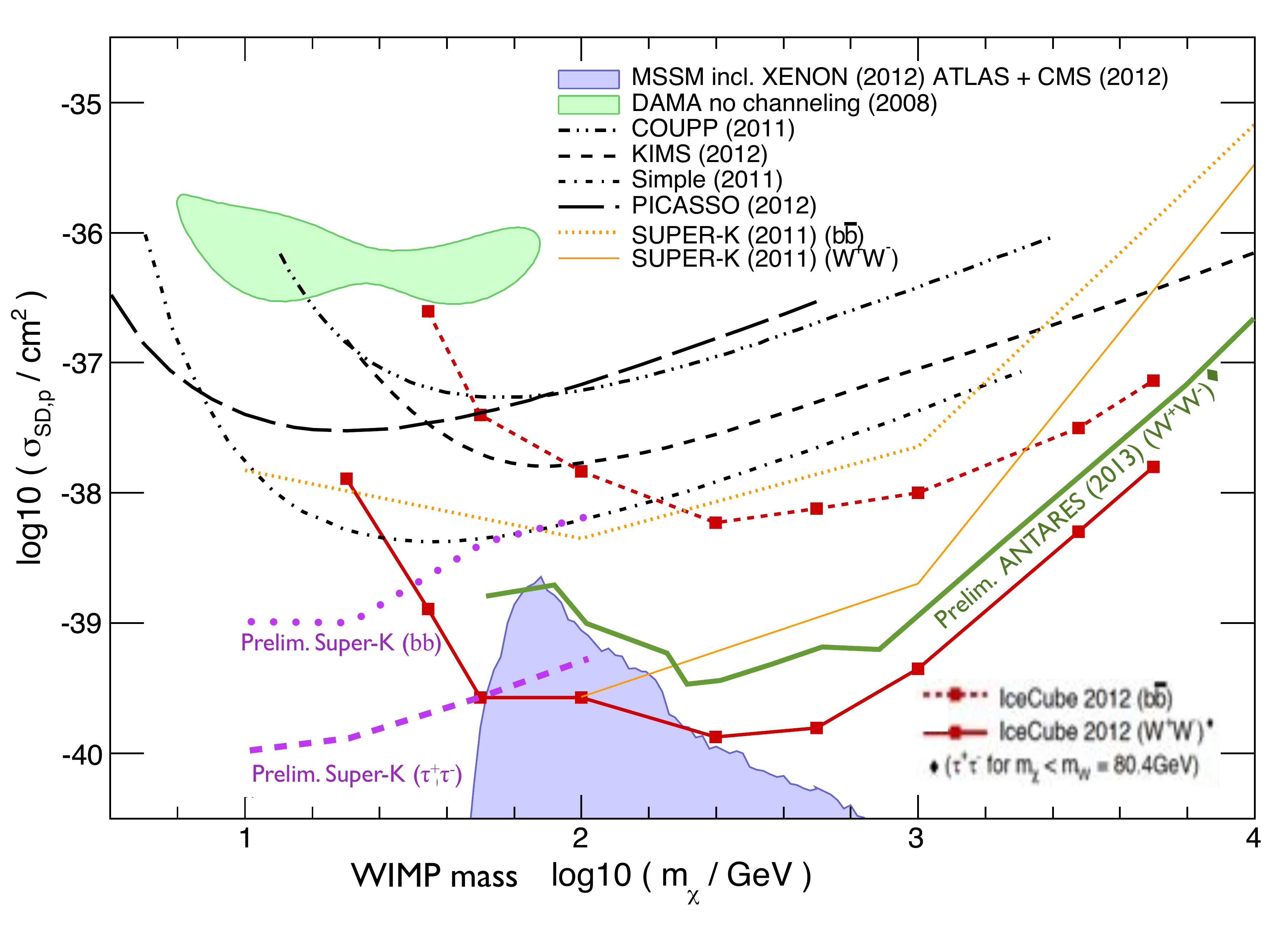}
      \caption{Constraints on the spin-dependent WIMP-proton
        scattering cross section.}
   \label{fig:SIConstraints}
\end{center}
\end{figure}

As a specific example, we present predictions for an IceCube/DeepCore (IC/DC)
search for neutralino dark matter. Our analysis closely follows that presented
in \cite{Cotta:2011ht}. In the results presented here we assume that each
SUSY model results in a neutralino relic density
given by the usual thermal calculation. We use
DarkSUSY 5.0.6 \cite{Gondolo:2004sc} to simulate the (yearly-average) signal
$\nu_{\mu}$/$\bar{\nu}_{\mu}$ neutrino flux spectra incident at the detector's
position and convolve with preliminary $\nu_{\mu}$/$\bar{\nu}_{\mu}$ effective
areas for muon events contained in DeepCore\footnote{These are the same
effective areas that were used in \cite{Cotta:2011ht}, referred to there as
``SMT8/SMT4."}. We consider a data set that includes $\sim 5$yr of data that is
taken during austral winters (the part of the year for which the sun is in the
northern hemisphere) over a total period of $\sim 10$yrs\footnote{In practice
the IceCube/DeepCore treatment of data is more sophisticated, classifying events as
through-going, contained and strongly-contained, and allowing for some
contribution from data taken in the austral summer.  We expect that inclusion
of this data would affect our results at a quantitative, but not qualitative,
level.}. An irreducible background rate of $\sim 10$ events/yr is expected from
cosmic-ray interactions with nuclei in the sun. Here we will take (as discussed
at greater length in \cite{Cotta:2011ht}) a detected flux of $\Phi=40$
events/yr as a conservative criterion for exclusion.

The basic results of this analysis are presented in Figure~\ref{icdc}. In the
figure we show all pMSSM models in our set using grey points, while
highlighting WMAP-saturating models with mostly bino, wino, Higgsino or mixed
($\leq$80\% of each) LSPs in red, blue, green and magenta, respectively.
Detectability is tightly correlated with the elastic scattering cross-sections
($\sigma_{SI}$ and $\sigma_{SD}$) while having little correlation with the
annihilation cross-section $\langle\sigma\upsilon\rangle$, as expected.

The biggest difference between these results and those of the previous analysis
\cite{Cotta:2011ht}, which used a set of pMSSM models that were chosen to have
relatively light ($\leq 1\tev$) sparticles, is that a much smaller percentage
of the current pMSSM models are able to reach capture/annihilation equilibrium
in the sun. This is due to the fact that so many of these models are nearly
pure wino or Higgsino gauge eigenstates (which have both low relic density and
small capture cross-sections) and that the LSPs in this model set tend to be
much heavier than those in the previous set.  If one defines out-of-equilibrium
models as those with solar annihilation rates less than 90\% of their capture
rates, we find that no such models can be excluded by IceCube/DeepCore. In contrast,
relatively light LSPs composed of a mixture of gaugino and Higgsino eigenstates
have large scattering and annihilation cross sections and are highly detectable
by IceCube/DeepCore.  We observe that all such WMAP-saturating well-tempered neutralinos
with masses $m_{LSP}\leq 500\gev$ should be excluded by the IceCube/DeepCore search
(\emph{c.f.}, the magenta points in Fig. \ref{icdc}).

\begin{figure}[hbtp]
    \centering
    \includegraphics[width=1.0\textwidth]{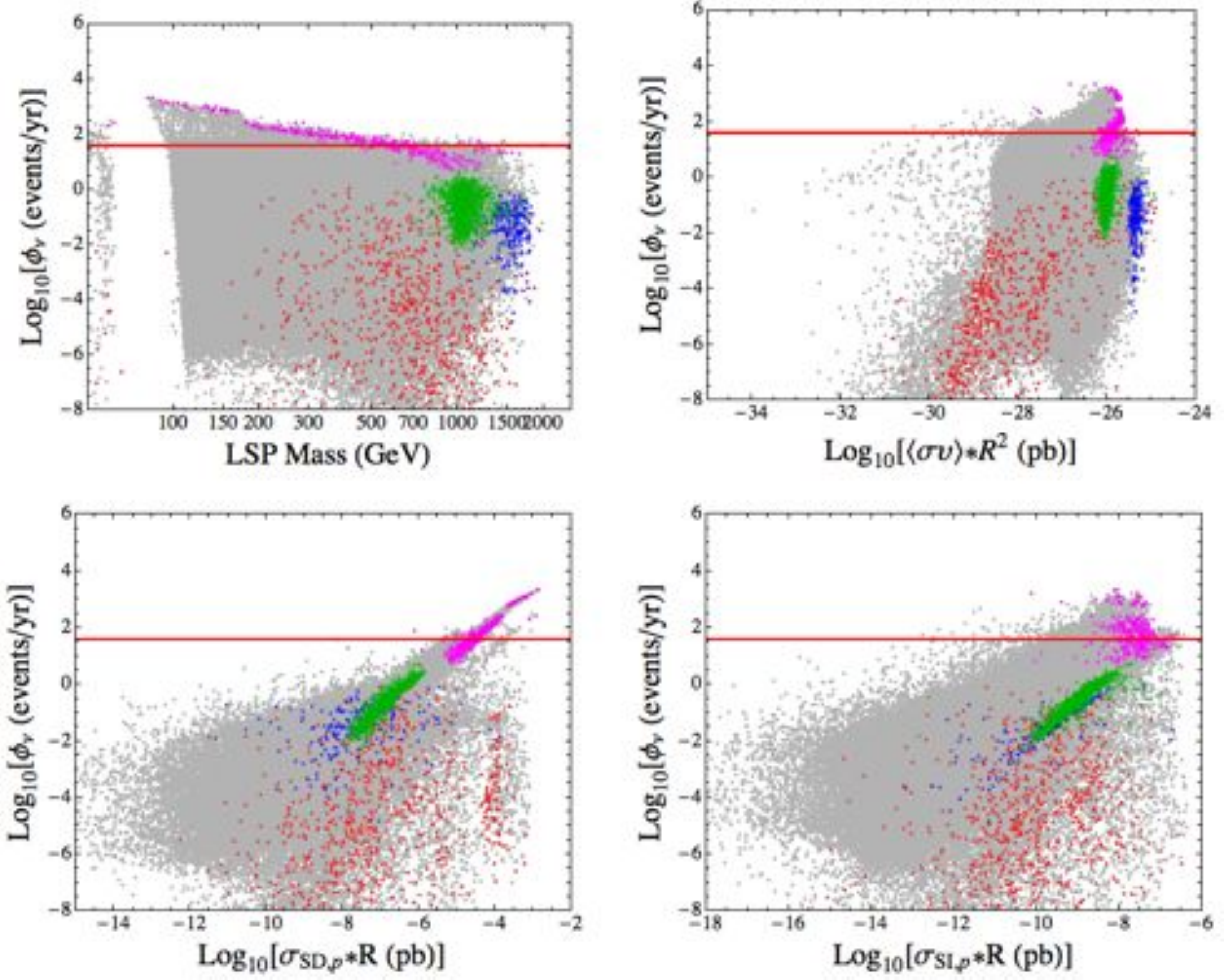}
    \caption{IceCube/DeepCore signal event rates as a function of LSP mass
      (upper-left), thermal annihilation cross-section
      $\langle\sigma\upsilon\rangle R^2$ (upper-right) and thermal
      elastic scattering cross-sections $\sigma_{SD,p}$ and
      $\sigma_{SI,p}$ (lower panels). In all panels the gray points
      represent generic models in our full pMSSM model set, while
      WMAP-saturating models with mostly bino, wino, Higgsino or mixed
      ($\leq$80\% of each) LSPs in are highlighted in red, blue, green
      and magenta, respectively.  The red line denotes a detected flux
      of $40$ events/yr, our conservative estimate for exclusion.}
    \label{icdc}
  \end{figure}

\subsubsection{Super-Kamiokande}

\label{sec:SuperKStatus}
The Super-Kamiokande detector began operation in 1996 and has
consisted of up to 11,146 modules instrumenting a 22~kton fiducial
volume of ultra-pure water.  Super-K has performed searches for solar
WIMPs using more than 3,000 live-days of data with sensitivity to WIMP
masses between 10-100~GeV~\cite{Tanaka:2011uf}(see
Fig.~\ref{fig:SIConstraints}).

SuperK has sensitivity to low-energy neutrinos that have recently been
identified as a complementary indirect detection probe of solar WIMP
annihilations~\cite{Rott:2012qb,Bernal:2012qh}.  Pions are produced abundantly
by the hadronization of annihilation products and by subsequent inelastic
interactions with the dense solar medium.  Stopped positive pions decay,
producing neutrinos at $\sim25-50$~MeV with known spectra, an energy range
covered by the currently-operating Super-K experiment and planned experiments
including Hyper-K and large liquid scintillator detectors such as LENA\@.  The
advantages of this detection channel are high multiplicity of signal neutrinos
(compared to the number of high-energy neutrinos from solar WIMP annihilation),
relatively low backgrounds, insensitivity of the signal spectrum to the
annihilation final states, and detection sensitivity that improves with
decreasing WIMP mass until evaporation becomes important at $\sim 4$~GeV.  The
low-energy neutrino channel is complementary to direct searches and indirect
searches with high-energy neutrinos, and is particularly valuable as an
indirect probe of the WIMP-nucleon scattering cross-section at low WIMP masses.

\subsection{Future Neutrino Experiments}\label{sec:futurenu}

\subsubsection{HyperK}

The Hyper-Kamiokande has been proposed as a next-generation
underground water Cherenkov detector~\cite{Abe:2011ts}.  The design
calls for a fiducial mass of 0.56 million metric tons, about 25 times
larger than that of Super-K, viewed by 99,000 20-inch PMTs.  This will
give the detector 20\% photo-cathode coverage, about half that of
Super-K.  A diagram of the Hyper-K detector is shown in
Fig.~\ref{fig:HyperKDetector}.

\begin{figure}[tbh]
   \begin{center}
      \includegraphics[width=0.7\hsize]{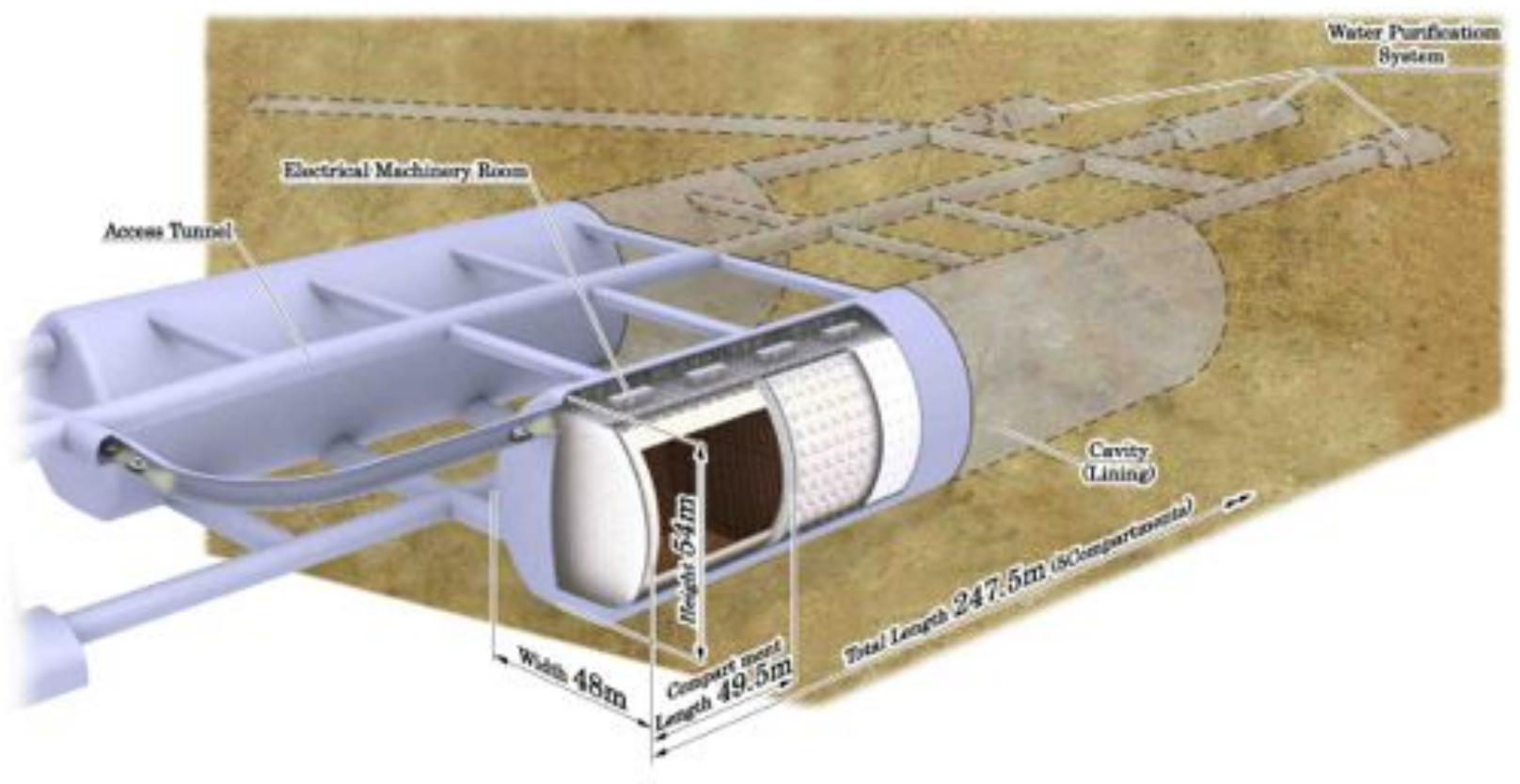}
      \caption{The Hyper-Kamiokande neutrino detector is planned to
        instrument 0.56 million metric tons of water with 99,000 PMTs.
        It will be housed in cavities with 1,750~m water equivalent
        overburden near Super-K in Gifu Prefecture, Japan.}
   \label{fig:HyperKDetector}
\end{center}
\end{figure}
Although focused primarily on other physics goals, the
Hyper-Kamiokande detector will have sensitivity to neutrinos from
solar, earth and galactic halo WIMP annihilations.  For solar WIMPs
annihilating in the soft-channel mode, Hyper-K's cross section
sensitivity could reach down to $10^{-39}~{\rm cm}^2$ at
10~GeV~\cite{Abe:2011ts}.

\subsubsection{IceCube/PINGU}

The Precision IceCube Next Generation Upgrade (PINGU) detector will be
proposed as a new in-fill array for IceCube.  The final geometry for
the detector is still under study but will probably be comprised of
between 20-40 new strings with 60-100 modules per string.  PINGU will
instrument an effective volume ranging from 2-4 million metric tons at
energies from 5-15~GeV, respectively, with $\cal{O}$(1000) modules.

Like Hyper-K, PINGU is focused on other physics goals but has
sensitivity to neutrinos produced by WIMP annihilations.  The two
detectors will be able to probe a WIMP mass region that is of
considerable interest due to intriguing results from other experiments
that are consistent with a WIMP mass of a few GeV.  WIMP properties
motivated by DAMA's annual modulation signal~\cite{Savage:2008er} and
isospin-violating scenarios~\cite{Feng:2011vu} motivated by DAMA and
CoGeNT signals would be testable by detectors with sensitivity to WIMP
masses at the few GeV scale.  The predicted sensitivity of Hyper-K and
PINGU is shown in Fig.~\ref{fig:HyperKPINGUSensitivity}.
\begin{figure}[tbh]
   \begin{center}
      \includegraphics[width=0.7\hsize]{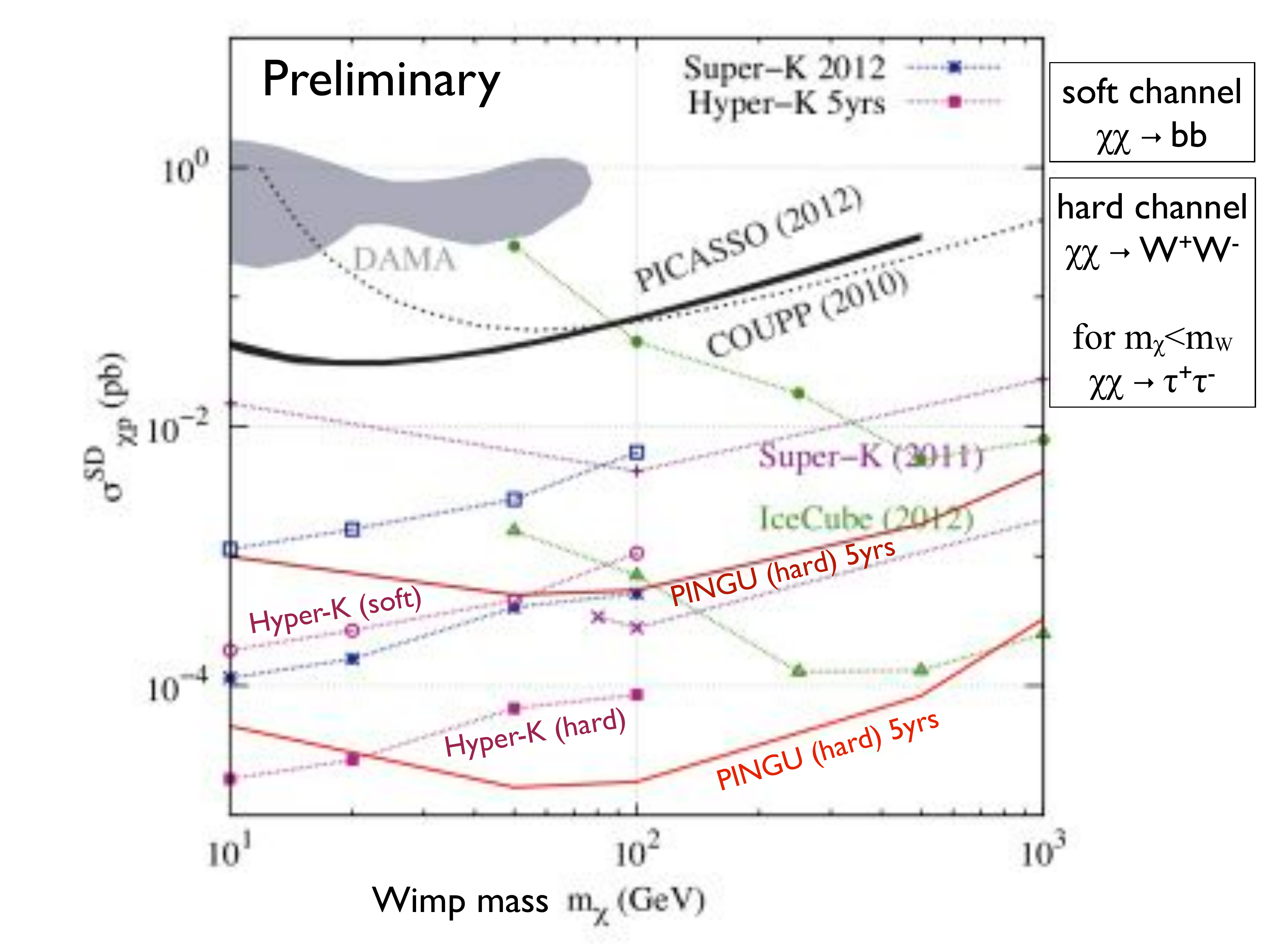}
      \caption{Constraints on the spin-dependent WIMP-proton
        scattering cross section from various current experiments and
        future sensitivities for Hyper-Kamiokande and PINGU with five
        years of data.}
   \label{fig:HyperKPINGUSensitivity}
\end{center}
\end{figure}

\subsection{Astrophysical Multiwavelength Constraints}

\label{sec:multi}

The pair-annihilation of WIMPs yields gamma rays from the two photon
decay of neutral pions, and, in the high energy end of the spectrum,
from internal bremsstrahlung from charged particle final
states. Concerning lower frequencies, a conspicuous non-thermal
population of energetic electrons and positrons results from the
decays of charged pions produced by the hadronization of strongly
interacting particles in the final state, as well as from the decays
of gauge bosons, Higgs bosons and charged leptons. This non-thermal
$e^\pm$ population looses energy and produces secondary radiation
through several processes: synchrotron in the presence of magnetic
fields, inverse Compton scattering off starlight and cosmic microwave
background photons and bremsstrahlung in the presence of ionized
gas. This radiation can actually cover the whole electromagnetic
spectrum between the radio band to the gamma-ray band.

The computation of the multi-wavelength emissions from WIMP-induced
energetic $e^\pm$ involves several steps: (1) the assessment of the
spectrum and production rate of electrons and positrons from dark
matter pair annihilations; 
(2) the computation of the effects of propagation and
energy losses, possibly leading to a steady state $e^\pm$
configuration; (3) the computation of the actual emissions from the
mentioned equilibrium configuration; (4) the evaluation of eventual
absorption of the emitted radiation along the line of sight to derive
fluxes and intensities for a local observer.

The CF2 group recommends that benchmark models be adopted for the
calculation of multi-wavelength emission from WIMP annihilation, at
least in key astrophysical environments such as the Milky Way,
selected nearby galaxy clusters and local dwarf spheroidal
galaxies. The benchmark should include choices for diffusion, magnetic
field intensity and spatial distribution, the background light spatial
distribution and intensity.

Several recent studies have made it clear that radio
\cite{Storm:2012ty} and X-ray frequencies \cite{Jeltema:2011bd} have
the potential to reach a sensitivity to the relevant WIMP parameter
space comparable, and in some instances broader than and complementary
to, gamma-ray experiments. A detailed performance comparison however
critically depends upon assumption on the propagation and energy
losses and, generically, the astrophysical environment where the
secondary radiation is emitted. This makes the accurate definition of
benchmarks 
all the 
more important for this field. A detailed
discussion of the uncertainties stemming from the magnetic field
parameters is given for example in figures 7 and 8 of
\cite{Storm:2012ty}.

At radio frequencies, the DM-induced emission is dominated by the
synchrotron radiation.  The power for synchrotron emission takes the
form \cite{1979rpa..book.....R}:
\begin{equation}
 P_{syn} (\vec x,E,\nu)=
\frac{\sqrt{3}\,e^3}{m_e c^2} \,B(\vec x) F(\nu/\nu_c)\;,
\end{equation}
where $m_e$ is the electron mass, the critical synchrotron frequency is defined
as $\nu_c \equiv 3/(4\,\pi) \cdot {c\,e}/{(m_e c^2)^3} B(\vec x) E^2$, and
$F(t) \equiv t \int_t^\infty dz K_{5/3}(z)$ is the standard function setting
the spectral behavior of synchrotron radiation. Radio emission from dark matter
was studied in a variety of recent works, including \cite{Gondolo:2000pn,radio2,
radio3}.  Near term and future radio facilities will give both large gains in
low frequency capabilities (e.g.~LOFAR, LWA) as well as increased sensitivity
(e.g.~ASKAP, MeerKAT, SKA).

The peak of the inverse-Compton emission from electrons and positrons produced
by dark matter annihilation or decay for a large class of dark matter models
falls in the hard X-ray band \cite{Jeltema:2011bd}. Future hard X-ray
facilities will feature significantly better imaging capabilities. While in the
observation of distant extragalactic objects the instrumental angular
resolution is much less critical than the instrument field of view in the
search for a signal from dark matter, there are other astrophysical
environments where the opposite might be true. A clear example is the Galactic
center, where source confusion plagues the possibility of a solid
discrimination between a diffuse signal from dark matter versus multiple point
sources with gamma-ray observations. Hard X-ray data on the Galactic center
region will greatly help clarify the nature of the high-energy emission from
that region, in particular in connection with a possible dark matter
interpretation.

\begin{figure}[tbh]
   \begin{center}
     \includegraphics[width=0.45\hsize]{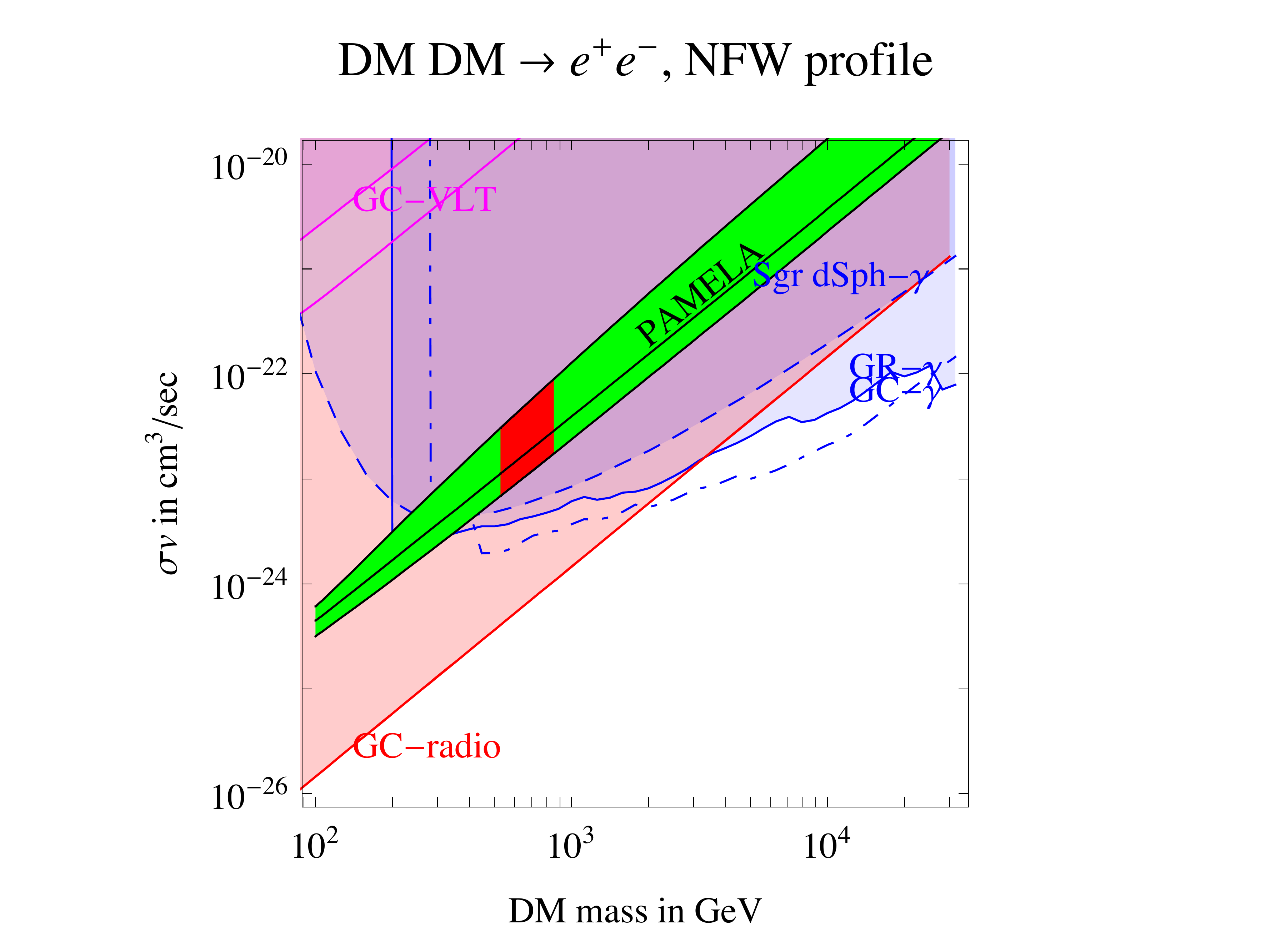}
     \includegraphics[width=0.45\hsize]{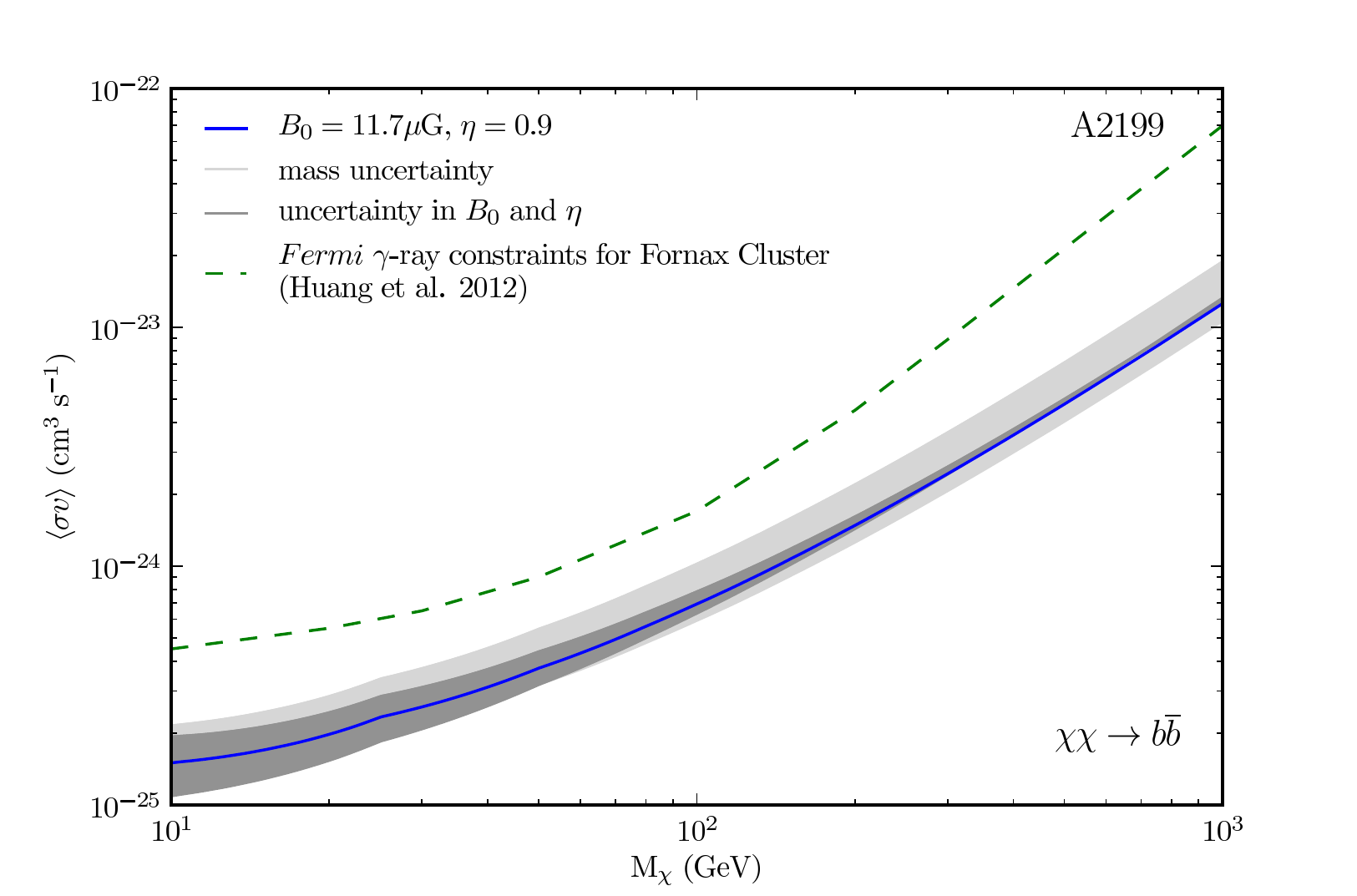}
      \caption{{\it Left:} Gamma-ray and radio constraints on
leptophyllic dark matter from galactic center observations, from 
Ref.~\cite{Bertone:2008xr}{\it Right:} Example radio
constraints on dark matter annihilation using the non-detection of diffuse
radio emission in the A2199 cluster \cite{Storm:2012ty}.  These are compared to
the best current gamma-ray constraints from {\em Fermi} observations of clusters. In
both cases, an NFW only profile is considered; the addition of substructure
would improve the constraints in both cases. Effects of uncertainty in cluster
mass and magnetic field parameters for A2199 are also shown.} 
  \end{center} 
\end{figure}

The multi-wavelength emission from dark matter annihilation was studied in
detail in the seminal works of \cite{baltzwai} for galactic dark matter clumps,
in \cite{Colafrancesco:2005ji} for the case of the Coma cluster and in
\cite{coladraco} for the dwarf spheroidal galaxy Draco. Other recent studies
include X-ray and radio observations of clusters of galaxies
\cite{Storm:2012ty, ophiuchus, colabullet} and an analysis of the broad-band
dark-matter annihilation spectrum expected from the galactic center region
\cite{ullioregis}. In addition, radio emission from $e^+e^-$ produced in dark
matter annihilation was considered as a possible source for the ``WMAP haze''
in the seminal paper of \cite{haze1}, and subsequently analyzed in detail in
\cite{haze2}, \cite{haze3} and \cite{2008arXiv0801.4378H}. Other studies have
also previously addressed synchrotron radiation induced by dark matter
annihilation (e.g., \cite{Gondolo:2000pn,radio3,radio2}).

Among the possible targets for mulit-wavelength searches for a signature of dark
matter annihilation, clusters of galaxies stand out as excellent candidates for
several reasons. They are both massive and dark matter dominated, and in terms
of searches for secondary IC or synchrotron radiation signals the effect of
spatial diffusion in clusters is typically negligible due to the fact that the
high energy electrons and positrons produced in dark matter annihilation or
decay events lose energy much faster than the time needed for them to diffuse
through cluster scales \cite{Colafrancesco:2005ji}. Limits on dark matter
annihilation were derived by \cite{Storm:2012ty} for a sample of nearby
clusters using upper limits on the diffuse radio emission, low levels of
observed diffuse emission, or detections of radio mini-haloes. It was found
that the strongest limits on the annihilation cross section are better than
limits derived from the non-detection of clusters in the gamma-ray band by a
factor of $\sim 3$ or more when the same annihilation channel and substructure
model, but different best-case clusters, are compared. The limits on the cross
section depend on the assumed amount of substructure, varying by as much as 2
orders of magnitude for increasingly optimistic substructure models as compared
to a smooth NFW profile.

Local dwarf spheroidal (dSph) galaxies are also potentially good targets for
multi-wavelength searches; unlike the galactic center region or galaxy clusters,
no significant diffuse radio, X-ray or gamma-ray emission is expected: the
gravitational potential well of dSph galaxies is too shallow for them to host
any sizable thermal bremsstrahlung emission at X-ray frequencies, and, more
importantly, the gas densities appear to be extremely low
(e.g.,\cite{1998ARA&A..36..435M}).  Constraints on particle dark matter models
from observations of dSph with X-rays were given in \cite{jeltemaprofumoxray}.
The hard X-ray regime was analyzed more recently in \cite{hardxray}.

Future studies and future
observations hold much promise for even tighter constraints on dark
matter from multi wavelength, and specifically radio, observations.

\section{Experimental Methods and Key Enabling Technologies}

\label{sec:technique}

\subsection{Atmospheric Cherenkov Detectors}

An Imaging Atmospheric Cherenkov Telescope (IACT) is essentially a
wide-field optical telescope consisting of a large reflector
(typically in a short focal-length f/0.6 to f/1.5 optical system) with
a high-speed camera in the focal plane.  To date, most ACT cameras use arrays of hundreds
of photomultiplier tubes for their cameras, although new cameras are under
development that make use of solid-state photomultipliers (SiPMs) where each
pixel is made up of an array of small $\lsim$100$\mu$m geiger-mode avalanche
photodiode (APD) cells.  Very large reflectors, and short exposures
($\lsim 20$nsec) are required to detect the faint flashes of Cherenkov
light against the Poisson fluctuation in the night sky background.
The signal to noise-ratio for detecting Cherenkov light flashes is
proportional to the square root of the mirror area $A_{\rm m}$ times
the reflectivity of the optics and quantum efficiency of the photodetectors
$\epsilon$ and inversely proportional to the square root of the
signal-integration timescale $\tau$ and solid angle of the pixels
$\Omega_{\rm pix}$.  Since the energy threshold is inversely
proportional to the $SNR$ it is advantageous to maximize the mirror
area and throughput of the optical system to minimize the threshold.
Operation at a dark site (and on moonless nights) is also important.
Higher-speed photodetectors and electronics are required to minimize the
integration time, ideally reducing this down to the shortest intrinsic
timescale of the Cherenkov light wavefront (a few nanoseconds).  Waveform
digitizers are often employed in the camera electronics to handle the
variation in the time profile of events with increasing impact parameter,
and to buffer the event data for a long enough time to allow for the development
 of the trigger. The
minimum angular size of the shower is determined by the angular extent
of the core of the lateral distribution which is roughly $0.1^\circ$
(full width) for a few-hundred GeV shower viewed at the zenith.
To resolve the structure of a shower image, the angular resolution of
the telescope, angular diameter of the pixels and the pointing
accuracy of the telescope mount should all ideally be $\lsim$0.1~deg.
While the angular resolution requirement is considerably more relaxed
than for an optical telescope (with $\lsim$ arcsec optics), the field
of view is substantially larger than most optical telescopes, with a
$\gsim 3.5^\circ$ FoV required to contain shower images from impact
parameters $\sim 120$m.  Larger fields of view (up to 8$^\circ$ diameter)
are being considered for the future CTA telescopes to provide a larger
field of view for more efficient survey of the Galaxy, and better sensitivity
to extended sources.

Images of the electromagnetic showers initiated by gamma rays have a compact elliptic shape,
and the major axis of the ellipse indicates the shower axis projected
onto the image plane.  In contrast, the image of showers produced by
cosmic-ray protons show complex structure due to hadronic interactions
that produce neutral pions (and electromagnetic sub-showers initiated
by the prompt decays of $\pi^0$s to gamma-rays) as well as penetrating
muons from the decay of charged pions.  The images of Cherenkov light
observed by an individual IACT are typically analyzed by using a
moment-analysis of the images to derive a set of quantities that
characterize the roughly elliptical shape of the images.  These are
referred to as the \textit{Hillas parameters} after Michael Hillas, who
first proposed the present definition of these parameters
\cite{Hillas1985}.  Gamma rays are selected based on cuts on these parameters.

Arrays of telescopes provide stereoscopic imaging, with multiple images that
better constrain the direction and energy of the gamma-ray shower.  For
analysis of data from multiple telescopes, one typically extends the moment
analysis approach and derives weighted combinations of the width and length
parameters.  A favored method is to calculate the mean-scaled-parameters (i.e.,
the mean-scaled-width {\emph MSW}) and mean-scaled-length ({\emph MSL}) from
the Hillas parameters of the individual telescopes(\cite{Konopelko95, Daum97}):

IACTs achieve their gamma/hadron separation based on intrinsic
rejection of background at the trigger level, data selection based on
the shape of the shower image (MSL and MSW) and on angular
reconstruction and restriction of the point of origin of each
candidate \gr event to the source region.  IACTs are triggered when
several pixels in a camera exceed threshold, within some time
coincidence window.  By requiring a high multiplicity $m$ for the
coincidence, and a narrow time window $\tau$ the accidental trigger
rate from the night sky background $R_{\rm acc}$ can be reduced well
below the single pixel trigger rate $R_1$ according to the relation
$R_{\rm acc} \propto R_1^m \tau^{m-1}$. 
Topological
constraints (such as strict adjacency of triggering pixels, or
requirements that the pixels lie within camera sectors) further reduce
the accidentals rate from night-sky-background (and associated PMT
after-pulsing).  The local trigger signals from the individual
telescopes are delayed (based on the position of the source) and
brought into coincidence in an \emph{array trigger} with a
coincidence window $\sim$40-100 nsec. 
Cosmic-ray showers are also rejected at the trigger level due to their
lower Cherenkov light yield than gamma-rays since for every neutral
pion that gives rise to electromagnetic showers through
$\pi^0->\gamma\gamma$ decay, roughly two thirds of the available
energy goes into the formation of charged pions and penetrating muons
and neutrinos that effectively reduce the available energy for the
cascade.  

Current generation experiments like VERITAS have typical data rates of
about $50 TB/year$.  With an order of magnitude more pixels, lower energy
thresholds, and at least an order of magnitude more telescopes, future 
ground-based gamma-ray experiments like CTA could easily result in several
orders of magnitude higher data rates, if realtime (pipelined) compression
and image processing is not done at the experiment.

Key enabling technologies for IACTs are high-quantum efficiency UV-blue
photodetectors, low-cost large area mirrors, and high-speed waveform digitizers
and data acquisition systems to handle the prodigious data rates.

\subsection{Neutrino experiments}

The IceCube neutrino observatory consists of 5160 optical sensors or Digital
Optical Modules (DOMs) installed on 86 strings between 1450 m and 2450 m below
the surface. Fifteen of the 86 strings were deployed at higher module density
and comprise the sub-array DeepCore.  DeepCore was designed to provide
sensitivity to neutrinos at energies over an order of magnitude lower than
initially envisioned for the original array, as low as about 10~GeV.  The
In-Ice array is complemented by a surface array, called IceTop. IceTop consists
of 160 ice-tanks, in pairs, near the top of each string. Each tank has two DOMs
for redundancy and extended dynamic range.

Each string is instrumented with 60 DOMs capable of detecting signals over a
wide dynamic range, from a single photon to several thousands arriving within a
few microseconds of each other.  Strings are deployed in a triangular grid
pattern with a characteristic spacing of 125 m (72~m for DeepCore) enclosing an
area of 1 km$^2$. Each hole cable, which carries 60 DOMs, is connected to a
surface junction box placed between the two IceTop tanks. The IceTop DOMs are
also connected to the surface junction box. A cable from the surface junction
box to the central counting house carries all DOM cables and service wires.
Signals are digitized and time stamped in the modules. Times and waveforms from
several modules are used to reconstruct events from the Cherenkov light emitted
by charged particles in the deep ice and in the IceTop tanks.

\subsection{Technology overlaps with other CF subgroups and Frontiers}
\label{sec:technology}

The main instrumentation issue confronting the high energy gamma-ray
and neutrino detector community is the cost of photon detection
technology. Traditional PMTs have the requisite performance to extract
the physics but their cost is prohibitive.  A less expensive photon
detection technology with comparable performance to PMTs would open up
a number of highly interesting opportunities in fundamental neutrino
physics, supernova neutrino burst detection, proton decay and indirect
dark matter detection.

Development of UV-blue sensitive photodetectors is not only important for
Atmospheric Cherenkov Telescopes (like CTA), but also for
HEP detectors (Cherenkov and
new high-yield scintillators), and liquid noble detectors.  
The 
wavelengths of interest are roughly 125nm for liquid Ar fluorescence, 
175nm for Liquid Xe fluorescence,
~320nm for ground-based instruments, and (as an example of a modern
high-yield scintillator) ~380nm for LaBr3.  Despite
progress in Silicon-based solid state detectors (e.g., Geiger-mode APDs)
the surface of Silicon has a poor response in this regime.  
Development of photodetectors based on III-V (nitride) semiconductors
(AlN, GaN - InN) offers potential for this wavelength regime. 

Neutrino and gamma-ray DM-indirect detection (ID) detectors as well
as liquid-Noble DM-Direct Detection experiments rely on high speed
(>100 Msps) waveform digitizers.  Development in this area would be of
broad benefit.  For neutrino and gamma-ray experiments, analog
pipeline ASICs with deep memory (up to $\sim$ several $\mu$secs)
are likely to be a key enabling
technology.  For DD, waveform digitizers with even deeper memory (up
to millisec depth) and wider dynamic range (up to 14 bits) are
desirable and currently only FADCs are viable.  Outside of the Cosmic Frontier, 
other detector developments (e.g., psec-timing devices for particle
ID) also rely on high-speed waveform digitizers (e.g, multi Gsps
analog pipelines as developed for the LAPPD project) with similar
architecture.

Microchannel plates (MCPs) and MCP based photodetectors
also play a unique roll in photodetection, providing higher gain,
lower electronic read noise than other solid state detectors- while
still providing high resolution imaging.  The LAPPD project has
developed ALD coated detectors suitable for low counting rate
applications.  The Berkeley SSL group has demonstrated that
cross-strip detectors can be used to achieve mega-pixel resolution by
centroiding the MCP charge cloud, giving these devices the potential for
combining single photon detection, essentially unlimited frame
rates, but very high spatial resolution.  However, current MCPs are not
suitable many of the most interesting applications in the cosmic
frontier (e.g, atmospheric Cherenkov detectors, widefield optical
surveys) where the MCP is exposed to a high level of night-sky
background.  This results in the need for higher bias currents,
resulting in thermal runaway or short lifetimes due to the limits of
the MCP secondary emission coating.  ALD offers a way to address these
issues, making a new class of imaging, photoncounting, fast timing
detectors possible.   Efforts to produce MCP-based photodetectors with
lower gains, and longer lifetimes are key advances needed to find broader
applications of MCP photodetectors to CF experiments.

For detectors with ~megaton and larger fiducial volumes, like IceCube and the
possible HyperK, KM3NeT, PINGU and MICA detectors, better than ns-scale timing
resolution is not needed.  The Cherenkov photons being detected have generally
traveled through tens of meters of medium (ice or water) and experienced delays
due to scattering.  Their timing residuals--the difference in arrival time for
a given photon undergoing scattering vs that same photon traveling in a
straight line--easily reach ns-scales. Also, segmentation of the photosensitive
element can improve the directionality these large detectors.  For instance,
instead of housing a single large PMT inside each pressure vessel, KM3NeT is
likely to put dozens of much smaller (~3") PMTs inside, each facing outward in
a different direction.  They can then determine from which direction a given
photon came by using their knowledge of which PMT or PMTs got hit.
Segmentation is thus readily achieved by adding more photosensitive elements,
something that is feasible provided the cost of individual elements is low. 
For future neutrino telescopes, the key enabling technology would be 
low-cost blue-sensitive, large area photodetectors.   Preferably such devices
would be available from a number of vendors to reduce risk (technical risk,
budgetary risk and schedule risk) for this critical component of these 
experiments.

\section{Tough Questions}\label{sec:toughquestions}

As part of this planning process, tough questions were solicited from the particle physics community, and addressed to the various subgroups.  Here, we 
explicitly answer two tough questions that were addressed to the CF2 subgoup.

(1) Tough question CF11 ``{\sl Can dark matter be convincingly discovered by
indirect searches given astrophysical and propagation model uncertainties? Do
indirect searches only serve a corroborating role?}'' 

As discussed extensively in \S\ref{sec:positron} uncertainties in propagation
models have a significant impact on the interpretation of the positron excess.
However, a sharp spectral feature could stand out
above the background, producing a strong indication of a dark matter signal.
Some have argued for a dark matter interpretation in the AMS signal from the
observed high level of spatial isotropy in the signal \cite{Aguilar:2013qda},
the argument being that an astrophysical source (such as a nearby Pulsar) would
reveal an observable anisotropy.   However, this argument is
problematic, since a secondary source of positrons (from cosmic-ray
interactions) could also yield a high level of isotropy.  In general, an
observable positron signal (at the level of the current detection) would
require a large astrophysical or particle physics boost.  An astrophysical
boost due to a nearby dark matter clump would result in an anisotropy similar
to a pulsar or other point source.   

However, the detection of a high energy neutrino signal from the sun, or
a narrow gamma-ray line could provide a smoking-gun signal, and more than
just a corroborating evidence.   Even if the LHC or Direct Detection experiment
sees a signal from dark matter, gamma-ray measurements would go further than
just confirming such a detection: the observed spectrum would provide much
better constraints on the dark matter particle mass, and details of the spectrum could help to identify the primary annihilation channels.  Indirect detection,
in this sense, is better described as {\it complementary} rather than 
{\it corroborating}.

(2) Tough Question CF12 ``{\sl Given large and unknown astrophysics
uncertainties (for example, when observing the galactic center), what is the
strategy to make progress in a project such as CTA which is in new territory as
far as backgrounds go? How can we believe the limit projections until we have a
better indication for backgrounds and how far does {\em Fermi} data go in terms of
suggesting them? What would it take to convince ourselves we have a discovery
of dark matter?}''

\begin{figure}[tbh]
\begin{center}
\includegraphics[width=0.8\hsize]{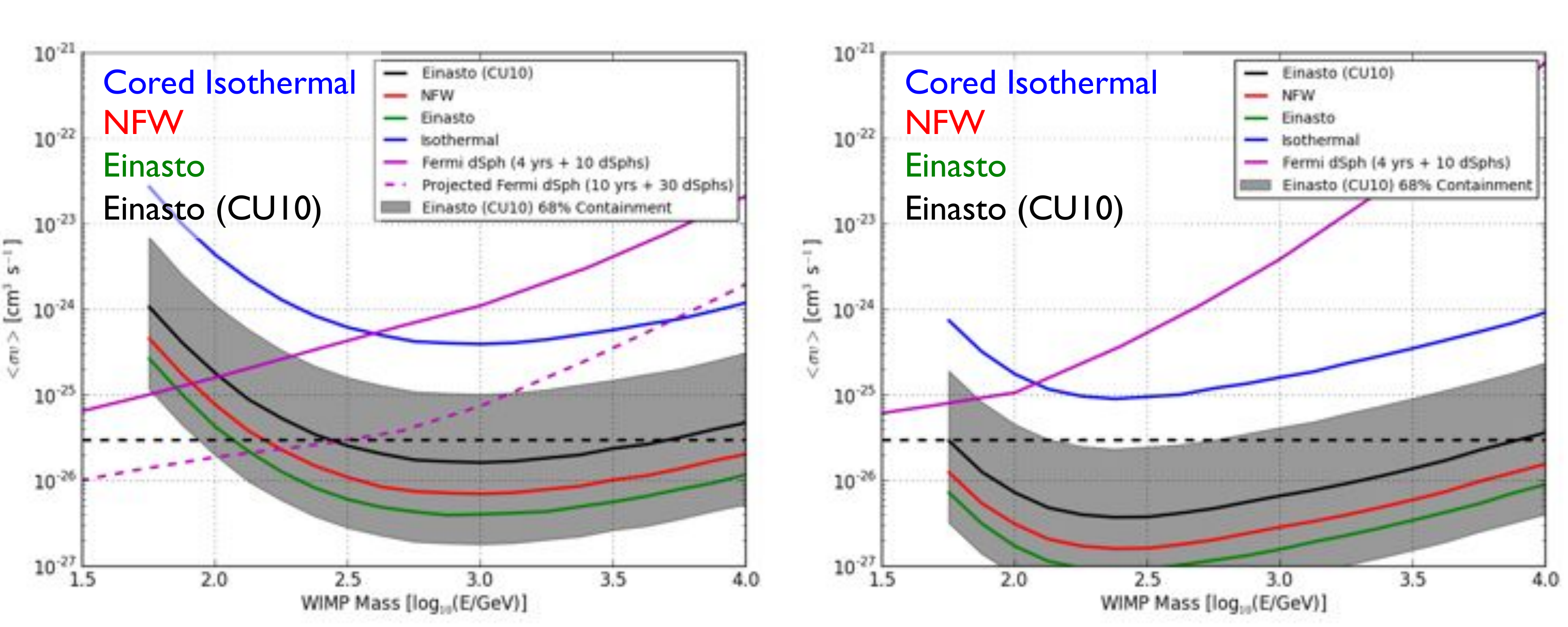}
\caption{Estimated sensitivity of the enhanced CTA experiment to dark
matter for Galactic center observations, taking into account the most likely
and most conservative halo models.
}
\label{fig:gchalouncert}
\end{center}
\end{figure}

It should be emphasized that nearby dwarf satellite galaxies have no known
high energy ($\gsim$GeV) backgrounds, and predictions for the annihilation
signals are robust and relatively insensitive to astrophysical uncertainties.  However, it is true that the astrophysical
backgrounds for the Galactic center (GC) are quite large and uncertain.
Similarly, models for the halo profile in the inner Galaxy have large
uncertainties: as discussed in this report, baryonic matter could erase the central cusp leaving a cored
halo, or adiabatic growth of the central black hole could result in
a very steep inner halo profile that leads to a very large enhancement in the
gamma-ray signal.  The dominant gamma-ray
background at the GC comes from the bright point source at the position of
Sgr~A*. Since the diffuse gamma-ray background
falls as a fairly steep function of energy, the confusion from the strong
diffuse gamma-ray backgrounds that makes measurements at GeV energies (with
{\em Fermi}) very difficult are significantly reduced at energies above 100 GeV (
as detected with
ground-based gamma-ray instruments).  Combined with the excellent angular
resolution of ground based instruments $\lsim 0.1^\circ$, it is possible to
show that the GC source is consistent with a point source to within this
resolution.  It is also possible to determine the distribution of other sources
along the galactic plane.  Sensitivity estimates for the Galactic center region
take this knowledge of source locations and create exclusion regions.   DM
upper limits produced with H.E.S.S., e.g., \cite{hessgclimit} exclude these regions but
do not make any further attempt to subtract (or perform a joint fit) with other
diffuse backgrounds, and hence are robust.  These upper limits already reach a
level only an order of magnitude above the GC.   Even with the diffuse
backgrounds present in the {\em Fermi} data, one can 
estimate that
{\it with no background subtraction} the integral {\em Fermi} flux between 1-3~GeV
and within $1^\circ$ or the GC ($F\approx 1\times 10^{-7}{\rm
cm}^{-2}{\rm s}^{-1}$) gives an upper limit of $\langle\sigma\, v\rangle\approx
1.6\times 10^{-25}{\rm cm}^3{\rm s}^{-1}$ less than an order of magnitude above
the natural cross-section (T. Linden, private communication).  

The best information about the extent of this point
source and other sources actually comes from ground-based observatories, not
{\em Fermi}.  It is true
that there is some uncertainty in the GC fluxes resulting from assumptions
about the halo profile in the inner Galaxy.  Moreover, neither N-body
simulations or dynamical measurements (the rotation curve of the Galaxy) {\it
directly} constrain the Galactic halo profile on scales much less than ~1kpc.
But given the bright point source at the GC, estimates already exclude the
innermost part of the halo profile and rely on an annulus around the galactic
center, somewhat mitigating this error.
To quantitatively address this real systematic error, in
Fig.~\ref{fig:gchalouncert} we show the effect of uncertainties in the GC halo 
model on the dark matter constraints.  This figure
clearly indicates that the GC measurements with an augmented CTA instrument
are still sensitive
enough to constrain the all but the most conservative (and physically
unrealistic) isothermal distribution.  

\section{Complementarity}\label{sec:complementarity}

While a much more detailed study of complementarity of Dark Matter detection
techniques is given in the CF4 report, here we summarize some of the most
important observations made from our scans using the pMSSM benchmark models.
Fig.~\ref{figxx} shows the
survival and exclusion rates resulting from the various dark matter
searches and their
combinations in the LSP mass-scaled SI cross section plane (coordinates
most relevant for direct detection experiments, but illustrative of the
overall complementarity of techniques). In the upper left
panel we compare these for the combined direct detection (DD = XENON1T +
COUPP500) and indirect detection (ID = \Fermi + CTA) DM searches.  While
the distribution of model points, and priors for these distributions are
a subject for religion rather than science, these scans provide some sense
of how the different experimental techniques constrain the model space.
For example, we see
that 11\% (15\%) of the models are excluded by ID but not DD (excluded by DD
but not ID) while 8\% are excluded by both searches. On the other hand, we also
see that 66\% of the models survive both sets of DM searches; 41\% of this
subset of models, in turn, are excluded by the LHC searches. Note that the DD-
and ID-excluded regions are all relatively well separated in terms of mass and
cross section although there is some overlap between the sets of models
excluded by the different experiments. In particular we see that the ID
searches (here almost entirely CTA) are covering the heavy LSP region even in
cases where the SI cross section is very low and likely beyond the reach of any
potential DD experiment. Similarly, in the upper right panel we see that 22\%
(0\%) of the models are excluded uniquely by DD (IceCube) only while 1\% can be
simultaneously excluded by both sets of searches and 77\% would be missed by
either search set. In the lower left panel, ID and LHC searches are compared
and we see that 17\% (32\%) of the models would be excluded only by the ID
(LHC) searches. However, 2\% (50\%) of the models are seen to be excluded by
(or would survive) both searches. The strong complementarity between the LHC,
CTA and XENON1T experiments is evident here as CTA probes the high LSP mass
region very well where winos and Higgsinos dominate, a 1T Xenon detector chops
off the top of the distribution where the well-tempered neutralino LSP states
dominate, while the LHC covers the relatively light LSP region (independent of
LSP type) rather well. Of course the strength of the LHC coverage will
significantly improve in the future as here only 7 and partial 8 TeV analysis
results have been employed. In the lower right panel the relative contributions
arising from the LHC and CTA searches to the model survival/exclusion are
shown. Here the intensity of a given bin indicates the fraction of models
excluded there by the combination of both CTA and the LHC while the color
itself indicates the weighting from CTA (blue) and LHC (red).  It is again
quite clear that CTA completely dominates for large LSP masses (which
correspond to the mostly wino and Higgsino LSPs) and also competes with the LHC
throughout the band along the top of the distribution where the LSP thermal
relic density is approximately saturating WMAP/Planck. The LHC covers the rest
of the region but is not yet as effective as CTA (excluding a smaller fraction
of models in the relevant bins). In the future we will examine how including
the new 8 TeV LHC analyses and our extrapolation to 14 TeV will enhance the
effectiveness of the LHC searches; we expect these improvements to be
substantial.

\section{Conclusions}\label{sec:conclusions}

We have described the primary methods of Indirect Detection including
observation of charged cosmic-ray antimatter, searches for high energy
neutrinos from annihilation in the sun, astrophysical signatures such as radio
and hard X-ray emission and gamma-ray measurements in the center of our own
Galactic halo, in nearby Dwarf galaxies and in clusters.   Just as the next
generation of direct detection experiments is approaching the natural
cross-section range, and reaching the limit of the technique (from the
irreducible background of pp and atmospheric neutrinos), the light (or 
perhaps dark) at the
end of the tunnel is now clearly in view for
gamma-ray searches as well.   Given the close relation of the gamma-ray
production cross-section to the total annihilation cross section, the bulk of
the likely parameter space for SUSY WIMPs (or, in fact, any weakly interacting
massive thermal relic) is within reach.   A future experiment like CTA, with
the U.S. enhancement (to increase sensitivity by a factor of 3, increase 
the field of view and enhance angular
resolution) could reach most of the majority of the
parameter space through observations of the GC for all but the most pessimistic
assumptions about the halo profile, or new astrophysical backgrounds. While the
most conservative estimates of sensitivity to Dwarf galaxies (assuming a boost
of one) will still fall short for the most conservative
halo models, a small boost in any one of these objects leads to the potential
for discovery of a clean signal.   If SUSY is not to be, and if some other
model for low-mass WIMPs prevails, antideuteron experiments or 
hard X-ray measurements may provide another
avenue for discovery.   The best future space-based instrument for these
searches is the continued operation of {\em Fermi} for at least 5 more years.  An
extension of {\em Fermi} leads to a linear (not square-root) improvement in the DM
limits with time, since the observations are currently signal not background
limited.   

\begin{figure}[!h]
\centerline{\includegraphics[width=3.5in]{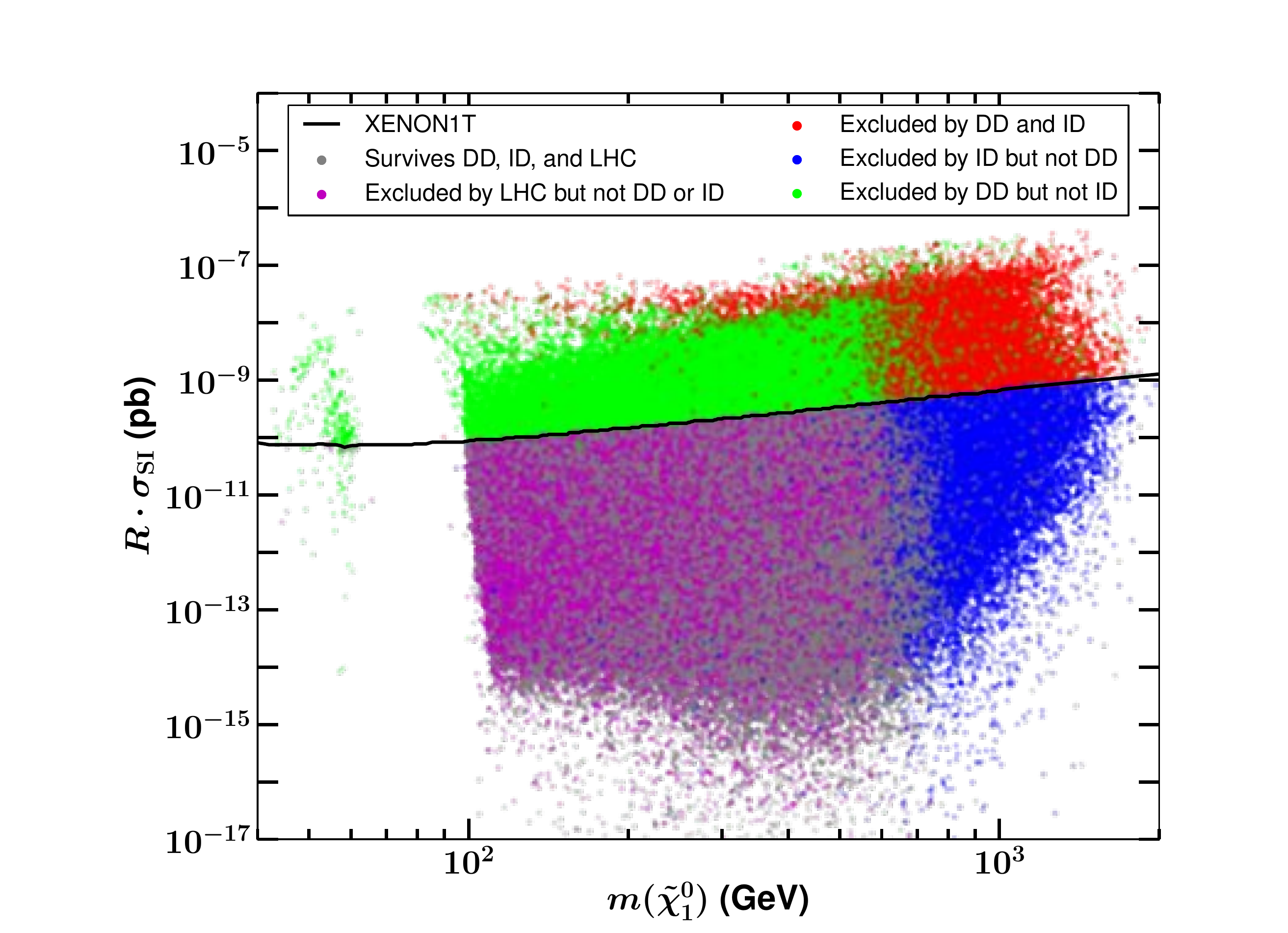}
\hspace{-0.50cm}
\includegraphics[width=3.5in]{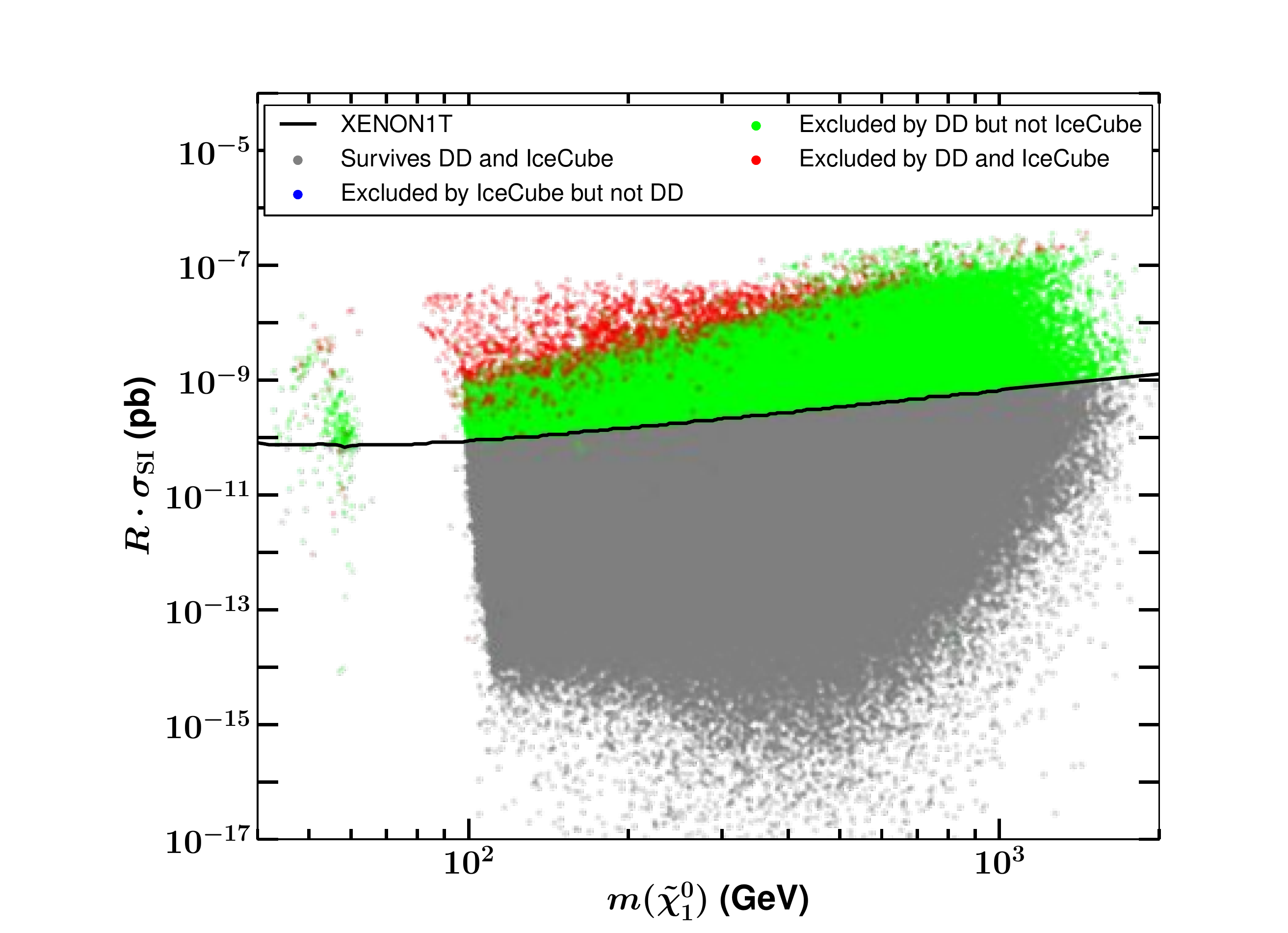}}
\vspace*{0.50cm}
\centerline{\includegraphics[width=3.5in]{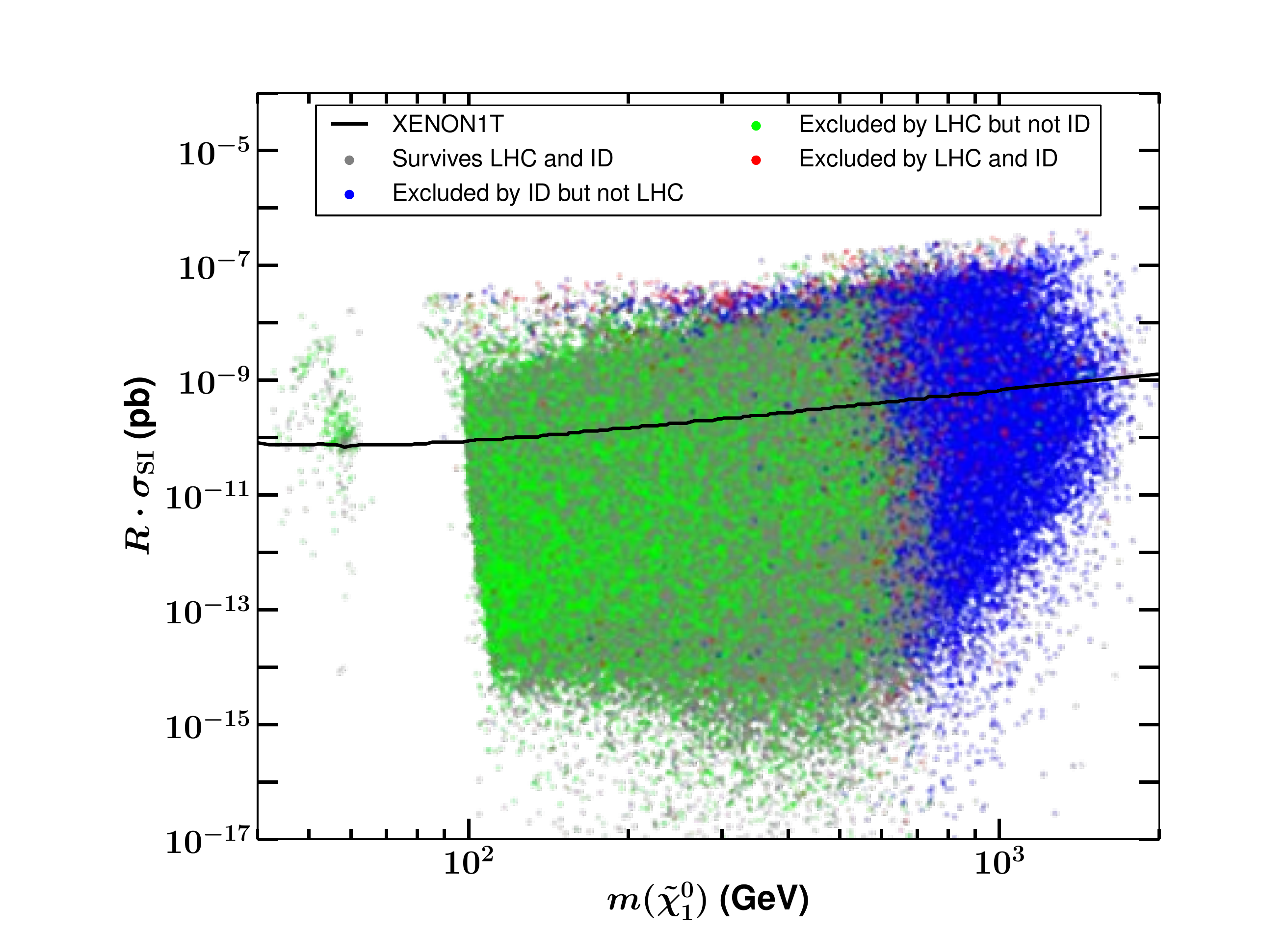}
\hspace{-0.50cm}
\includegraphics[width=3.5in]{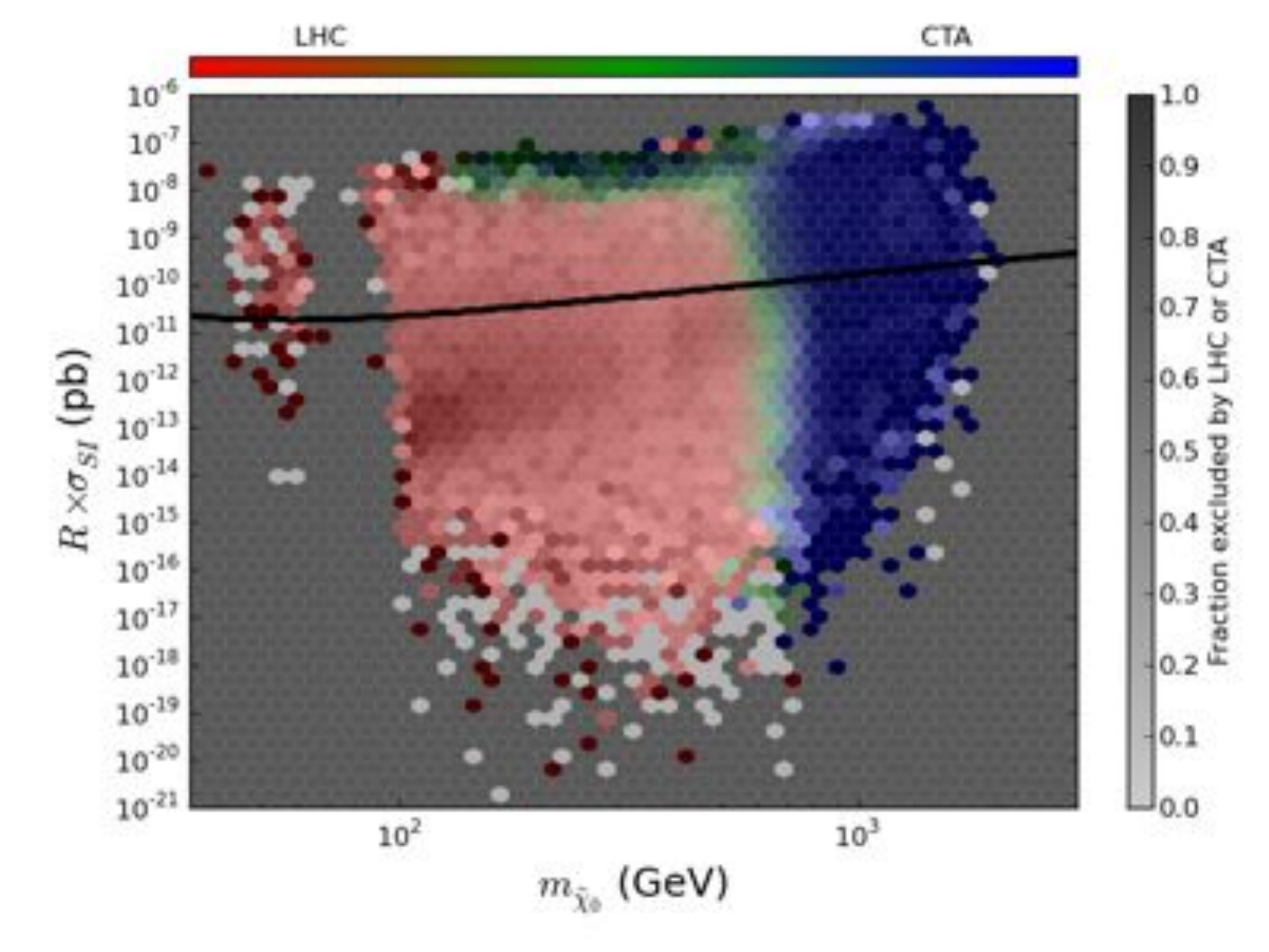}}
\caption{Comparisons of the models surviving or being excluded by the
  various searches in the LSP mass-scaled SI cross section plane as
  discussed in the text. The SI XENON1T line is shown as a guide to
  the eye.}
\label{figxx}
\end{figure}

Clearly, without indirect detection we are missing a key component of
a comprehensive dark matter program.  Perhaps it is important to
consider various scenarios for how we might finally solve the dark
matter problem to see just what we might be missing: Say that the G2
(or even G3) detectors (now limited by irreducible backgrounds from
diffuse neutrinos) see a hint of a signal in the form of an
exponential tail of events, suggesting a WIMP with a mass greater than
that of the recoil nucleus, but little more information.  And say that
the LHC continues to find no evidence for dark matter up to energies
of 1 TeV.  Would we have a discovery of dark matter?  But perhaps a
future gamma-ray experiment could observe a peculiar hard spectrum
that cuts off at several TeV both in the Galactic center, and in one
of a few of the most promising dwarf spheroidal galaxies (at precisely the
same energy, with the same spectral shape).  If this would not provide
the smoking gun detection of dark matter, wouldn't this provide a
great motivation to push the energy frontier to still higher energies,
and build a larger accelerator to definitively measure DM properties
in the lab?  Or, in another scenario, say that the LHC sees evidence
for missing energy or missing momentum, only weakly constraining the
mass of the new particle and with no clear link of this with the dark
matter - wouldn't it be desirable to build the biggest, most capable
gamma-ray experiment to identify this particle on the sky, improve the
mass measurement, and help identify the branching ratios through the
spectral shape?  Or perhaps the future upgrades of IceCube will result
in a highly significant signal of high energy neutrinos from the sun,
with energy that corresponds to the cutoff in the gamma-ray spectrum
of an extended halo around the galactic center.  Clearly this would be
both a solution to the dark matter problem, and would provide a number
of key measurements necessary to identify the particle properties and
to map out the inner halo, beyond the resolution limits of dynamical
measurements or N-body simulations.  One could go on, but the message
should be clear; indirect detection stands as one of the key pillars
on which the U.S. dark matter program should be constructed.

Solar WIMP searches are quite competitive with direct searches
\cite{icecube2013,superk2011,rott2012,bernal2012}, currently beating
out direct searches by orders of magnitude for spin-dependent
WIMP-proton interactions and $m_\chi \gsim 10$ GeV.  If large direct
experiments like COUPP go forward, solar WIMP neutrino searches will
cede some ground in the near term, but future upgrades to neutrino 
experiments (like PINGU) may once agian make these indirect searches
the most sensitive for probing spin-dependent cross sections.
Stacked Milky Way dwarf searches are already excluding highly
interesting regions of WIMP parameter space---in particular, at small
WIMP masses ($m_\chi \lsim 30$ GeV but above a few GeV, driven by
{\em Fermi}-LAT sensitivity).  Limits are likely to improve as {\em Fermi} runs
longer and more Milky Way dwarf galaxies are discovered in
the southern hemisphere with the Dark Energy Survey.  These analyses
typically assume a boost factor of 1, which is  generically a 
conservative choice.

\Acknowledgements

We are grateful to Jonathan Feng and Steve Ritz for coordinating the
Cosmic Frontier working group, and for all of the contributions from
individuals and Collaborations in the indirect detection community.

\bibliography{CF2/bibCF2}{}



\end{document}